\documentclass[12pt]{JHEP3}
\input epsf.tex

\usepackage{epsfig}
\usepackage{graphicx,amsmath,amssymb}
\usepackage{epsfig,multicol}
\allowdisplaybreaks

\usepackage{bbm,bm,amsmath,amssymb}
\def\blfootnote{\xdef\@thefnmark{}\@footnotetext}

\long\def\symbolfootnote[#1]#2{\begingroup%
\def\thefootnote{\fnsymbol{footnote}}\footnote[#1]{#2}\endgroup}

\newcommand{\be}{\begin{eqnarray}}
\newcommand{\ee}{\end{eqnarray}}
\newcommand{\ben}{\begin{eqnarray*}}
\newcommand{\een}{\end{eqnarray*}}

\newcommand{\bcent}{\begin{center}}
\newcommand{\ecent}{\end{center}}
\newcommand{\benum}{\begin{enumerate}}
\newcommand{\eenum}{\end{enumerate}}
\newcommand{\bdesc}{\begin{description}}
\newcommand{\edesc}{\end{description}}
\newcommand{\bitem}{\begin{itemize}}
\newcommand{\eitem}{\end{itemize}}
\newcommand{\bquote}{\begin{quote}}
\newcommand{\equote}{\end{quote}}
\newcommand{\bhalfp}{\begin{minipage}{0.45\textwidth}}
\newcommand{\ehalfp}{\end{minipage}}
\newcommand{\bhead}{\begin{center}\bf \Large}
\newcommand{\ehead}{\end{center}\bigskip}

\def\be{\begin{equation}}
\def\ee{\end{equation}}
\def\ba{\begin{eqnarray}}
\def\ea{\end{eqnarray}}

\newcommand{\roughly}[1]{\mathrel{\raise.3ex\hbox{$#1$\kern-0.85em
\lower1ex\hbox{$\sim$}}}}

\def\2pi{\left(2\pi\right)}

\def\beq{\begin{equation}}
\def\eeq{\end{equation}}
\def\bg{\begin{eqnarray}}
\def\nd{\end{eqnarray}}
\def\bea{\begin{eqnarray}}
\def\eea{\end{eqnarray}}

\def\D3{\overline{\mbox{D3}}}

\title{Non-Extremality, Chemical Potential and the Infrared limit of Large $N$ Thermal QCD}

\author{Mohammed Mia$^1$, Fang Chen$^2$, Keshav Dasgupta$^2$, Paul Franche$^2$, Sachindeo Vaidya$^3$\\
${}^1$ Department of Physics, Columbia University,\\ 538 West 120th Street,
New York, 10027, USA\\
${}^2$ Ernest Rutherford Physics Building, McGill University,\\ 3600 University
Street, Montr{\'e}al QC, Canada H3A 2T8\\
${}^3$ Centre for High Energy Physics, Indian Institute of Science,\\
Bangalore, 560012, India
\vskip.07in
{\tt fangchen, keshav, franchep@hep.physics.mcgill.ca, mm3994@columbia.edu, vaidya@cts.iisc.ernet.in}}
\date{January 2012}

\abstract{Non-extremal solution with warped resolved-deformed conifold background is important to study the infrared limit of large $N$
thermal QCD. Earlier works in this direction have not taken into account
all the back-reactions on the geometry, namely from the branes, fluxes, and black-hole carefully.
In the present work we make some progress in this direction by
solving explicitly the supergravity equations of motions in the presence of the backreaction from the black-hole.
The backreactions from the
branes and the fluxes on the other hand and to the order that we study, are comparatively suppressed.
Our analysis reveal,
among other things, how the resolution parameter would depend on the horizon radius and how the RG flows of the coupling constants
should be understood in these
scenarios, including their effects on the background three-form fluxes.
We also study the effect of switching on a chemical potential in the background and, in a particularly simplified scenario, compute the actual
value of the chemical potential for our case.}

\maketitle

\begin{document}

\section{Introduction}

Much of what is known about the phases of strongly coupled gauge
theories (and in particular, QCD) comes from a variety of
techniques, each of which accompanied by its attendant limitations.
Perturbative (i.e. weak coupling) computations can probe a large
part of the parameter space of the theory, like allowing one to deal
with varying number of colors $N$, flavors $N_f$. However, these
results are valid only at temperatures well above the deconfinement
temperature $T_c$, and at large values of the baryon number chemical
potential $\mu$ in order for the QCD coupling to be small, and thus
the perturbation valid. These exclusions put almost all of the
interesting region of the parameter space explored by RHIC data
beyond the reach of perturbative computations.

Lattice gauge theory, which provides a rigorous non-perturbative
starting point for QCD, is not without its limitations
 as well. It is difficult to incorporate realistic quark masses, and results from the traditional lattice simulations are limited to the regime near $T_c$, and $\mu$
 very small\footnote{Recent improvements in lattice simulations allow one to
access temperature as high as $5 T_c$, see for example the review
\cite{1203.5320}. We thank the referee for pointing this out to us.}.
Nonetheless, a combination of such conventional methods of analysis
(including insights from effective theories like chiral models)
suggest that the gauge theory possesses a {\it color superconductivity}
phase at asymptotically large value of the baryon number chemical
potential $\mu$. The literature is replete with conjectures for
the phase diagram of QCD in the ($T, \mu$) plane, especially for
small values of $T$ and large values of $\mu$ (see for example
\cite{kogutbook, 1203.5320}).

In recent years, there has been a considerable advance in
understanding the behavior of $U(N)$ gauge theories at finite
temperature using the gauge/gravity duality. That this development
is more than timely is beyond dispute, as the new and interesting
results from RHIC have provided a glimpse into a wide variety of
interesting phenomena arising in the strong coupling regime of QCD.
For instance, the quark-gluon plasma (identified as a new state of
matter) displays many properties of a fluid with low (shear)
viscosity, explanations for which are difficult to obtain from
traditionally available tools in perturbative QCD.

Many (perhaps most!) analytic results coming from gauge-gravity
duality are derived for gauge theories with ${\cal N}=4$
supersymmetry and with $N$ very large, and in the limit of the
theory possessing exact conformal symmetry. One may thus genuinely
be concerned about their applicability to QCD for which all these
are not true. Recent progress in this area, however, has provided us
with strong hints to overcome these limitations, and move towards
models of gauge-gravity duality that are not supersymmetric, and are
non-conformal (in a sense that will be made precise later).

The first set of models that managed to expand the original AdS/CFT construction
to incorporate renormalization group runnings are \cite{9904017, 9906194} that connected conformal fixed points at IR and UV, and \cite{9909047} that connected the 
UV ${\cal N} = 4$ conformal fixed point to a ${\cal N} = 1$ confining theory. The next set of models, that we would be mostly interested in, do not have any fixed points
(or fixed surfaces) in the paths of the RG flows. The key example in this set  
is the Klebanov-Strassler (KS)
model \cite{Klebanov:2000hb} (with an extension by Ouyang \cite{Ouyang:2003df} to incorporate fundamental matters)
that provided an IR dual of, although not exactly QCD, but
at least its closest cousin: large $N$ supersymmetric QCD. The UV of the original
Klebanov-Strassler model is now known to have some issues, like the divergences of the
Wilson loops at high energies, and additional Landau poles once fundamental matters have been introduced. This means that
UV completion is necessary, and to have the full gravity dual of the corresponding gauge theory that behaves well at
high energies, the KS geometry has to be augmented by a proper asymptotic manifold.

Other extensions to the original KS model quickly followed. For example in \cite{KT-non-ex, thorimal} the cascading picture of the original KS model
was extended to incorporate black-hole without any fundamental matter, which was then further extended to incorporate matter in \cite{cotrone}.
However none of the above models actually considered the full UV completion as most of the analysis of these works were directed towards
unravelling the IR physics. Therefore issues like UV divergences of Wilson loops and Landau poles were not investigated.

In a series of works \cite{Mia:2009wj, jpsi1, Mia:2010zu, Mia:2011iv} done over the last couple of years, we tried to address these concerns. Our
aim therein was to incorporate the backreactions from the black-hole, fluxes, and branes consistently so as to have a well defined UV completion
that not only allow us to get rid of all the poles etc., but also give us a model that could come {\it closest} to what we
might have expected from a large $N$ thermal QCD. We did manage to at least successfully generate such a UV completed dual picture, but many of the
backreactions turned out to be too difficult to incorporate fully. One aim of this paper is to make progress in this direction. In sec. \ref{chap2.1}
we will show how exactly to incorporate the backreactions from the black-hole in IR regime of our theory to lowest orders in string
coupling, and color-to-flavor ratio. Interestingly, to this order, the backreactions from branes and fluxes could be consistenly ignored.
We will demonstrate this
in sec. \ref{chap2.1.1}, and relevant EOMs will be solved in sec. \ref{chap2.1.2} and in {\bf Appendix} \ref{bgcomp}.

Although to the order that we study the IR regime of our theory allows us to ignore the backreactions of the fluxes, we will in-fact work out the
detailed fluxes in sec. \ref{actfluxes}. The backreactions from the black hole and the flavor seven-branes will be fully incorporated in the
fluxes. It will also be clear from sec. \ref{actfluxes} as to how these backreactions conspire to make the three-form fluxes non-ISD.

One persistent problem associated with thermal QCD is the interpretation of beta function of the theory. There is a long history on the subject
starting with the classic work by Collins and Perry \cite{colperry}. In sec. \ref{chap2.3} we discuss briefly how we should interpret the running of
couplings from the gravity dual. Our work strongly suggest one framework, although alternative interpretation could possibly be entertained.

We end sec. \ref{chap2} by taking a short detour. In sec. \ref{chap2.4} we study the dipole and non-commutative deformations of the seven-brane theory
on the gravity side. This detour is not without its merit. Dipole deformations are known to {\it increase} the masses of the KK states, so that a dipole-deformed
theory will have a slightly different KK spectra. We again make very brief speculations of the underlying physics, leaving most of the details for
future work.

Thus sec. \ref{chap2} prepares us with a backreacted metric, and with backreacted fluxes. One may now go on with this to fill up rest of the
missing steps left in \cite{Mia:2009wj, jpsi1, Mia:2010zu, Mia:2011iv}. These issues however will be addressed in future works. Here we aim for more
modest return. In sec. \ref{chap3} we present a small computation on the chemical potential. The reason for choosing this computation over other
possible interesting ones is two-fold. One: its simple enough computation that carries sufficiently interesting physics, and two: we use this to
show in sec. \ref{chap3.2} how an alternative way of getting the chemical potential, via say duality chasing, may be inherently flawed. In sec. \ref{chap3.3}
we compute the chemical potential for our model. In {\bf Appendices} \ref{TBT}, and \ref{funJ} we speculate more on the duality chasing techniques. We
conclude in sec. \ref{cono} with some discussions.

\section{Analysis of the background \label{chap2}}

Before discussing the details of the UV complete dual geometry, let us first review the Klebanov-Tseytlin (KT), Klebanov-Strassler (KS) and
Ouyang-Klebanov-Strassler (OKS) geometry without the black hole. The supergravity description arises as the low energy limit of brane excitations
placed in conifold geometry. In particular, the Klebanov-Tseytlin geometry arises from the following brane configuration:
 Embed $N$ D3 branes and $M$ D5 branes in ten dimensional manifold
with the metric 
\bg
ds_{10}^2=-dt^2+d\overrightarrow{x}^2+ds_6^2
\nd
where $ds_6^2$ is given by 
\bg \label{conemt}
ds^2_{6}=dr^2+r^2 ds^2_{T^{1,1}}
\nd
The metric of the base $T^{1,1}$ is given by 
\bg \label{mt11}
ds^2_{T^{1,1}}=\frac{1}{9}\left[d\psi +\sum_{i=1}^2 {\rm cos}\theta_i d\phi_i\right]^2
+\frac{1}{6}\sum_{i=1}^2\left[d\theta_i^2 +{\rm sin}^2\theta_1 d\phi_i^2\right]
\nd
That is we have four dimensional Minkowski space along with six dimensional conifold. The D3 branes live in the flat four
dimensional space and and are placed at the tip of the conifold at fixed radial location $r=0$. The D5 branes wrap the shrinking two cycle 
$S^2$ at the tip of the conifold  and extend 
in four Minkowski directions.  

The excitations of the massless open strings ending on these D branes are described by 
gauge fields and complex matter fields $A_i,B_i$, $i=1,2$ which
transform as bi-fundamental fields under the gauge group $SU(N+M)\times SU(N)$. Note that the  matter fields $A_1, A_2$ transform 
under global $SU(2)$ and so do $B_1,B_2$ under another $SU(2)$ and we also have global $U(1)$ phase rotation. Thus we have
$SU(2)\times SU(2)\times U(1)$ global symmetry, which is also the symmetry of the conifold.This is not surprising as these
fields describe motion of the D branes \cite{Witten-DBExc} and  the branes move in the
conifold direction. Thus the fields $A_i, B_i$ are really coordinates of the conifold

At the lowest energies, the entire setup of branes and their interaction with the gravitons can be captured by supergravity with only fluxes
and metric and {\it no branes}. This geometry arising from supergravity is referred to as the {\it dual geometry}. For the brane configuration just described, the dual geometry
is the warped regular cone with the following metric

\bg\label{KT_sol}
ds^2 &=& {1\over \sqrt{h_{\rm KT}}}
\left(-dt^2+dx^2+dy^2+dz^2\right)+\sqrt{h_{\rm KT}}\Big[dr^2+r^2 ds^2_{T^{1,1}}\Big]\nonumber\\
h_{\rm KT}&=&\frac{L^4}{r^4}
\Bigg[1+\frac{3g_sM^2}{2\pi N}{\rm log}r\Bigg]
\nd
The above warped geometry is known as the Klebanov-Tseytlin (KT) solution \cite{Klebanov:2000nc}. Right away, one observes that the warp factor $h_{\rm KT}$ becomes
negative for small $r$ and the geometry is not well defined. In fact classical gravity description breaks down
for small $r$ and the solution (\ref{KT_sol}) is only valid for large $r$. 

Since $r$ is related to the energy scale of the gauge theory, to
understand what happens for small $r$, we can look at the IR limit of the gauge group $SU(N+M)\times SU(N)$. If $N=kM$, $k$ is a natural
number, then at the IR, the gauge theory
cascades down to $SU(M)$ under a Seiberg duality cascade. At the IR, the gauge theory develops a non-perturbative superpotential and the
vacuum solution gives non-trivial expectation values for the gauge invariant operator $N_{ij}=A_i B_j, i,j=1,2$. This means ${\rm det}
N_{ij}=\epsilon^2 \neq 0$.
Since $A_i,B_j$ are also cone coordinates, ${\rm det} N_{ij}=\epsilon^2$  also gives the cone embedding equation. However, $\epsilon\neq 0$
just means we no longer have a regular cone, but a deformed cone. This way the field theory analysis indicates that the dual geometry must be a
deformed cone. 

Thus to resolve the small $r$ singularity of the metric (\ref{KT_sol}), we must replace the warped regular cone with the {\it deformed} warped
cone. This is the essence of the Klebanov-Strassler (KS) proposal \cite{Klebanov:2000hb} and the dual geometry of the warped deformed cone has the following metric:
\bg\label{metd}
ds^2 &=&\frac{1}{\sqrt{h_{\rm KS}}}
\Big[-dt^2+dx^2+dy^2+dz^2\Big] +\sqrt{h_{\rm KS}}\;\bar{g}_{mn} dx^m dx^n\nonumber\\
h_{\rm KS}(\rho)&=&\alpha \frac{2^{2/3}}{4}\int_{\rho}^\infty dx \frac{x\; {\rm coth}x-1}{{\rm sinh^2x}}\left({\rm sinh}(2x)-2x\right)^{1/3}
\nd
where $\alpha={\cal O}(g_s^2M^2)$ and $\bar{g}_{mn}$ is the metric of the deformed cone
\bg\label{inmateda}
&&\bar{g}_{mn} dx^m dx^n ~ = \frac{1}{2}a^{4/3} K(\rho)\Big[\frac{1}{3K^3(\rho)}\left(d\rho^2+(g^5)^2\right)+{\rm
cosh}^2\left(\frac{\rho}{2}\right)\left[(g^3)^2+(g^4)^2\right]\nonumber\\
&&+ {\rm
sinh}^2\left(\frac{\rho}{2}\right)\left[(g^1)^2+(g^2)^2\right]\Big]
\nd
where $a$ is a constant, $g^i,i=1,..,5$ are one forms given by

\bg\label{oneforms}
&&g^1=\frac{e^1-e^3}{\sqrt{2}},~~~~g^2=\frac{e^2-e^4}{\sqrt{2}}\nonumber\\
&&g^3=\frac{e^1+e^3}{\sqrt{2}},~~~~g^4=\frac{e^2+e^4}{\sqrt{2}},~~~g^5=e^5\nonumber\\
&& e^1\equiv-{\rm sin}\theta_1 \;d\phi_1, ~~~~ e^2\equiv d\theta_1\nonumber\\
&& e^3\equiv {\rm cos}\psi \;{\rm sin}\theta_2 \;d\phi_2-{\rm sin}\psi \;d\theta_2,\nonumber\\
&& e^4\equiv {\rm sin}\psi \;{\rm sin}\theta_2\; d\phi_2+{\rm cos}\psi\; d\theta_2,\nonumber\\
&& e^5\equiv d\psi +{\rm cos}\theta_1\; d\phi_1+{\rm cos}\theta_2 \;d\phi_2
\nd
and $K(\rho)$ is defined as:
\bg
K(\rho)=\frac{\left({\rm sinh}(2\rho)-2\rho\right)^{1/3}}{2^{1/3}{\rm sinh}\rho}
\nd   
For $\rho$ large, we can make the following transformation  $r^3\sim a^2 \;e^\rho$. Then one obtains that the metric (\ref{metd}) and
(\ref{KT_sol}) become identical as $r$ becomes very large. While KT solution has singularities at small $r$, KS geometry is regular for all
radial distances and becomes KT geometry for large radial distance. 

\vspace{1cm}
\begin{table}
\begin{tabular}{|l|p{1.5in}|p{1.3in}|p{1.2in}|} \hline
\em Field &$SU(N+M)\times SU(N)$ &$SU(N_f)\times SU(N_f)$ &$SU(2)\times SU(2)$\\\hline
$q$ &({$\bf N+M$, 1})&({$\bf N_f$, 1})&({\bf 1,1})\\
$\tilde{q}$ &($\overline{\bf N+M}, 1$)&({1,$\bf N_f$})&({\bf 1,1})\\
$Q$ &(1,${\bf N+M}$)&($\overline{\bf N_f}$,1)&({\bf 1,1})\\
$\tilde{Q}$ &(1,$\overline{\bf N+M}$)&($1,\overline{\bf N_f}$)&({\bf 1,1}) \\
$A_{1,2}$ &(${\bf N+M}$,$\overline{\bf N+M}$ )&($\overline{\bf N_f}$,$\bf N_f$)&({\bf 2,1})\\
$B_{1,2}$ &($\overline{\bf N+M}$,${\bf N+M}$ )&($\bf N_f$,$\overline{\bf N_f}$)&({\bf 1,2})\\\hline
\end{tabular}
\caption{{The field content and their representation under symmetry groups.}}
\end{table}  
In both KS and KT solutions, there are no fundamental matter. To introduce fundamental matter, one has to embedd D7 branes in the geometry and
compute the backreaction of the axio-dilaton field sourced by the D7 branes. For holomorphic embedding of D7 branes in KT geometry or
equivalently embedding D7 branes in large $r$ regime of KS geometry, the effect of the axio-dilaton field on the metric was computed by
Ouyang \cite{Ouyang:2003df}. The resulting Ouyang-Klebanov-Strassler (OKS) metric up to linear order in $g_sN_f$ is,   
\bg\label{OKS_sol}
ds^2 &=& {1\over \sqrt{h_{\rm OKS}}}
\left(-dt^2+dx^2+dy^2+dz^2\right)+\sqrt{h_{\rm OKS}}\Big[dr^2+r^2 ds^2_{T^{1,1}}\Big]\nonumber\\
h_{\rm OKS}&=&\frac{L^4}{r^4}\Bigg[1+\frac{3g_sM^2}{2\pi N}{\rm log}r\left\{1+\frac{3g_sN_f}{2\pi}\left({\rm
log}r+\frac{1}{2}\right)+\frac{g_sN_f}{4\pi}{\rm log}\left({\rm sin}\frac{\theta_1}{2}
{\rm sin}\frac{\theta_2}{2}\right)\right\}\Bigg]\nonumber\\
\nd    
In addition to the bi-fundamental fields $A_i,B_i$, introduction of the D7 branes give rise to flavor symmetry group $SU(N_f)\times SU(N_f)$ and  matter fields
$q,\tilde{q},Q,\tilde{Q}$
which transform as fundamental under the gauge group $SU(N+M)\times SU(N)$ \cite{Ouyang:2003df}. In Table {\bf 1.1}, we list the various
matter fields and their representation under local and global symmetry groups for the OKS model.

Therefore with a clear understanding of the distinction between KT, KS and OKS geometry, let us come back to the model that we studied in 
\cite{Mia:2009wj}. The IR physics is captured by the
Ouyang-Klebanov-Strassler-black-hole
(OKS-BH) geometry, namely, the small $r$ physics is determined by a warped resolved-deformed
conifold with fluxes, seven-branes and a black hole in the ten-dimensional spacetime.
On the other hand the UV physics is conformal, and is captured by an asymptotically AdS
geometry with fluxes and seven-branes.

As discussed in \cite{jpsi1}, these two geometries, namely the asymptotic AdS and OKS, can be connected by an
intermediate configuration with brane sources and fluxes. These branes sources were elaborated in details in \cite{jpsi1},
although many coefficients in the background geometry were left undetermined therein.
In the following we will fill up some of these missing steps.

Let us begin with the basic ansatze for the metric in the three regions. For all the three regions we assume that the
radial coordinate $r$ spans $b < r < r_{\rm min}$ for Region 1 where we expect all the confining dynamics to take place;
$r_{\rm min} < r < r_o$ for the intermediate region
called Region 2; and $r_o < r < \infty$ for Region 3 which captures the asymptotically conformal region. The minimum
radius $r = b$, which signifies the cut-off coming from the
blown-up $S^3$ (as well as $S^2$, although for most of the calculations in this paper we will only consider a warped resolved conifold instead of
a warped resolved-deformed conifold),
maps to the expectation of the gluino condensates of the dual gauge theory at zero temperature.
Considering all these regions, the non-extremal metric takes
the following form:
\bg\label{bhmet1}
ds^2 &=& {1\over \sqrt{h}}
\Big[-g_1(r)dt^2+dx^2+dy^2+dz^2\Big] +\sqrt{h}\Big[g_2(r)^{-1}dr^2+ d{\cal M}_5^2\Big]\nonumber\\
& \equiv & -e^{2A+2B} dt^2 + e^{2A} \delta_{ij} dx^idx^j + e^{-2A-2B} {\widetilde g}_{mn} dx^m dx^n
\nd
where $g_i(r)$ are the black-hole factors and we have taken $g_1 = g_2$, the components go as $i, j = 1, 2, 3$ and
$m, n = 4, ..., 9$, the warp factors $A, B$ are defined as:
\bg\label{abdef}
A ~= ~ -{1\over 4} {\rm log}~h,~~~~~ B ~ = ~  {1\over 2} {\rm log}~g_1
\nd
$d{\cal M}_5^2$ is typically
the metric of warped resolved-deformed conifold and $h$ is the warp factor that behaves differently in the three
regions as shown in \cite{jpsi1}.

Observe that in the extremal limit, $g_1=g_2 \approx 1$ and the extremal metric is dual to the low temperature confining phase of the gauge theory. To
see this, note that in
the absence of any seven branes, Region 1 of the geometry of \cite{jpsi1} in the extremal limit is identical to the IR geometry of
Klebanov-Strassler (KS) model \cite{Klebanov:2000hb}. If seven branes are placed far away from Region 1, that is $r_{\rm min}\gg b$, we can neglect
their back-reactions and consider the axion-dilaton
field to be effectively constant as in  \cite{Klebanov:2000hb}. Hence in the extremal limit, Region 1 of \cite{jpsi1} is identical to the IR
region of KS which, in turn,
is dual to the low temperature confining phase of the $SU(M)$ gauge theory wherein chiral symmetry is broken. The extremal
geometry can incorporate temperature of the field theory once we analytically continue to Euclidean signature with $it\rightarrow \tau$ and
impose periodic and anti-periodic boundary conditions for the bosons and fermions on the closed time circle. Furthermore, in
extremal case the entropy will vanish. This is expected as the entropy from
the dual geometry arises from the fluxes which are at least ${\cal O}(N_{\rm eff})$, where $N_{\rm
eff}$ is effective  brane charge. As the deformed cone represents confinement of charge, we expect  to get $N_{\rm eff}=0$ from the dual
geometry.  This is indeed what happens as energy scale for a thermal field theory is set by the temperature and at low temperature,
only the IR degrees of freedom are
excited. This means in the dual geometry, all we need is the region near $r\sim b$ of the deformed cone $-$ but in this region
the five-form flux vanishes \cite{Klebanov:2000hb}
and we get $N_{\rm eff}=0$.

As the temperature is increased, we expect that the non-extremal solution will have less free energy than the extremal solution,
just as in the case for the AdS-black
holes \cite{Witten:1998zw}, and Hawking-Page phase transition will take place \cite{{Hawking:1982dh}}.
The
focus of this work will be to analyze the non-extremal solution which is dual to the deconfined
phase of large $N$ thermal QCD,
while a  detailed analysis of phase transitions will be presented in
a follow up paper\cite{fangmiamik}.

The non-extremal solutions we present in this paper are precisely dual to the high temperature
regime of the gauge theory $-$ where chiral symmetry is restored and the light degrees of freedom are deconfined. However, heavy quarkonium
states arising
from the seven branes placed in the UV region can coexist with the chirally symmetric phase above the deconfinement temperature.
But as temperature is raised even further, the heavy quarkonium states will eventually melt \cite{Rey:1998bq, Mia:2010zu}.

For both extremal and non-extremal cases, typically $h$ would have logarithmic factors in Region 1
whereas it would have inverse $r$ behavior in Region 3. In the intermediate region, the warp factor will
typically have both the logarithmic and the inverse $r$ behavior. Therefore to summarise, the background should
satisfy the following properties:

\vskip.1in

\noindent $\bullet$ Fluxes are non imaginary self-dual i.e non-ISD, and become ISD once the black-hole factors $g_i$
in the metric
are removed. Therefore the deviation for ISD property is proportional to the horizon radius $r_h$.

\vskip.1in

\noindent $\bullet$ The gravity dual of the deconfined phase is given by resolved warped-deformed conifold with a black-hole. In the limit
where the deformation parameter is small, the background can be succinctly captured by a resolved conifold with fluxes
and black hole.

\vskip.1in

\noindent $\bullet$ The resolution parameter is no longer constant because of the various back-reactions. In fact the
resolution parameter becomes function of $r_h/r$ as well as $g_sN_f$, and $g_s M^2/N$
where $g_s$ is the string coupling, $M$ is the number of bi-fundamental matter, $N$ is the number of colors, and $N_f$ is the
number of fundamental flavors.

\vskip.1in

\noindent {}From the above set of arguments, we can use the following ansatze for the internal metric:
\bg\label{inmate}
&&{\widetilde g}_{mn} dx^m dx^n ~ = ~dr^2 + r^2 e^{2B} \Big[A(d\psi + {\rm cos}~\theta_1 d\phi_1 +
{\rm cos}~\theta_2 d\phi_2)^2 + {\cal O}(g_sM^2/N, r_h^4/r^4) \nonumber\\
&&~~~~~~~~~~~+ B(d\theta_1^2 + {\rm sin}^2\theta_1 d\phi_1^2) +
{1\over 6}(1 + F)(1 + {\cal G})\left(\frac{d\theta_2^2}{1 + {\cal G}} + {\rm sin}^2\theta_2 d\phi_2^2\right)\Big]\\
&& ~~~~~~ + 2f_b\Big[{\rm cos}~\psi(d\theta_1d\theta_2 + {\rm sin}~\theta_1~{\rm sin}~\theta_2 d\phi_1 d\phi_2) -
{\rm sin}~\psi({\rm sin}~\theta_2 d\phi_2 d\theta_1 - {\rm sin}~\theta_1 d\phi_1 d\theta_2)\Big]\nonumber
\nd
where we will only consider the resolved conifold limit, with $F$ being related to the resolution parameter (whose value
will be determined later). In other words, we take:
\bg\label{wetake}
&&f_b ~ \to ~ 0, ~~~~~~~~ F ~ \equiv ~ {6a^2\over r^2}, ~~~~~~~~ {\cal G} ~ \to ~ 0\nonumber\\
&&A ~ = ~ {1\over 9} + {\cal O}(g_sM^2/N, r_h^4/r^4), ~~~~~ B ~ = ~ {1\over 6} + {\cal O}(g_sM^2/N, r_h^4/r^4)
\nd
where the numerical factor of $6$ is inserted to bring certain expressions in a better format. As we will see, this
$F$ (or equivalently $a$) determines the {squashing} factor {\it between} the two spheres, and we can consistently keep the
second squashing factor, ${\cal G}$, to be zero.

The resolution parameter discussed above needs a bit more elaboration. First of all, as we mentioned earlier,
 $a^2$ is not a constant in our model. As we will show in \eqref{respagi},
the resolution parameter takes the following form:
\bg\label{resupagi}
a^2 ~ = ~ a_0^2 + r_h^2 {\cal O}(g_sM^2/N) + r_h^4 {\cal O}(g_s^2M^2N_f/N)
\nd
where we have switched on a {\it bare} resolution parameter $a^2_0$ to allow for the theory to have a baryonic branch \cite{seiberg}. However even if we switch off
the bare resolution parameter, the background EOMs will still generate a resolution parameter proportional to the horizon radius $r_h$. This not a contradiction with the
result of \cite{candossa} wherein it was argued that one may not be able to simultaneously resolve and deform a Calabi-Yau cone. The fact that our metric is
non-K\"ahler takes us away from the constraints imposed in \cite{candossa}.

In the following section we will argue for these parameters and their dependences on
the horizon radius by analysing the non-extremal limit of the warped resolved-deformed conifold background\footnote{We will continue
calling this background as the Klebanov-Strassler background as they all fall in the same class of supergravity solution.}.

\subsection{Derivation of the non-extremal BH solution for the Klebanov-Strassler model \label{chap2.1}}

We first compute the non-extremal metric arising from Type IIB
supergravity action given, in the notations of \cite{Giddings:2001yu}, in the following way\footnote{Although in this section we will use the Einstein frame to express the
metric, we will however not distinguish between the two frames in later sections because the dilaton will be considered constant, unless mentioned otherwise.}:
\bg\label{tubra}
S_{\rm IIB} &=& \frac{1}{2\kappa_{10}^2}\int d^{10}x\;
\sqrt{-g}\Bigg[R-\frac{\partial_a \tau \partial^a \bar\tau}{2|{\rm
Im}\tau|^2}-
\frac{G_3\cdot\bar{G}_3}{12{\rm Im}\tau}-\frac{\widetilde{F}_5^2}{4\cdot 5!}\Bigg]\nonumber\\
&+& \frac{1}{8i\kappa_{10}^2}\int \frac{C_4\wedge G_3 \wedge
\bar{G_3}}{{\rm Im}\tau } +S_{\rm loc}
\nd
where $S_{\rm loc}$
is the action for all the localized sources in ten dimensional
geometry i.e five-branes and seven-branes mostly from Region 2 onwards.
Our aim is to re-analyse the non-extremal Klebanov-Strassler solution. Recall that
for
Klebanov-Tseytlin model the non-extremal solutions were analyzed in \cite{KT-non-ex},
while in \cite{Mia:2009wj} there have been studies of gravity duals of finite temperature
cascading gauge theory with fundamental matters\footnote{See also \cite{cotrone-et-al} where somewhat similar analyses
were also done.}.
However in \cite{Mia:2009wj} precise background fluxes and the warp factors taking into the
backreactions of the BH geometry were only conjectured. Here we
 will derive the non-extremal metric dual to a UV complete gauge theory that mimics features of large N QCD at the
lowest energies, justifying the proposals made in \cite{Mia:2009wj, jpsi1}. One immediate outcome of this would
be the verification of the conjectured dependence of the resolution parameter $a^2$ on the horizon radius $r_h$.

 Our ansatz for the metric is \eqref{bhmet1}.
 We look for solutions with
regular Schwarzschild horizon at $r=r_h$. This is achieved by imposing
$e^{B(r_h)}=0$ and considering solutions to $A$ such that
$e^{A(r_h)}\neq 0$, which guarantees a non-singular horizon
\cite{KT-non-ex}. By solving the Einstein equations along with the
flux equations with these boundary conditions, we will find the
non-extremal solutions with regular horizons.

Observe that we have warped Minkowski four directions, a
non-compact radial direction $r$ and a compact five manifold ${\cal
M}_5$. The back reactions of the fluxes $G_3, \widetilde{F}_5$ and
axion-dilaton field $\tau$ will modify the warp factor $e^{A+B}$
while $\widetilde{g}_{mn}$ will be altered due to the presence of a black hole
and the
various sources. In particular $\widetilde{g}_{mn}$ will be a warped resolved-deformed conifold with a bare resolution parameter $a_0$.
Note however that only the warp factor $e^{A+B}$ will be
essential to analyze the confinement/deconfinement mechanism for the
boundary field theory \cite{jpsi1}. The linear confinement of quarks
and the string breaking mechanism which eventually describes the
deconfinement of $Q\overline{Q}$ pair, is only sensitive to the warp
factor.
The exact solutions for the internal metric in the non-extremal limit taking into account the
back reaction of the various fluxes is not essential to study free energy of
the $Q\overline{Q}$ pair. Nevertheless we will find the exact form of the
internal metric up to linear order in resolution function $F$.

We restrict to fluxes and axion-dilaton field $\tau$ which only
depend on $x^m$ and not on the Minkowski coordinates $x^\mu$. Then
the Einstein equations can be written as
\bg \label{ricci_T}
R_{\mu\nu}&=&-g_{\mu\nu} \left[\frac{G_3 \cdot \bar{G_3}}{48\; {\rm
Im}\tau}+\frac{\widetilde{F}_5^2}{8\cdot 5!}\right]+\frac{\widetilde{F}_{\mu
abcd}\widetilde{F}_\nu^{\;abcd}}{4 \cdot 4!}
+\kappa_{10}^2 \left(T_{\mu\nu}^{\rm loc}-\frac{1}{8} g_{\mu\nu} T^{\rm loc}\right)\nonumber\\
R_{mn}&=&-g_{mn} \left[\frac{G_3 \cdot \bar{G_3}}{48 \;{\rm
Im}\tau}+\frac{\widetilde{F}_5^2}{8\cdot 5!}\right]+\frac{\widetilde{F}_{m
abcd}\widetilde{F}_n^{\;abcd}}{4 \cdot 4!}+\frac{G_m^{\;bc}\bar{G}_{nbc}}{4\;{\rm Im}\tau}
+\frac{\partial_m \tau \partial_n \bar\tau}{2\;|{\rm Im}\tau|^2}\nonumber\\
&+&\kappa_{10}^2 \left(T_{mn}^{\rm loc}-\frac{1}{8} g_{mn} T^{\rm
loc}\right) \nd
where $\widetilde{F}_5$ is given by the following self
dual form
\bg \label{5form} \widetilde{F}_5=(1+\ast_{10})d\alpha\wedge
dx^0\wedge dx^1\wedge dx^2\wedge dx^3 \nd with $\alpha = e^{4A}$
and $T^{\rm loc}$ being the
trace of \bg T^{\rm loc}_{ab}=-\frac{2}{\sqrt{-g}}\frac{\delta
S_{\rm loc}}{\delta g^{ab}} \nd
Using the form  of the five-form flux (\ref{5form}), the first
equation in (\ref{ricci_T}) becomes
\bg\label{Ricci_Min}
R_{\mu\nu}&=&-g_{\mu\nu}
\left[\frac{G_3 \cdot \bar{G_3}}{48\; {\rm Im}\tau}+\frac{e^{-8A-2B}
\partial_m\alpha \partial^m\alpha}{4}\right]+\kappa_{10}^2
\left(T_{\mu\nu}^{\rm loc}-\frac{1}{8} g_{\mu\nu} T^{\rm loc}\right)
\nd
On the other hand, the Ricci tensor in the Minkowski direction takes
the following simple form
\bg \label{ricci_Min}
R_{\mu\nu}&=&-\frac{1}{2} \left[\partial_m (g^{mn} \partial_n
g_{\mu\nu})+ g^{mn} \Gamma^M_{nM}\partial_m g_{\mu\nu}-
g^{mn}g^{\nu'\mu'}\partial_m g_{\mu'\mu}
\partial_n g_{\nu'\nu}\right]
\nd
where $\nu',\mu'=0,..,3$ and $\Gamma^M_{nM}$ is the Christoffel
symbol. Now using the ansatz (\ref{bhmet1}) for the metric,
(\ref{ricci_Min}) can be written as \bg\label{ricci_Min_a}
R_{tt}&=&e^{4(A+B)}\left[\widetilde{\triangledown}^2(A+B)-3\widetilde{g}^{mn}\partial_nB\partial_m(A+B)\right]\nonumber\\
R_{ij}&=&-\eta_{ij}
e^{2(2A+B)}\left[\widetilde{\triangledown}^2A-3\widetilde{g}^{mn}\partial_nB\partial_mA\right]
\nd where we have defined the Laplacian as:
\bg\label{laplabe}
\widetilde{\triangledown}^2=\widetilde{g}^{mn}\partial_m\partial_n
+\partial_m\widetilde{g}^{mn}\partial_n+\frac{1}{2}\widetilde{g}^{mn}\widetilde{g}^{pq}\partial_n\widetilde{g}_{pq}
\partial_m
\nd
The set of equations can be simplified by taking the trace of the first equation in (\ref{ricci_T}) and using
(\ref{ricci_Min_a}). Doing this we get
\begin{eqnarray} \label{warp_eq_1}
\widetilde{\triangledown}^2(4A+B)-3\widetilde{g}^{mn}\partial_nB\partial_m(4A+B)
&=&
e^{-2A-2B}\frac{G_{mnp}\bar{G}^{mnp}}{12\textrm{Im}\tau}+e^{-10A-4B}\partial_m\alpha\partial^m\alpha
\nonumber\\
&&+\frac{k^2_{10}}{2}e^{-2A-2B}(T^m_m-T^{\mu}_{_\mu})^{loc}
\end{eqnarray}
 On the other hand  using
(\ref{Ricci_Min}) in (\ref{ricci_Min_a}), one gets
\bg \label{BHfactorA1}
R_t^t-R_x^x=0
\nd
which in turn would immediately imply
\bg\label{BHfactorA}
\widetilde{\triangledown}^2 B-3\widetilde{g}^{mn} \partial_m B \partial_n B=0
\nd
Minimizing the action (\ref{tubra}) also gives the  Bianchi identity for the five-form flux, namely \bg
\label{bianchi5} d\widetilde{F}_5= H_3\wedge F_3+2\kappa_{10}^2  T_3
\rho_3 \nd where $\rho_3$ is the D3 charge density from the
localized sources \cite{Giddings:2001yu}. Using (\ref{5form}) in
(\ref{bianchi5}) and subtracting it from (\ref{warp_eq_1}) one gets the following
\begin{eqnarray}\label{GKP_BH}
\widetilde{\triangledown}^2(e^{4A+B}-\alpha)&=&\frac{e^{2A-B}}{6\textrm{Im}\tau}|{i}
G_3- \ast_6G_3|^2+e^{-6A-3B}|\partial (e^{4A+B}-\alpha)|^2\nonumber\\
&&+3e^{-2A-2B}\partial_mB\partial^m(e^{4A+B}-\alpha)+\textrm{local
source}
\end{eqnarray}
The Ricci tensor, on the other hand, for the $x^m, m=4,..,9$ directions takes the following form
\bg \label{Rmn_fangA}
R_{mn}&=&\widetilde{R}_{mn} + \widetilde{g}_{mn} \widetilde{\triangledown}^2\left(A + B\right)-3\widetilde{g}_{mn}\widetilde{g}^{\lambda k}\partial_\lambda B
\partial_k
\left(A + B\right) \nonumber\\
&+& 3\widetilde{\triangledown}_m\partial_n B + \partial_mB\partial_nB
-8\partial_mA\partial_nA - 2\partial_{(m}A\partial_{n)}B
\nd
where $\widetilde{\triangledown}_m$ is the covariant derivative given by
\bg
\widetilde{\triangledown}_mV_c=\partial_m V_c-\widetilde{\Gamma}_{mc}^b V_b
\nd for any vector $V_b$.
Here $\widetilde{R}_{mn}$ is the Ricci tensor and $\widetilde{\Gamma}_{mc}^b$ is the Christoffel symbol for the metric $\widetilde{g}_{mn}$.
The equation for $\widetilde{R}_{mn}$ is given by:
\begin{eqnarray}\label{wideR}
\widetilde{R}_{mn}&=&-g_{mn}\frac{G_3\cdot\bar{G}_3}{24\textrm{Im}
\tau}+\frac{G_{mab}\cdot\bar{G}_n^{ab}}{4\textrm{Im}\tau}+\frac{\partial_m\tau
\partial_n\bar{\tau}}{2\mid\textrm{Im}\tau\mid^2}\nonumber\\
&&+\frac{F_{mabcd}F_n^{abcd}}{4\cdot 4!}+g_{mn}\frac{F_{\mu
abcd}F^{\mu abcd}}{16\cdot 4!}+8\partial_m A\partial _n A\nonumber\\
&&-3\widetilde{\triangledown}_m\partial_n B - \partial_mB\partial_nB
 + 2\partial_{(m}A\partial_{n)}B
\end{eqnarray}
which means, in general, this could lead to twenty different equations in six-dimensions (including another one for the trace). On the other hand
the equation of motion for $G_3$ can be expressed in terms of a seven-form $\Lambda_7 \equiv \ast_{10}G_3-iC_4\wedge G_3$ in the following way:
\begin{equation}\label{g3eom}
d\Lambda_7 +\frac{i}{\textrm{Im}\tau}d\tau\wedge \textrm{Re}\Lambda_7 ~ = ~ 0
\end{equation}
where typically $\Lambda_7$ would study the deviations from the ISD behavior. For example, using our metric ansatz we can express $\Lambda_7$ as
\bg\label{lambdaeqon}
\Lambda_7 ~ = ~ \left[e^{4A+B}\ast_{6}G_3-i\alpha G_3\right] \wedge dt\wedge dx\wedge dy\wedge dz
\nd
The above choice of $\Lambda_7$ leads us to three different classes of solutions from the $G_3$ EOM \eqref{g3eom}. These three classes can be tabulated in the
following way:
\vskip.1in

\noindent $\bullet$ If $\alpha=e^{4A+B}$ in \eqref{lambdaeqon} and
$\Lambda_7 = d\Lambda_7 = 0$ then $G_3$ must be ISD. When $B=0$ then
this is the same as GKP solution \cite{Giddings:2001yu}, and in this
case $\tau$ is not restricted\footnote{One can find solutions for
$\alpha=e^{4A+B}$ case when $B\neq 0$, but this solution doesn't
have correct conformal limit, i.e. when we switch off $G_3$, it
doesn't reduce to the KW solution. In the dual gauge theory the
charge obviously varies with the temperature which is not the case
in the ordinary gauge theory.}.

\vskip.1in

\noindent $\bullet$ If $\alpha\neq e^{4A+B}$ then we can take $\Lambda_7\neq 0$ but keep
$d\Lambda_7=0$ and $d\tau=0$. This means $\Lambda_7$ is closed but not necessarily exact, and $\tau$ is a constant\footnote{Or $\tau = d\lambda_{-1}$
i.e $d$ of a ($-1$)-form. The functional form for the ($-1$)-form is non-trivial, so this option is more cumbersome to use.}.

\vskip.1in

\noindent $\bullet$ If $\alpha\neq e^{4A+B}$ then we can again take $\Lambda_7 \ne 0$ but now $d\Lambda_7\neq 0$ and
$d\tau\neq 0$ such that \eqref{g3eom} is satisfied. This means both axion and the dilaton could run in this scenario.

\vskip.1in

\noindent In this paper we are taking $\alpha=e^{4A}$, so we have to consider the last two cases.
Expressing $\Lambda_7$ as $\Lambda_7 = T_3 \wedge dt\wedge dx\wedge dy\wedge dz$ we have
$e^B\ast_6 G_3-iG_3=T_3$ where $T_3$ is non-zero as long as $B$ is
non-zero. A simple solution then would be to restrict oneself to the second case, i.e
\bg\label{case2} dT_3 ~ = ~ 0, ~~~~~~~~ \tau ~ = ~ {\rm constant} \nd
Notice also that at far infinity, i.e $r\rightarrow \infty$, $B\rightarrow 0$,
therefore $T_3\rightarrow 0$ as well\footnote{This is of course {\it without} considering the UV completion. With UV completion the large $r$ behavior is
non-trivial as discussed in \cite{Mia:2009wj, jpsi1}.}.
Using the above argument, $G_3$ can then be expressed in terms of $T_3$ as
\begin{equation}\label{supreason}
G_3=\frac{e^B\ast_6 T_3+iT_3}{1-e^{2B}}
\end{equation}
Since $\tau = {\rm constant}$, this means the closure of $G_3$ will involve a non-trivial constraint connecting the internal metric components with $B$ and $T_3$. However
in this paper we will not be solving these equations explicitly but
approximating $G_3$ by Ouyang-Klebanov-Strassler flux $G_3^{(0)}$ which is ISD in their metric. This approximation suffices for our case,
as we show below.

Let us substitute $G_3 = G_3^{(0)}$  and $\alpha = e^{4A}$ into \eqref{lambdaeqon}. This will convert $\Lambda_7$ to a simpler seven-form in the following way:
\begin{eqnarray}
\Lambda_7 &= & e^{4A}(e^B\ast_6 G_3^0-iG_3^0) \wedge  dt \wedge dx \wedge dy \wedge dz \nonumber\\
&\thickapprox &
3e^{4A}(e^{2B}-1)g^{rr} \epsilon_{rabc}^{\quad de}G_{rde}^{(0)} ~ dx^a \wedge dx^b \wedge dx^c \wedge dt \wedge dx \wedge dy \wedge dz \nonumber\\
&\thickapprox &
\frac{3r_h^4}{N} g^{rr} \epsilon_{rabc}^{\quad de}G_{rde}^{(0)} ~ dx^a \wedge dx^b \wedge dx^c \wedge dt \wedge dx \wedge dy \wedge dz
\end{eqnarray}
At large $N$ the right hand side is small and therefore deviation from OKS flux is of ${\cal O}(r_h, 1/N)$
so one may consider $\Lambda\thickapprox 0$. This means
$G_3=G_3^{(0)}$ is a good approximation.
Additionally, since
$F_5$ is
self-dual, $\widetilde{R}_{mn}$ can be simplified as
\begin{eqnarray}\label{wideRmn}
\widetilde{R}_{mn}&=&-g_{mn}\frac{G_3\cdot\bar{G}_3}{24\textrm{Im}
\tau}+\frac{G_{mab}\cdot\bar{G}_n^{ab}}{4\textrm{Im}\tau}+\frac{\partial_m\tau
\partial_n\bar{\tau}}{2\mid\textrm{Im}\tau\mid^2}+8(1-e^{-2B})\partial_m A\partial _n
A\nonumber\\
&&-3\widetilde{\triangledown}_m\partial_n B -
\partial_mB\partial_nB
 + 2\partial_{(m}A\partial_{n)}B
\end{eqnarray}
We see the first two terms are suppressed by $g_sM^2/N$ and the third term is removed because
$\tau$ is a constant. So we can ignore these contributions for the time being.
Then, assuming
$A$ and $B$ only depends on $r$, \eqref{wideRmn} will lead to
\begin{eqnarray}
\widetilde{R}_{rr}&=&8(1-e^{-2B})\partial_r A\partial _r
A-3\widetilde{\triangledown}_r\partial_r B -
\partial_rB\partial_rB
 + 2\partial_{(r}A\partial_{r)}B\nonumber\\
\widetilde{R}_{ab}&=&-\frac{3}{2}\partial_r\widetilde{g}_{ab}\partial_rB
\end{eqnarray}
where ($a, b$) denote the angular directions. We now see that for $r > r_h$, the $\widetilde{R}_{ab}$ contribution is suppressed equivalently as
the $\widetilde{R}_{rr}$ contribution, therefore
we need to keep both the parts. This conclusion can also be
extended to $R_{mn}$ in \eqref{Rmn_fangA}, which implies that the $R_{rr}$ and $R_{ab}$
contributions are equally suppressed. All this then further implies that we need to solve the twenty-one metric equations.
This is a formidable exercise. Is there a way by which we can avoid doing this?

A possible way out would be to study the relative suppressions of various terms in the system of equations.
This criteria was already anticipated in \cite{Mia:2009wj}. For example, as we discussed in
\cite{Mia:2009wj}, we can equivalently take:
\bg\label{limits}
(g_s, N, M, N_f) ~ \to ~ (\epsilon^{c}, \epsilon^{-a}, \epsilon^{-b}, \epsilon^{-d})
\nd
This would clearly show that ($g_sN, g_s M$) are very large but ($g_s N_f, g_sM^2/N, g_s^2 M N_f$) as well as $M/N$ are suppressed in the following way:
\bg\label{suppression}
&&(g_sN, g_s M) ~ \to ~ (\epsilon^{c-a}, \epsilon^{c-b})\nonumber\\
&&(g_sN_f, g_sM^2/N, g_s^2 M N_f, M/N) ~ \to ~ (\epsilon^{c-d}, \epsilon^{c-2b+a}, \epsilon^{2c-b-d}, \epsilon^{a-b})
\nd
provided ($a, b, c, d$) satisfy the following
inequalities\footnote{A solution to the inequalities is $a = 8, b = 3, c = 5/2, d = 1$, as given in \cite{Mia:2009wj}. One can of course allow other
values of ($a, b, c, d$) that satisfy the inequalities. \label{limfoot}}:
\bg\label{inequa}
a ~>~b ~>~c ~>~d, ~~~~~~~~ a+c ~ > ~ 2b, ~~~~~~~~ 2c ~ > ~ b + d
\nd
Let the smallest scale in our problem be the ratio $M/N$. Then if the argument of the relative suppressions of various terms in $R_{mn}$ has to make sense, one
would require the precise range of $r$ where our approximations hold water. This gives us:
\begin{equation}\label{rrange} r ~ \ge ~ r_h \left({N\over M}\right)^{1/4} \end{equation}
Thus if we are in this range, we can see that the curvature terms simplify drastically. This would give us a hint that if we solve the simplest
trace equation along with the flux equations  (\ref{BHfactorA}), (\ref{bianchi5}), and (\ref{GKP_BH}) we would be reasonably close to the correct answer because the
other twenty component equations would only change the results\footnote{This in particular means that not only the coefficients of all the terms of the internal metric
will change to  ${\cal O}(r_h^4/r^4)$ but also any {\it new} component
will appear to  ${\cal O}(r_h^4/r^4)$. This is exactly how we choose our initial
metric ansatze \eqref{inmate} and therefore the system is self-consistent.}
to ${\cal O}(r_h^4/r^4)$.
So once we are in the range \eqref{rrange} the only corrections to our
simplified trace equation will be to ${\cal O}(g_sM^2/N)$ and  ${\cal O}(r_h^4/r^4)$.
This is not so bad because if we
choose $\epsilon$ in \eqref{limits} to be $\epsilon = 0.1$, then
\bg\label{bgdata} N ~ = ~ 10^8, ~~~ M ~ = ~ 10^3, ~~~ N_f ~ = ~ 10, ~~~ g_s ~=~ 0.0032, ~~~ r ~ > ~ 17.78 r_h \nd
which means for $r$ beyond $17.78 r_h$ the contributions coming from the individual component equations to the solution generated using only the trace equation will
not be too drastic.

Therefore, once the dust settles, tracing the second equation in (\ref{ricci_T}), using (\ref{warp_eq_1}), (\ref{BHfactorA}) and (\ref{Rmn_fangA}),  we get
\bg \label{TrRmnA}
&&\frac{\widetilde{R}}{6}+\frac{4}{3}\widetilde{g}^{mn}\partial_mA \partial_nA \left(e^{-2B}-1\right)+\frac{\widetilde{g}^{mn}}{6}\left(
3\widetilde{\triangledown}_m\partial_nB+\partial_mB\partial_nB\right)\nonumber\\
&& ~~~~~-\frac{\widetilde{g}^{mn}}{3} \partial_{(m} A \partial_{n)} B
=\frac{\widetilde{g}^{mn}\partial_m\tau \partial_n\bar\tau}{12|{\rm Im}\tau|^2}
\nd
where $\widetilde{R}=\widetilde{g}^{mn}\widetilde{R}_{mn}$ and we have ignored all local sources.

Our goal now is to solve the system of four equations
(\ref{BHfactorA}), (\ref{bianchi5}), (\ref{GKP_BH}) and
(\ref{TrRmnA}) and find solutions for the warp factors $A, B$, the
internal metric $\widetilde{g}_{mn}$ and the fluxes. In obtaining
the  solutions, we will be working in the limit where there is no
local sources, $G_3$ is closed while the explicit form of the fluxes
that solve the flux equations are described in the following
subsection\footnote{It is of course possible to consider additional
sources to obtain a UV complete solution as done in \cite{jpsi1}.
But for the purpose of the current section, which is to analyze the
non-extremal limit for the IR geometry, we will ignore local sources
and discuss their effects briefly towards the end.}. As we mentioned
earlier, if we choose  $\alpha=e^{4A}$, (\ref{GKP_BH}) will imply
that $G_3$ is ISD, in the extremal limit i.e $e^{B}=1$. On the other
hand, $G_3$ is not ISD on a deformed cone in the presence of a black
hole, and the terms in $G_3$ which make it non-ISD are precisely
proportional to the blackhole horizon and the deformation function
$F$ that appears in $\widetilde{g}_{mn}$. With these considerations
and our choice of internal metric $\widetilde{g}_{mn}$ we get
\bg\label{G3^2} |{i}G_3-\ast_6G_3|^2 ~ = ~
\Big|\frac{i\ast_6T_3+T_3}{1+e^B}\Big|^2 ~ \sim ~ {\cal O}(F^2,
r_h^8/r^8) \nd Thus with a choice of $\alpha= e^{4A}+{\cal O}(F^2)$,
(\ref{GKP_BH}) can be solved exactly. But if $F\ll 1 $, we can
ignore ${\cal O}(F^2)$ terms which means up to linear order in $F$,
(\ref{GKP_BH}) becomes
\begin{eqnarray}\label{GKP_BH_M^2A}
\widetilde{\triangledown}^2(e^{4A+B}-e^{4A})&=&e^{-6A-3B}|\partial
(e^{4A+B}-e^{4A})|^2+3e^{-2A-2B}\partial_mB\partial^m(e^{4A+B}-e^{4A})\nonumber\\
\end{eqnarray}
Thus ignoring ${\cal O}(F^2)$ in (\ref{GKP_BH})\footnote{The term in (\ref{G3^2}) appearing in (\ref{GKP_BH}) contributes as
$\sim {\cal O}(F^2(g_sM^2/N)^l), l\ge 1$ which can be easily obtained by using
$e^{-4A}\sim {\cal O}(g_s N)\left[1+{\cal O}(g_sM^2/N)\right]$. As $g_sM^2/N\ll 1$, we can ignore ${\cal O}(F^2 g_sM^2/N)$ terms.
See also \eqref{limits}, \eqref{suppression} and \eqref{inequa} for more details on the various scaling limits.},
we are essentialy   solving  (\ref{BHfactorA}), (\ref{bianchi5}), (\ref{TrRmnA}) and (\ref{GKP_BH_M^2A}). In fact we will show that (\ref{bianchi5}) dictates $F\ll 1$ and our explicit numerical solutions will also be consistent with this assumption, justifying our perturbative analysis.

  Now only considering up to linear order terms in $F$, we get $\alpha=e^{4A}$ which relates the warp factor to the five-form
field strength which in turn depends on $G_3$ by the Bianchi identity
(\ref{bianchi5}). Thus $e^A$ depends on the non-ISD $G_3$
as $G_3$ is modified in the presence of a black hole. But the choice of $\alpha=e^{4A}$ also means that the
dependence of $G_3$ on blackhole horizon $r_h$ appears in the form of
a resolution parameter $a = a(r_h)$, a crucial fact that was first conjectured in
\cite{Mia:2009wj} and will be further illustrated in the next subsection.

As already mentioned, equation (\ref{bianchi5}) determining $e^{A}$
also depends on the internal metric $\widetilde{g}_{mn}$. In the absence
of any flux and axion-dilaton field, $\widetilde{g}_{mn}$ is the metric
of base of the deformed conifold $T^{1,1}$ which has the topology of
$S^2\times S^3$. In the presence of a black hole horizon and various
sources, the internal metric will be modified in the following way:
\bg\label{modmet}
\widetilde{g}_{mn}=\widetilde{g}_{mn}^{[0]}+\widetilde{g}_{mn}^{[1]} \nd
  where $\widetilde{g}_{mn}^{[0]}$ is the metric of a resolved deformed cone (or more appropriately, here, the resolved cone) with base $T^{1,1}$ and therefore
  $\widetilde{g}_{mn}^{[1]}$ denotes all the corrections due the black hole and
 all other sources. This means that  $\widetilde{g}_{mn}^{[1]}$ contains all the informations of the resolution
factor and its subsequent dependence on the horizon radius etc.
Note also that, as we have a horizon at $r=r_h$ with $M$ units of fluxes\footnote{In the intermediate region, i.e Region 2 of the geometry,
we will also have ($p, q$) five-brane sources.}
and $N_f$ number of seven branes,
 $\widetilde{g}_{mn}^{[1]}$ must at least be of ${\cal O}(M,N_f,r^4_h/r^4)$. We will
 evaluate $e^A$ and  $\widetilde{g}_{mn}^{[1]}$ to lowest order in
 ${g_sM^2\over N}$ and $g_sN_f$ which in turn will drastically simplify our analysis. Our choice of
  $\widetilde{g}_{mn}^{[0]}$ and $\widetilde{g}_{mn}^{[1]}$
will be such that we have \eqref{inmate} for the internal metric.

The Bianchi identity for the five-form flux, in the absence of any three-brane sources,
reads
\bg \label{bianchi} d\widetilde{F}_5=H_3\wedge F_3 \nd where $F_3$
and $H_3$ are the RR and the NS three-form fluxes. They are given as
\bg\label{onguflux1}
F_3 ~=~ F_3^{(0)} + {\cal O}(F), ~~~~~~~ H_3 ~= ~ H_3^{(0)} + {\cal O}(F)
\nd
where $F_3^{(0)}, H_3^{(0)}$ are the
fluxes in the absence of any squashing, that is for $F=0$ (we expressed this earlier as $G_3^{(0)} \equiv F_3^{(0)} - \tau H_3^{(0)}$).
For the regular cone, taking into account the running of the $\tau$ field,
 $F_3^{(0)}$ and $G_3^{(0)}$ are exactly the Ouyang fluxes \cite{Ouyang:2003df}, while the exact form of the fluxes in a deformed conifold were discussed in \cite{Dasgupta:2008hw}  \cite{Klebanov:2000hb}. Now from the form of the fluxes on deformed cone\footnote{See section \ref{actfluxes} for more details.}
one gets  that
\bg\label{onguflux}
F_3 ~\sim ~  M[1+ {\cal O}(F)],~~~~~~~~ H_3 ~ \sim ~ g_sM[1+ {\cal O}(F)]
\nd
Using this and (\ref{onguflux1}) one readily gets that
\bg\label{ouyang+res}
H_3\wedge F_3= F_3^{(0)}\wedge H_3^{(0)} + {\cal O}(M^2F)
\nd
An immediate question is:
what can be said about the squashing function $F$? In the absence of the three-form fluxes, i.e $M=0$, there is no squashing as
the Klebanov-Witten solution \cite{klebwit} with running dilaton \cite{Ouyang:2003df} needs no squashing. This remains
true even when we introduce temperature. To see this, observe that the non-extremal limit of Klebanov-Witten(KW) model does not require any modification of the internal space: which means $F=0$ with $e^{2B}=1-\bar{r}_h^4/r^4$ and the internal space is exactly $T^{1,1}$.
 There could be squashing due to the running of $\tau$ field in the KW blackground, but squashing would be at
${\cal O}(g_s^2N_f^2)$, so we can ignore it as we will only consider up to linear order in
$g_sN_f$. These behaviors indicate that $F$ must be at least proportional to $M$. In the following subsection, we will justify this claim.

\subsubsection{Behavior of $F$ and various scaling limits \label{chap2.1.1}}

Let us go to the case when there is no blackhole but we have non-zero three-form flux i.e $M\neq 0$. For this case
we are back to Klebanov-Strassler-Ouyang background with  no squahing and $F=0$. This means, $F$ must also be proportional to the blackhole
horizon $r_h$.
Combining this with the form of the Ouyang fluxes, taking into account of the back reactions of the seven branes, we expect
\bg\label{fgdef}
F\sim {\cal O}(a_0^2, r_h g_s M^\alpha/N^\beta, r_hg^2_s N^2_f)
\nd
with $a_0$ being the bare resolution parameter discussed earlier and ($\alpha, \beta$) are some integers.
Notice that we have inserted a suppression factor of $N^{-\beta}$ assuming $\beta > 0$ in anticipation of a possible perturbative expansion.
   Therefore using our ansatz \eqref{fgdef} in \eqref{ouyang+res} gives us
\bg\label{ouyang-res-1}
F_3\wedge H_3=F_3^{(0)}\wedge H_3^{(0)}+ {\cal O}(a_0^4, r_h^2g_s^2M^{\alpha + 2}/N^{\beta + 1},r_h^2g_s^2N_f^2)
\nd
implying that up to
quadratic order in $M$, we only need Ouyang fluxes to solve
(\ref{bianchi}). But to guarantee that we only need to consider up to quadratic order in $M$, we must show that higher order
i.e ${\cal O}(g_s^2 M^{\alpha + 2}/N^{\beta + 1})$ terms are small compared to the $g_sM^2/N$ terms coming from  $F_3^{(0)}\wedge H_3^{(0)}$.
This will indeed be the case once we solve
(\ref{bianchi}) up to ${\cal O}(M^2)$
\footnote{If the solution to  (\ref{bianchi}) up to ${\cal O}(M^2)$ tells us that $F > 1$, then
we cannot ignore the second term in (\ref{ouyang-res-1}) and therefore have to include ${\cal O}(M^3)$ and higher in solving (\ref{bianchi}).
But, as we will argue soon, our solutions show that $F\ll 1$, which justifies our truncation.}. We will see  $F\sim M/N$ where
$N\gg M$ and this justifies ignoring the second term in (\ref{ouyang-res-1}). In fact solving (\ref{bianchi}) with our ansatz for the warp factor shows that ${1\over g_sN}(F_3\wedge H_3)$ is the relevant term that enters into the equaton of motion (see {\bf Appendix \ref{bgcomp}}).
Hence in solving (\ref{bianchi}) with our choice of warp factor, we are really ignoring
${\cal O} (g_s M^3/N^2) $ and keeping terms only up to ${\cal O}(g_s M^2/N)$. This truncation is consistent for $N\gg M$ which is achievable as we
showed in \eqref{limits} and \eqref{suppression}.
However
one might question the suppression terms in \eqref{fgdef} and in \eqref{ouyang-res-1} if ($\alpha, \beta$) exponents are arbitrary compared to the
range that \eqref{limits} would impose. That this will not be the case will become apparent from the following discussions.


To start then we shall continue using the following five-form flux:
\bg\label{F5}
\widetilde{F}_5=(1+\ast_{10}) d\alpha\wedge d^4x
\nd
With this form of $\widetilde{F}_5$ and $\alpha=e^{4A}=1/h$,  (\ref{bianchi}) becomes an equation involving $h$, $e^{2B}$ and $F$.
We already know that in the AdS limit $e^{2B}=1-\bar{r}_h^4/r^4$. In our non-AdS geometry we expect:
\bg\label{BHfactora}
e^{2B}= 1-\frac{\bar{r}_h^4}{r^4}+ G
\nd
where $G$ is at least ${\cal O} (M,N_f)$.
Using this expansion for $e^{2B}$, along with the precise form of the Ouyang three-form fluxes
$F_3^{(0)}, H_3^{(0)}$ and only considering up to
${\cal O}(M^2)$ terms \footnote{Again in ignoring higher order terms in $M$, we are assuming that
$F, G \sim {\cal O}(M/N)< 1$, which will be consistent with our solution. On the
 other hand, the ${\cal O}(M^2)$ term that enters into (\ref{EQA}) from the Ouyang warp factor should be understood to be of ${\cal O}(g_sM^2/N)$.
Terms of ${\cal O}(M^3)$ in (\ref{EQA})
come from products of $g_sM^2/N$ with $F $ and since $F \sim {\cal O}(M/N)< 1$, the ${\cal O}(M^3)\ll {\cal O}(M^2)$ can be
ignored. Thus we have sometimes ignored the $1/N$ factor or $g_s/N$ factor, but they can always be inserted back in appropriate context.},
(\ref{bianchi}) reads
\bg \label{EQA}
\Bigg[\partial_r\partial_r h^1+ \frac{1}{g}\partial_{\theta_i}\left(\bar{g}_0^{\theta_i\theta_i}\partial_{\theta_i} h^1\right)
+ \frac{r_h^4/r^4}{g}\partial_{\theta_i}\left(\bar{g}_0^{\theta_i\theta_i}\partial_{\theta_i} h^0\right)\Bigg]
r^5 + 5r^4\partial_r h^1=4L^4 \partial_r F~
\nd
where $\bar{g}_0^{mn}$ is proportional to the deformed conifold metric
(see {\bf Appendix \ref{bgcomp}}), $h=h^0+h^1$ with $h^0$ being the Ouyang warp factor
\bg\label{owf}
h^0&=&\frac{L^4}{r^4}\Bigg\{1+\frac{3g_sM^2}{2\pi N} \;{\rm log}r \left[1+\frac{3g_s N_f}{2\pi}\left({\rm log}r
+\frac{1}{2}\right)\right]\nonumber\\
&+&\frac{3g_s^2 M^2 N_f}{8\pi^2N}\;{\rm log}r\; {\rm
log}\left(\rm{sin}\frac{\theta_1}{2}\rm{sin}\frac{\theta_2}{2}\right)\Bigg\}
\nd
and $h^1$ is the contribution due to the presence of the black hole.

We can readily see from (\ref{EQA}) why $F\sim M/N$. First note that the
non-extremal limit of Klebanov-Witten model has an exact solution, $h=L^4/r^4$
with $h^1=0$. $h^1$ is only non-trivial due to the presence of three form
fluxes, the black hole and  other sources. Thus, $h^1\sim {\cal O}(M, g_s N_f,
r^4_h/r^4)$. On the other
hand  $L^4=g_s N\alpha'^2$ and thus  one gets from (\ref{EQA}) that
 $F\sim {\cal O}(M/N, g_sM^2/N)$. But $L^4/\alpha'^2\gg 1$ and we can choose it large enough such that $N\gg M$ which guarantees that $F\ll 1$. This is
of course consistent with \eqref{limits}\footnote{Note that the third term in \eqref{EQA}, because of the $\theta_i$ derivative, is suppressed as
$g_s^3 M^2 N_f$. Using \eqref{limits} and footnote \ref{limfoot}
this would go to zero as $\epsilon^{1/2}$. Also comparing this term with $g_sM$, the fall-of is $g_s^2M N_f$
which from \eqref{suppression} goes to zero as $\epsilon$. Therefore from all criteria in \eqref{EQA}, $h^1\sim {\cal O}(M)$ seems consistent.}.

The key point in the above argument came from $L^4/\alpha'^2\gg 1$ appearing in the Ouyang solution, which is on a regular cone while we have a deformed cone. How can we use the form of $h^0$ as given by (\ref{owf}) for the case of a deformed cone? The answer lies in the fact that for large radial distances, the deformed cone coincides with the regular cone. The Klebanov-Strassler solution in the large $r$ regime behaves as the Klebanov-Tseytlin solution,
i.e the warp factor for KS model becomes
\bg
h_{KS}&\sim& \frac{\alpha'^2}{r^4}\left[g_s^2M^2{\rm log}\left(\frac{r}{r_*}\right)\right]\nonumber\\
&=& \frac{\alpha'^2}{r^4}\left[g_s^2M^2{\rm log}~b + g_s^2M^2{\rm log}\left(\frac{r}{br_*}\right)\right]\nonumber\\
&=&  \frac{L^4}{r^4}\left[1+\frac{g_sM^2}{N}{\rm log}\left(\frac{r}{r_0}\right)\right]
\nd
where $b$ is some scale and $L^4=g_sN\alpha'^2$ with $N=g_s M^2 {\rm log}~b, r_0= br_*$.
The above expansion shows that the KS warp factor in the deformed cone can really
coincide with the Klebanov-Tseytlin solution. Once back-reactions of the flavor D7 branes are taken into account, KS solution in the deformed
cone background will take the form of the Ouyang solution.
We can of course choose ${\rm log} ~ b \gg 1$ such that $L^4/\alpha'^2\gg 1$, so our argument that $M/N\ll 1$
holds even if we started with KS solution and not the Ouyang solution\footnote{Incidentally, using \eqref{limits}, we would require $b$ to
go to infinity as ${\rm exp}\left(\epsilon^{-9/2}\right)$.}.
Hence it is justified to use the Ouyang solution even for the deformed cone.

Also note that, although there were no D3 branes in the KS solution, an effective $N= g_s M^2 {\rm log}~b$
reappears in the warp factor of  KS model in the large $r$
region.  This $N$ can be identified with the $N$ appearing in the Ouyang solution which also justifies using the Ouyang solution on the
deformed cone background for large $r$ region.
For small radial distances, we cannot use the $h^0$ as given in (\ref{owf}) $-$ hence the non-extremal solutions we
consider are only valid for  large radial distances. This also means, we are considering large horizon $r_h$ and the geometry is dual to the high
temperature regime of the gauge theory. A conclusion that is consistent with our earlier works.

\subsubsection{Analysis of the full background with backreactions \label{chap2.1.2}}

Once the behavior of $F$ and the suppression orders for various terms are laid out, we are ready to tackle the backreactions to order
$g_sN_f$ and $g_sM^2/N$. We start from the equation of motion for $\tau$ given in the following way:
\bg
\widetilde{\triangledown}^2\tau\sim \widetilde{g}^{mn}\partial_m\tau \partial_n\bar\tau
\nd
However, the underlying F-theory picture \cite{vafaF} on which we based our solution \cite{jpsi1},
dictates that $\partial \tau\sim {\cal O}(g_s N_f)$  and therefore we will ignore terms of ${\cal O}(g_s^2N_f^2)$.
So the precise form of $\tau$ will not appear in any of the
equations (\ref{EQA}), (\ref{BHfactorA}), (\ref{TrRmnA}) and (\ref{GKP_BH_M^2A}).

Thus with our ansatz for the metric (\ref{bhmet1}), (\ref{inmate}) and choice of fluxes, we have {\it four} equations
(\ref{EQA}), (\ref{BHfactorA}), (\ref{TrRmnA}) and (\ref{GKP_BH_M^2A}) that we need to solve
and {\it three} unknown functions $h^1, G$ and $F$. However, it is more
convenient to write $h^1\sim A^1L^4/r^4$ and then from (\ref{EQA}) one readily sees that
\bg\label{h1h0}
A^1 ~ \sim ~ {\cal O}(M/N) \ll 1, ~~~ {\rm with} ~~~ F\sim {\cal O}(M/N) + {\cal O}(g_sM^2/N) + {\cal O}(g_s^2M^2 N_f/N) \ll 1\nonumber\\
\nd
and so the third term in $F$ is even more suppressed.
Now what can we say about $G$? As already pointed out, $G\sim {\cal O}(M, g_s
N_f)$. But using the form of $F$ as given above in (\ref{h1h0}), one readily
gets from expanding  (\ref{BHfactorA}), that
\bg\label{glimit}
G= {\cal O}(F)\sim  {\cal O}(M/N) + {\cal O}(g_sM^2/N) + {\cal O}(g_s^2M^2 N_f/N) \ll 1
\nd
Thus it is reasonable to consider only up to linear order terms in  $A^1,G$ and $F$. But  (\ref{GKP_BH_M^2A}) is a trivial equation up to linear order
(see {\bf Appendix \ref{bgcomp}}) and hence the only non-trivial equations we
are solving are (\ref{EQA}), (\ref{BHfactorA}) and  (\ref{TrRmnA}). Thus we
have a system of three equations and three functions $A^1, G $ and $F$ $-$ which can be easily solved.

Note that once the above three equations are solved, the corrections from ${\it all}$ the other Einstein equations are automatically
suppressed, as long as we are in the range \eqref{rrange},
and the precise functional form of the axion-dilaton field $\tau$ and the non-ISD three-form flux $G_3$ do not influence the four equations
up to linear order in $A^1, G$ and $F$. This is because (\ref{EQA}) is obtained from (\ref{bianchi5}) which is identical to (\ref{warp_eq_1}) (up to linear order in  $A^1,G$ and $F$) which in turn  is  obtained by tracing Einstein equations in the Minkowski directions. On the other hand, (\ref{TrRmnA}) is obtained from tracing the Einstein equations in the internal directions. Hence a solution to    (\ref{EQA}) and  (\ref{TrRmnA}) along with the background Ouyang warp factor $h^0$ and three form fluxes $G_3$ minimizes the action (\ref{tubra}) where only Ricci scalar and the flux strength appear for the radial range \eqref{rrange}.
Thus  solving   (\ref{EQA}) and  (\ref{TrRmnA}) really means putting the action on shell which guarantees that individual Einstein equations change the metric only to
order $r_h^4/r^4$ as depicted in \eqref{inmate}.

The form of the solutions to the three equations along with the boundary conditions that dictate the behavior of the warp factor $A, B$ near the
horizon is discussed in {\bf Appendix \ref{bgcomp}}. Here we only quote the functional form of the solutions
\bg \label{ansatz-1A}
&&h^1~=~ \frac{L^4}{r^4}\left(A_0+A_1 ~{\rm log}~r+A_2~ {\rm log}^2r\right)\nonumber\\
&&e^{2B}~\equiv ~ g=1-\frac{\bar{r}_h^4}{r^4} +G\equiv1-\frac{\bar{r}_h^4}{r^4} + g_0+g_1~ {\rm log}~r+g_2~ {\rm log}^2r\nonumber\\
&&F ~= ~F_0+F_1~ {\rm log}~r+F_2~ {\rm log}^2r
\nd
where $A_i, g_i, F_i$ for $i=0,1,2$ are in general functions of $r$ and the internal coordinates $\theta_j, \phi_j, \psi$,
with $j=1,2$. In
{\bf Appendix \ref{bgcomp}} we have worked out the simplest case where $A_i, g_i, F_i$ are assumed to be
functions of $r$ only by neglecting ${\cal O}(g_sN_f)$ terms\footnote{It should also be clear that $A_i \sim {\cal O}(M/N)$ from \eqref{h1h0}.}.
This is a
reasonable assumption for small number of flavors. Furthermore, the thermodynamics of the field theory is dictated by the behaviour of the
dual geometry near the black hole horizon \eqref{rrange}.
If we keep all the seven branes away from the black hole, we can ignore running of $\tau$ near the
black hole. On the other hand, for constant $\tau$ we expect a Klebanov-Strassler type solution  which essentially means the warp factors
$A, B$ and squashing factor $F$ are only functions of $r$. Hence, as long as we are dealing with the light degrees of freedom that
arise from the deformed cone ignoring the back reaction of seven branes, we
can  neglect the contributions from the seven branes far away from the black hole and consider the solution in (\ref{ansatz-1A}) to be functions
of $r$ only.

To account for the heavy quarks, we have to include ${\cal O}(g_sN_f)$ terms but our ansatz (\ref{ansatz-1A}) remains the same with the
understanding that now $A_i,g_i,F_i$ are additionally funtions of the internal coordinates. Interestingly,
however, to analyze the melting of the heavy
quarkonium states, we can consider string world sheets that are {\it fixed} in the internal directions which results in evaluating the warp factors
$A, B$ only for fixed values of the angles  $\theta_j, \phi_j, \psi$. This means our above analysis would suffice.
Hence, even for the study of linear confinement and melting of heavy $Q\overline{Q}$ pairs, it is
sufficient enough to treat the solutions in (\ref{ansatz-1A}) as being functions of the radial coordinate only (see \cite{brambilla, boschi} for related
works in this direction).

In {\bf Figures 1, 2} and {\bf 3} we have plotted $g(u), A_0(u)$ and $F_0(u)$ where $u\equiv r/\bar{r}_h$  using the numerical solutions to equations (\ref{EQA}),
(\ref{BHfactorA}) and (\ref{TrRmnA}). As discussed in  {\bf Appendix \ref{bgcomp}},
at the lowest order of perturbation, only keeping up to linear order terms in $g_s M^2/N$,  equations  (\ref{EQA}), (\ref{BHfactorA})
and (\ref{TrRmnA}) drastically simplify. We obtain a solution with only  $A_0, g_0$ and $F_0$ non trivial while  $A_1=A_2=g_1=g_2=F_1=F_2=0$.
For the plots, we have chosen
$3g_sM^2/2\pi N= 1/2$ and the following boundary conditions\footnote{Let us assume, for simplicity and for performing the numerical analysis,
$3g_sM^2/2 \pi N= 1/2$ to be the smallest scale in the theory (instead of $M/N$ that we took earlier). Then the
argument used earlier in \eqref{rrange} will imply that we should trust our result for $r > 1.19 r_h$. \label{lomlet}}
\bg
&&A_0(\infty)=0, ~~ A_0'(\infty)=0\nonumber\\
&& g_0(\infty)=0, ~~ g_0'(\infty)=0\nonumber\\
&& F_0(\infty)=0, ~~ F_0'(\infty)=0
\nd
Note that $g(1.04)\sim 0$, indicating that the horizon has shifted from the AdS black hole value of $\bar{r}_h$ and
we have obtained a larger black hole with horizon $r_h\sim 1.04 ~\bar{r}_h$. Our numerical results shows that $g_0,A_0,|F_0|< 1$ which validates our perturbative analysis. The fact
that the black hole is of larger size than the AdS limit is consistent with the underlying gauge theory
structure\footnote{Also note that the result is consistent with the first law of black hole thermodynamics which states that the increase in horizon radius is related to
the increase in the mass of the black hole. The addition of five-branes have increased the effective mass of the black hole compared to the AdS limit.}.
The presence of the
fractional branes has  increased the effective mass of the black hole. In fact, the black hole entropy is larger than the
corresponding AdS limit since  $A_0(r_h)>0$ and using Walds formula, one readily gets that $s/T^3\sim N_{\rm
eff}^2>N^2$ where we have defined $g_sN_{\rm eff}=r_h^4 h(r_h)$.

\begin{figure}[htb]\label{g(r)}
        \begin{center}
\includegraphics[height=6cm]{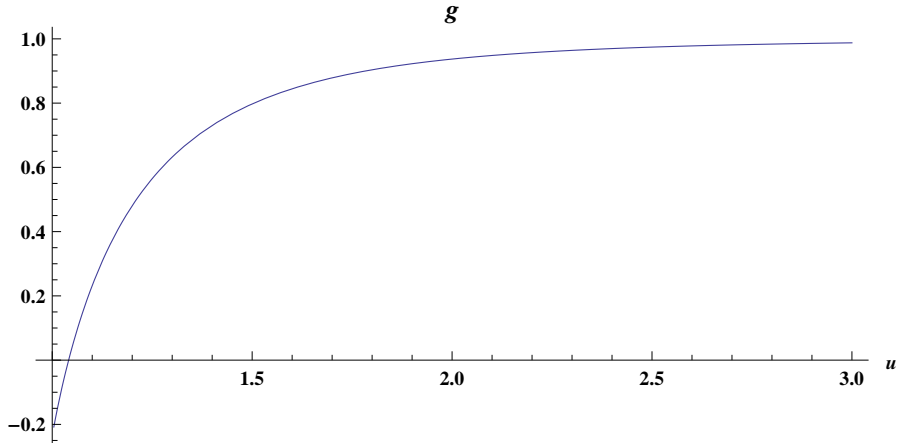}
        \caption{The blackhole factor $g$ as a function of $u\equiv r/\bar{r}_h$. We have plotted $g$ along the y-axis and $u$ along the x-axis. Using above considerations,
one may trust the result for $u > 1.02$.}
        \end{center}
        \end{figure}
\begin{figure}[htb]\label{A0}
        \begin{center}
\includegraphics[height=6cm,angle=0]{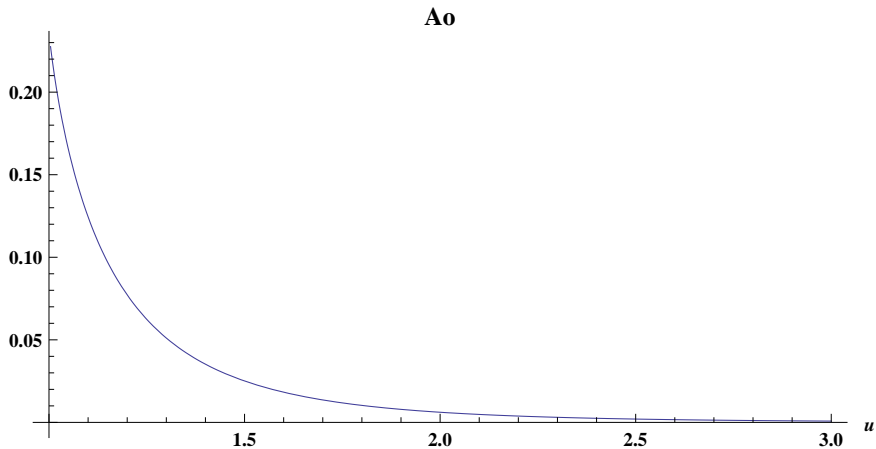}
        \caption{Non-extremal contribution to the warp factor given by $A_0$ plotted as a function of $u\equiv r/\bar{r}_h$. $A_0$ is plotted along y-axis, and $u$ is still along the x-axis and our analysis is valid for $u > 1.02$.}
        \end{center}
        \end{figure}
\begin{figure}[htb]\label{F0}
        \begin{center}
\includegraphics[height=6cm,angle=0]{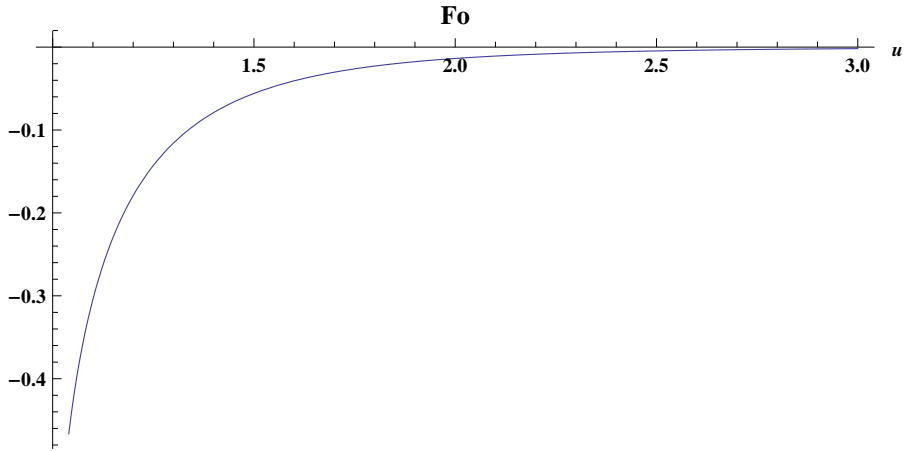}
        \caption{The squashing factor given by the resolution function $F_0$ as a function of $u\equiv r/\bar{r}_h$. Note that $F_0$ is always negative for all distances outside the black hole horizon. We plot $F_0$ along the y-axis, and
$u$ is along x-axis, as before with validity from $u > 1.02$.}
        \end{center}
        \end{figure}
 Finally, note that
the identification of $F$ with $a^2$ in \eqref{wetake} implies that the resolution parameter is given by
\bg\label{respagi}
a^2 = && a_0^2 + \frac{5g_s M^2 p_{11} r_h^2}{32\pi N} + \frac{g_s M^2}{N}\frac{r_h^2}{4\pi}
\left[p_{12}{\rm log}~r + p_{13}{\rm log}^2 r\right] \nonumber\\
&& + {1\over 4\pi}\left({g_sM^2\over N}\right)\left(g_sN_f\right) r_h^4
\left(p_{14}{{\rm log}~r\over r^2} + {p_{15} \over r^2}\right)
{\rm log}\left(\rm{sin}\frac{\theta_1}{2}\rm{sin}\frac{\theta_2}{2}\right)
\nd
where we show the {bare} resolution parameter\footnote{In the limit where the bare resolution parameter
vanishes, which is the Klebanov-Tseytlin solution, we see that the $g_sM^2/N$ corrections actually make the small
$r$ regions non-singular creating an apparent resolution parameter proportional to the horizon radius.}
in $F$ and $a^2$.
The coefficients $p_{ij}$ are constant numbers that could be determined from \eqref{ansatz-1A} and {\bf Appendix \ref{bgcomp}}.
The above representation of the resolution
parameter is perfectly consistent with our conjecture in \cite{Mia:2009wj}: the resolution parameter will pick up
dependence on the horizon radius $r_h$. Interestingly we now have managed to get the leading order ${g_sM^2\over N}{\rm log}~r$
corrections to the result.

However there is one issue that might be confusing the reader. From {\bf Figure 3} we see that $F_0$ is always negative for all values of $r$ in the range
$r_h \le r < \infty$. Our identification of $F$ with $a^2$ would then imply $a$ to be a purely imaginary number. However surprisingly this {\it does not}
create a problem. As we
will show in the next subsection, all the fluxes etc. are completely expressed in terms of $a^2$, so that $a$ does not appear anywhere.
Even terms with logarithms, for example \eqref{gdas}, appear
as ${\rm log}~\vert a^2\vert$, so that $a^2 < 0$ do not create any inconsistencies. This is of course shouldn't come as a surprise because the resolution parameter
appear in the metric \eqref{inmate} as $1 + F_0$ and since $\vert F_0\vert < 1$ it shouldn't lead to any inconsistencies no matter how we relate $F_0$ to $a$.

In our opinion the result that we presented above is probably the first time where the backreaction effects from black hole,
including the resolution factor, are taken into account in a self-consistent way to lowest orders in $g_sN_f$ and $g_sM^2/N$. To this order, as we showed
above, the backreactions from fluxes and branes could be consistently ignored in the near horizon limit \eqref{rrange}.
One may now take this background and
compute the IR effects for large $N$ thermal QCD. However before we go about studying these effects we would like to
dwell, just for the sake of completeness, on the corrections to the Klebanov-Strassler
three-form fluxes that arise from the backreactions of the black-hole, local brane sources, and the resolution parameter. Readers
wishing to know our results may however skip the next sub-section altogether and proceed on with the calculations of the RG flows and
the effects of the chemical potential.

\subsection{The three-form fluxes revisited \label{actfluxes}}

In the above subsection we managed to provide a detailed derivation of the non-extremal limit of the Klebanov-Strassler type solution
with a background warped resolved conifold. The ansatze that we used to solve for the fluxes was \eqref{onguflux} where we
divided the three-form fluxes into two pieces: one coming from the known Ouyang fluxes, and the other coming from the various backreactions.
The second piece, for both RR and NS three-form fluxes, received contributions from the bare resolution parameter $a_0^2$ and the
$g_sM^2/N$ terms in addition to the ${\cal O}(r_h)$ terms. In the following we will not only justify this but also provide the form of the
three-form fluxes including the above-mentioned corrections in the limit where the second squashing factor ${\cal G}$ in \eqref{inmate} is negligible.
Our analysis will also not be affected by the constraint \eqref{rrange} that we had to impose to solve EOMs in the above subsection. In particular this
means that the radial
coordinate may take all values above $r_h$.

Using the metric \eqref{bhmet1} and \eqref{inmate} with the condition \eqref{wetake}, the non-ISD
RR three-form flux $\widetilde{F}_3 \equiv F_3 - C_0 H_3$ takes the following
form\footnote{For the derivations of the three-form fluxes, the readers may want to look up our earlier papers \cite{Dasgupta:2008hw, Mia:2009wj,
jpsi1} where all the necessary
details are given. For example, the ISD fluxes on the resolved conifold are derived in \cite{Dasgupta:2008hw}, and their
extension to the non-ISD cases are argued in \cite{Mia:2009wj, jpsi1}. In the following we will elaborate more on the derivations of
\cite{Mia:2009wj, jpsi1} and show the consistency of the results presented therein.}:
\begin{eqnarray}
&&{\widetilde F}_3  = \left(\widetilde{a}_o - {3 \over 2\pi r^{g_sN_f}} \right)
\sum_\alpha{2M(r)c_\alpha\over r^{\epsilon_{(\alpha)}}}
\left({\rm sin}~\theta_1~ d\theta_1 \wedge d\phi_1-
\sum_\alpha{f_\alpha \over r^{\gamma_{(\alpha)}}}~{\rm sin}~\theta_2~ d\theta_2 \wedge
d\phi_2\right)\nonumber\\
&&~~ \wedge~ {e_\psi\over 2}-\sum_\alpha{3g_s M(r)N_f d_\alpha\over 4\pi r^{\sigma_{(\alpha)}}}
~{dr}\wedge e_\psi \wedge \left({\rm cot}~{\theta_2 \over 2}~{\rm sin}~\theta_2 ~d\phi_2
- \sum_\alpha{g_\alpha \over r^{\rho_{(\alpha)}}}~
{\rm cot}~{\theta_1 \over 2}~{\rm sin}~\theta_1 ~d\phi_1\right)\nonumber \\
&& -\sum_\alpha{3g_s M(r) N_f e_\alpha\over 8\pi r^{\tau_{(\alpha)}}}
~{\rm sin}~\theta_1 ~{\rm sin}~\theta_2 \left({\rm cot}~{\theta_2 \over 2}~d\theta_1 +
\sum_\alpha{h_\alpha \over r^{\delta_{(\alpha)}}}~
{\rm cot}~{\theta_1 \over 2}~d\theta_2\right)\wedge d\phi_1 \wedge d\phi_2\label{brend1}
\end{eqnarray}
with $\widetilde{a}_o = 1 + {3\over 2\pi}$ and is
defined in the intermediate region $r_{\rm min} < r < r_o$. The additional contributions to \eqref{brend1} are all
proportional to powers of $r_h$, as they vanish in the ISD case. Finally,
the quantity
$\epsilon_{(\alpha)}$ is defined in the following way:
\bg\label{epdef}
\epsilon_{(\alpha)} ~ = ~ \alpha + \sum_n {b_{\alpha n}\over r^n}
\nd
with $b_{\alpha n}$ are functions of $g_sN_f, M$ and the horizon radius $r_h$. In a similar fashion
$\rho_{(\alpha)}, \sigma_{(\alpha)}, \delta_{(\alpha)}$ etc are also defined.
The other coefficients, for example
$c_\alpha, ... h_\alpha$
would again be functions of $g_sN_f$ and $r_h$, but also of the resolution factor $a$ (including the internal
angular coordinates).
The resolution factor
appears from the gravity dual that we considered in \cite{Mia:2009wj} i.e a resolved warped deformed
confold with the resolution factor can be viewed as a function dependent on the horizon radius $r_h$. We will
argue this in details below. The coefficients $b_{\alpha n}$ can be represented in terms of the following matrix:
\bg\label{za}
b_{\{\alpha n\}} ~ \equiv ~ \begin{pmatrix} b_{00}& ~b_{01} & ~b_{02}& ~b_{03} & ....\\
b_{10} & ~b_{11} & ~b_{12} & ~b_{13} & ....\\
b_{20} & ~b_{21} & ~b_{22} & ~b_{23} & ....\\
b_{30} & ~b_{31} & ~b_{32} & ~b_{33} & ....\\
... & ... & ... & ... & ....\\
b_{m0} & ~b_{m1} & ~b_{m2} & ~b_{m3} & ....
\end{pmatrix}
\nd
The elements of the matrix $b_{\alpha n}$ can be determined in terms of the $c_0, c_1, c_2, ..$ coefficients that
appears in the expansion ${c_\alpha\over r^{\epsilon_{(\alpha)}}}$. This is one reason of writing the various
powers of $r$ using different symbols. For example $\sigma_{(\alpha)}$ will have a similar expansion as \eqref{epdef}
but with a different matrix. The various elements of the matrix will now be determined in terms of  $d_0, d_1, d_2,..$
etc as one would expect. For the first case, we have managed to determine $c_\alpha$ up to few terms. They are represented
as:
\bg\label{calpha}
c_0 ~ = ~ 1 + {\cal O}(r_h), ~~~ c_1 ~ = ~ {\cal O}(r_h), ~~~
c_2 ~ = ~ {9a^2 g_sN_f\over 2\pi \zeta^2} \left(1 - {3\over 2}{\rm log}~\zeta\right) + {\cal O}(r_h)
\nd
where $a$ is the resolution factor and $\zeta$ is a parameter whose importance will become apparent soon. Once we know
$c_\alpha$, it is not too difficult to get the relations between the various components of the matrix \eqref{za}. One
may now show that the components $b_{\alpha n}$ satisfy:
\bg\label{kdev}
&&b_{00} ~ = ~ b_{01} ~ = ~ b_{10} ~ = ~ {\cal O}(r_h) \nonumber\\
&& c_0 b_{02}~ + ~ c_1 b_{11}~ + ~ c_2 b_{20}~ = ~ -{27 a^2 g_s N_f\over 4\pi \zeta^2} ~ + {\cal O}(r_h)\nonumber\\
&&2 c_0 b_{00} b_{01}~ + ~ c_1 b_{10}~ = ~ {\cal O}(r_h)\nonumber\\
&&c_0 b^2_{01}~ + ~ 2c_0b_{00} b_{02}~ + ~ 2c_1 b_{10}b_{11}~ + ~ c_2 b_{20}~ = ~ {\cal O}(r_h)
\nd
Following the above set of relations one may show that:
\bg\label{relshow}
\sum_{\alpha} {c_\alpha\over r^{\epsilon_{(\alpha)}}} ~ &=& ~ 1 + {9 a^2 g_sN_f \over 2\pi \zeta^2 r^2}
\left(1 - {3\over 2} {\rm log}~\zeta\right) - {27 a^2 g_s N_f \over 4\pi \zeta^2} \cdot {{\rm log}~r\over r^2}
+ {{\cal O}(r_h) \over r^2} + {\cal O}(r_h)\nonumber\\
 & \equiv  & ~ 1 + {9g_sN_f\over 4\pi} \cdot {a^2(r_h, g_sN_f)\over
(\zeta r)^2}\cdot \left[2- 3 {\rm log}~(\zeta r)\right] + {\cal O}(r_h, g_s^2N_f^2)
\nd
which is consistent with what we discussed in \cite{Mia:2009wj}, namely, the resolution parameter $a$ can be thought
of as a function of ($r_h, M, g_sN_f$), including the radial and the angular directions, i.e
\bg\label{resfac}
a^2 ~ = ~ a^2(r_h, M, g_s N_f) ~ = ~ a^2_0 + \sum_{\alpha = 1}^\infty {b_\alpha g_s^\alpha[M(r)r_h]^{\alpha
+ 1}\over N^\alpha r^{\epsilon_{(\alpha)}}}
\nd
with $b_\alpha$ being functions of the angular directions so that this is
consistent\footnote{Its not too difficult to argue for \eqref{resfac} using \eqref{EQA} and \eqref{owf} if we say that $h^1$ goes as
${L^4\over r^4}[{\cal O}(M/N)+{\cal O}(g_S M^2/N)]$ from \eqref{h1h0}. The first term in $F$ is of ${\cal O}(M/N)$ from \eqref{h1h0}
which means $a_0^2$ is of ${\cal O}(M/N)$. This will again be shown later in this section using a slightly different argument.
Once this is established, comparing both sides of \eqref{EQA} then easily implies \eqref{resfac}.
To lowest order then $a^2$ will be a function of ($r_h^2, M/N, g_s M^2/N, g_s N_f$) as shown in \eqref{respagi}.}
with \eqref{respagi}.
These can be determined by comparing \eqref{resfac} with \eqref{respagi} derived in the
previous subsection (note that \eqref{respagi} implies $b_0 = 0$).
The other similar factors appearing in the flux \eqref{brend} are given in terms of the following series expansions
similar to \eqref{relshow} above:
\bg\label{monfuen}
&&\sum_{\alpha} {d_\alpha\over r^{\sigma_{(\alpha)}}} ~ = ~ 1
+ {18 a^2(r_h, M, g_sN_f) {\rm log}~(\zeta r)\over (\zeta r)^2} +
{\cal O}(r_h, M, g_sN_f)\nonumber\\
&&\sum_{\alpha} {e_\alpha\over r^{\tau_{(\alpha)}}} ~ = ~ 1
- {18 a^2(r_h, M, g_sN_f) {\rm log}~(\zeta r)\over (\zeta r)^2} +
{\cal O}(r_h, M, g_sN_f)
\nd
Note also that there are {\it squashing} factors given by ${f_\alpha\over r^{\gamma_{(\alpha)}}},
{g_\alpha\over r^{\rho_{(\alpha)}}}$ and ${h_\alpha\over r^{\delta_{(\alpha)}}}$. These squashing factors
distort the spheres and therefore affect the fluxes on them. Its easy to show that these factors are given by:
\bg\label{squash}
\sum_{\alpha} {f_{\alpha}\over r^{\gamma_{(\alpha)}}} ~& = &~ 1 - {729\over 32\pi^2} \cdot
{g_s^2 N_f^2 a^4(r_h, M, g_sN_f)\over (\zeta r)^4} \cdot {\rm log}~(\zeta r)\left[2-3 {\rm log}~(\zeta r)\right] +
{\cal O}(r_h, M, g_s^2N_f^2)\nonumber\\
&& + {81\over 8\pi} \cdot
{g_sN_fa^2(r_h, M, g_sN_f) {\rm log}~(\zeta r)\over (\zeta r)^2}\\
& = & ~ 1 + {81\over 2} \cdot
{g_s N_f a^2(r_h, M, g_sN_f) {\rm log}~(\zeta r) \over 4\pi r^2 + 9 g_s N_f a^2(r_h, M, g_sN_f) [2 -
3~{\rm log}~(\zeta r)]}
+ {\cal O}(r_h, M, g_s^2 N_f^2)\nonumber\\
\sum_{\alpha} {g_{\alpha}\over r^{\rho_{(\alpha)}}} ~& = &~ 1
+ {36 a^2(r_h, M, g_sN_f) {\rm log}~(\zeta r)\over (\zeta r)^3}
- {648 a^4(r_h, M, g_sN_f) {\rm log}^2(\zeta r)\over (\zeta r)^5} + {\cal O}(r_h, M, g_s^2N_f^2)\nonumber\\
& = & ~ 1 + {36 a^2(r_h, M, g_sN_f)~{\rm log}~(\zeta r) \over (\zeta r)^3 +
18 a^2(r_h, M, g_sN_f) \zeta r ~{\rm log}~(\zeta r)} + {\cal O}(r_h, M, g_s^2 N_f^2)\nonumber\\
\sum_{\alpha} {h_{\alpha}\over r^{\delta_{(\alpha)}}} ~& = &~ 1
+ {36 a^2(r_h, M, g_sN_f) {\rm log}~(\zeta r)\over (\zeta r)^2}
+ {648 a^4(r_h, M, g_sN_f) {\rm log}^2(\zeta r)\over (\zeta r)^4} + {\cal O}(r_h, M, g_s^2N_f^2)\nonumber\\
& = & ~ 1 + {36 a^2(r_h, M, g_sN_f)~{\rm log}~(\zeta r) \over (\zeta r)^2 +
18 a^2(r_h, M, g_sN_f)~{\rm log}~(\zeta r)} + {\cal O}(r_h, M, g_s^2 N_f^2)\nonumber
\nd
The far IR physics is then determined from \eqref{brend} and the squashing factors \eqref{squash} by making the
replacement $(\zeta r) ~ \to r$ to the radial coordinate. Note also that all the flux components are expressed in terms of $a^2$ and therefore the sign of $a^2$ can be
directly inserted here. Considering all the above,
this then gives us exactly the result that we had in
\cite{Mia:2009wj}, namely:
\begin{eqnarray}
{\widetilde F}_3 & = & 2M {\bf A_1} \left(1 + {3g_sN_f\over 2\pi}~{\rm log}~r\right) ~e_\psi \wedge
\frac{1}{2}\left({\rm sin}~\theta_1~ d\theta_1 \wedge d\phi_1-{\bf B_1}~{\rm sin}~\theta_2~ d\theta_2 \wedge
d\phi_2\right)\nonumber\\
&& -{3g_s MN_f\over 4\pi} {\bf A_2}~{dr\over r}\wedge e_\psi \wedge \left({\rm cot}~{\theta_2 \over 2}~{\rm sin}~
\theta_2 ~d\phi_2
- {\bf B_2}~ {\rm cot}~{\theta_1 \over 2}~{\rm sin}~\theta_1 ~d\phi_1\right)\nonumber \\
&& -{3g_s MN_f\over 8\pi}{\bf A_3} ~{\rm sin}~\theta_1 ~{\rm sin}~\theta_2 \left({\rm cot}~{\theta_2 \over 2}~d\theta_1 +
{\bf B_3}~ {\rm cot}~{\theta_1 \over 2}~d\theta_2\right)\wedge d\phi_1 \wedge d\phi_2\label{brend}
\end{eqnarray}
where we have taken $M(r) ~\to ~ M$ in the far IR, and
the various coefficients ${\bf A}_i, {\bf B}_i$ are related to \eqref{relshow} and \eqref{squash} as:
\bg\label{fiveclap}
&& \sum_\alpha {c_\alpha \over r^{\epsilon_{(\alpha)}}} ~ \equiv ~ {\bf A}_1, ~~~
\sum_\alpha {d_\alpha \over r^{\sigma_{(\alpha)}}} ~ \equiv ~ {\bf A}_2, ~~~
\sum_\alpha {e_\alpha \over r^{\tau_{(\alpha)}}} ~ \equiv ~ {\bf A}_3 \nonumber\\
&& \sum_\alpha {f_\alpha \over r^{\gamma_{(\alpha)}}} ~ \equiv ~ {\bf B}_1, ~~~
\sum_\alpha {g_\alpha \over r^{\rho_{(\alpha)}}} ~ \equiv ~ {\bf B}_2, ~~~
\sum_\alpha {h_\alpha \over r^{\delta_{(\alpha)}}} ~ \equiv ~ {\bf B}_3
\nd
As we mentioned before, the additional contributions to \eqref{fiveclap} are all
proportional to powers of $r_h$, as they vanish in the ISD case. Needless to say, the functional form for $F_3$ is consistent with \eqref{onguflux}.

The NS three-form flux $H_3$ is now more interesting. Unlike $\widetilde{F}_3$, it has to be closed. When the
resolution parameter $a$ and $M$
are just constants, it is easy to construct a closed $H_3$. In the presence of non-constant $a$ and $M(r)$, finding
a closed $H_3$ is more non-trivial. For our case
$H_3$ is given by\footnote{We correct a minor typo in \cite{jpsi1}.}:
\begin{eqnarray}\label{H33}
&&H_3 =  \sum_\alpha {6g_s M(r) k_\alpha \over r^{\beta_{(\alpha)}}}\Bigg[1+\frac{1}{2\pi} -
\frac{\left({\rm cosec}~\frac{\theta_1}{2}~{\rm cosec}~\frac{\theta_2}{2}\right)^{g_sN_f}}{2\pi r^{{9g_sN_f\over 2}}}
\Bigg]~ \left[dr + \sum_i {\cal O}(r_h)d\sigma_i\right]\nonumber\\
&&\wedge \frac{1}{2}\Bigg({\rm sin}~\theta_1~ d\theta_1 \wedge d\phi_1
-\sum_\alpha{p_\alpha \over r^{\kappa_{(\alpha)}}} ~{\rm sin}~\theta_2~ d\theta_2 \wedge d\phi_2\Bigg)
+\sum_\alpha \frac{3g^2_s M(r) N_f l_\alpha}{8\pi r^{\theta_{(\alpha)}}}
\Bigg(\frac{dr}{r}\wedge e_\psi -\frac{1}{2}de_\psi \Bigg)\nonumber\\
&& \wedge \Bigg({\rm cot}~\frac{\theta_2}{2}~d\theta_2
-\sum_\alpha{q_\alpha \over r^{\xi_{(\alpha)}}}~{\rm cot}~\frac{\theta_1}{2}
~d\theta_1\Bigg) +
g_s {dM(r) \over dr}
\left(b_1(r)\cot\frac{\theta_1}{2}\,d\theta_1+b_2(r)\cot\frac{\theta_2}{2}\,d\theta_2\right)\nonumber\\
&&\wedge e_\psi
\wedge dr
+{3g_s\over 4\pi} {dM(r) \over dr}\left(1+g_sN_f -{1\over r^{2g_sN_f}} + {9a^2g_sN_f\over r^2} + b_3(r)\right)
\log\left(\sin\frac{\theta_1}{2}\sin\frac{\theta_2}{2}\right)\nonumber\\
&& \sin\theta_1\,d\theta_1\wedge d\phi_1\wedge dr
-{g_s\over 12\pi}{dM(r) \over dr} \Bigg(2 -{27a^2g_sN_f\over r^2} + 9g_sN_f -{1\over r^{16g_sN_f}} -
{1\over r^{2g_sN_f}}\Bigg)\nonumber\\
&& \log\left(\sin\frac{\theta_1}{2}\sin\frac{\theta_2}{2}\right)\sin\theta_2\,d\theta_2\wedge d\phi_2\wedge dr -
{g_sb_4(r)\over 12\pi}{dM(r) \over dr}~ \log\left(\sin\frac{\theta_1}{2}\sin\frac{\theta_2}{2}\right)
\sin\theta_2\,d\theta_2\wedge d\phi_2\wedge dr\nonumber\\
\label{brend2}
\end{eqnarray}
with ($k_\alpha, ..., q_\alpha$) being constants, $\sigma_i \equiv (\theta_i, \phi_i)$
 and $b_n = \sum_m {a_{nm}\over r^{m + \widetilde{\epsilon}_m}}$
where $a_{nm} \equiv a_{nm}(g_sN_f, M, r_h)$ and $\widetilde{\epsilon}_m \equiv \widetilde{\epsilon}_m(g_sN_f, M, r_h)$.

Note that in addition to the simplest ${\cal O}(r_h)$ term that we mentioned above, there would be more terms
of the same order that vanish in the ISD limit. We might also wonder about terms of the form $da/dr$ and
$da/d\sigma_i$. From the form of \eqref{resfac} we see that these terms themselves are of ${\cal O}(M)$ so to this
order they could either be absorbed in $dM(r)/dr$ terms or in the ${\cal O}(r_h)$ terms. The various squashing
factor etc are now given by:
\bg\label{sqafac}
\sum_\alpha {k_\alpha\over r^{\beta_{(\alpha)}}} ~ &= & 1 - {3a^2(r_h, M, g_sN_f) \over (\zeta r)^2} +
{\cal O}(r_h, M, g^2_sN^2_f)\nonumber\\
\sum_\alpha {l_\alpha\over r^{\theta_{(\alpha)}}} ~ &= & 1 + {36 a^2(r_h, M, g_sN_f) ~{\rm log}~(\zeta r)
\over \zeta r} +
{\cal O}(r_h, M, g^2_sN^2_f)\nonumber\\
\sum_\alpha {p_\alpha\over r^{\kappa_{(\alpha)}}} ~ &= & 1 + {3 g_sa^2(r_h, M, g_sN_f) \over (\zeta r)^2} +
{9 g_sa^4(r_h, M, g_sN_f) \over (\zeta r)^4}
+ {\cal O}(r_h, M, g^2_sN^2_f)\nonumber\\
& = & 1 + {3g_s a^2(r_h, M, g_sN_f) \over (\zeta r)^2 - 3 a^2(r_h, M, g_sN_f)} + {\cal O}(r_h, M, g_s^2 N_f^2)\nonumber\\
\sum_\alpha {q_\alpha\over r^{\xi_{(\alpha)}}} ~ &= & 1 + {72 a^2(r_h, M, g_sN_f)~{\rm log}~(\zeta r)
 \over \zeta r} -
{2592 a^4(r_h, M, g_sN_f)~{\rm log}^2~(\zeta r) \over (\zeta r)^2}
+{\cal O}(r_h, M, g^2_sN^2_f)\nonumber\\
& = & 1 + {72 a^2(r_h, M, g_sN_f)~{\rm log}~(\zeta r) \over \zeta r + 36 a^2(r_h, M, g_sN_f)~ {\rm log}~(\zeta r)}
+ {\cal O}(r_h, M, g_s^2 N_f^2)
\nd
To study the far IR physics, we again consider $M(r) \to M$, with the various expansions in \eqref{sqafac}
are related to ${\bf A_4, A_5, B_4}$ and ${\bf B_5}$ respectively. Note again that the resolution parameter in all the coefficients appear as $a^2$.
This then reproduces again the far IR result of
\cite{jpsi1} as well as the expected ansatze \eqref{onguflux}, namely:
\begin{eqnarray}
H_3 &=&  {6g_s {\bf A_4} M}\Bigg(1+\frac{9g_s N_f}{4\pi}~{\rm log}~r+\frac{g_s N_f}{2\pi}
~{\rm log}~{\rm sin}\frac{\theta_1}{2}~
{\rm sin}\frac{\theta_2}{2}\Bigg)\frac{dr}{r}\nonumber \\
&& \wedge \frac{1}{2}\Bigg({\rm sin}~\theta_1~ d\theta_1 \wedge d\phi_1
- {\bf B_4}~{\rm sin}~\theta_2~ d\theta_2 \wedge d\phi_2\Bigg)
+ \frac{3g^2_s M N_f}{8\pi} {\bf A_5} \Bigg(\frac{dr}{r}\wedge e_\psi -\frac{1}{2}de_\psi \Bigg)\nonumber  \\
&& \hspace*{1.5cm} \wedge \Bigg({\rm cot}~\frac{\theta_2}{2}~d\theta_2
-{\bf B_5}~{\rm cot}~\frac{\theta_1}{2} ~d\theta_1\Bigg)
\end{eqnarray}
with the necessary ${\cal O}(r_h)$ terms that vanish when the horizon radius vanishes.

In the far IR the closure of $H_3$ is again non-trivial because the
resolution parameter $a$ is no longer a constant now, although $M(r)
\to M$. All the informations of non-constant $a$ are captured in the
coefficients ${\bf A_{4, 5}}$ and the squashing factors ${\bf
B_{4,5}}$. In the following section we will determine the resolution
parameter to ${\cal O}(M^2)$ in the IR. This means up to this order,
the closure of $H_3$ implies the following three conditions:
\bg\label{3cond} && (i)~~\alpha_2 {\rm cot}~{\theta_2\over 2}
~{\partial {\bf A_5}\over \partial \theta_1} + \alpha_2 ~{\bf
A_5}{\rm cot}~{\theta_1\over 2} ~{\partial {\bf B_5}\over \partial
\theta_1}
+ {\cal O}(r_h, M^3, g_sN_f) = 0\\
&& (ii)~~\alpha_1 {\rm sin}~{\theta_1} ~{\partial {\bf A_4}\over \partial \theta_2} +
\alpha_2 {\rm cos}~{\theta_1}~{\rm cot}~{\theta_2\over 2} ~{\partial {\bf A_5}\over \partial \theta_1}
- \alpha_3 {\rm sin}~{\theta_1}~{\rm cot}~{\theta_2\over 2} ~{\partial {\bf A_5}\over \partial r} \nonumber\\
&& ~~~~~~~~~~~~~~~ +\alpha_2 {\bf A_5}
{\rm cos}~{\theta_1}~{\rm cot}~{\theta_1\over 2} ~{\partial {\bf B_5}\over \partial \theta_2} +
{\cal O}(r_h, M^3, g_sN_f) = 0\nonumber\\
&&(iii)~~\alpha_1~{\bf B_4}~{\rm sin}~{\theta_2}~{\rm cot}~{\theta_1\over 2}~{\partial {\bf A_4}\over \partial \theta_1}
- \alpha_1 ~{\bf A_4}~{\rm sin}~{\theta_2}~{\partial {\bf B_4}\over \partial \theta_1}
-\alpha_2 ~{\bf A_5}~{\rm cos}~{\theta_2}~{\rm cot}~{\theta_1\over 2} ~{\partial {\bf B_5}\over
\partial \theta_2}\nonumber\\
&& ~~~~~~~~~~~~~~~- \alpha_3 ~{\bf A_5}~{\rm sin}~{\theta_2}~{\rm cot}~{\theta_1\over 2}
~{\partial {\bf B_5}\over \partial r}
+ \alpha_2 ~{\rm cos}~{\theta_2}~{\rm cot}~{\theta_2\over 2} ~{\partial {\bf A_5}\over \partial \theta_1}\nonumber\\
&&~~~~~~~~~~~~~~~~ -\alpha_3 ~{\bf B_5}~{\rm sin}~{\theta_2}~{\rm cot}~{\theta_1\over 2}
~{\partial {\bf A_5}\over \partial r} + {\cal O}(r_h, M^3, g_sN_f) = 0\nonumber
\nd
where $\alpha_1, \alpha_2$ and $\alpha_3$ are defined as:
\bg\label{alpde}
&&\alpha_1 = {3g_s M \over 2}\Bigg(1+\frac{9g_s N_f}{4\pi}~{\rm log}~r+\frac{g_s N_f}{2\pi}
~{\rm log}~{\rm sin}\frac{\theta_1}{2}~
{\rm sin}\frac{\theta_2}{2}\Bigg)\nonumber\\
&&\alpha_2 = {3g_s^2 MN_f\over 8\pi r}, ~~~~~~~~ \alpha_3 = -{3g_s^2 MN_f\over 16\pi} = - {r\alpha_2\over 2}
\nd
The RR three-form flux ${\widetilde F}_3 \equiv F_3 - C_0 H_3$ is not closed, but it satisfies the condition
$d{\widetilde F}_3 = -dC_0 \wedge H_3$, which is equivalent to the statement that $F_3$ is closed. Of course as
described in \cite{jpsi1}, the closure of $F_3$ is only in Region 1. In Region 2 there are anti five-brane sources
that make $F_3$ non-closed. The closure of $F_3$ in Region 1 implies the following nine conditions on the various
coefficients of the three-form fluxes:
\bg\label{condon3}
&& (iv) ~~\beta_1 {\partial {\bf A_1}\over \partial r}~{\rm sin}~\theta_1
- \beta_2 ~ {\partial {\bf A_2}\over \partial \theta_1}~{\rm cot}~{\theta_1\over 2}~{\rm sin}~\theta_1
- \beta_2 ~{\bf A_2}~ {\partial {\bf B_2}\over \partial \theta_1}~{\rm cot}~{\theta_1\over 2}~{\rm sin}~\theta_1
\nonumber\\
&& + k\alpha_1 ~{\bf A_4}~{\rm sin}~\theta_1
+ k\alpha_2 ~{\bf A_5}~ {{\bf B_5}}~{\rm cot}~{\theta_1\over 2}~{\rm cos}~\theta_1
+ k\alpha_2 ~{\bf A_5}~ {{\bf B_5}}~{\rm cot}~{\theta_1\over 2} + {\cal O}(r_h, M^3, g_sN_f)= 0\nonumber\\
&& (v)~~ - \beta_1 ~{\bf B_1}~{\partial {\bf A_1}\over \partial r}~{\rm sin}~\theta_2
+ \beta_1 ~{\bf A_1}~ {\partial {\bf B_1}\over \partial r}~{\rm sin}~\theta_2
+ \beta_2 ~{\partial {\bf A_2}\over \partial \theta_2}~{\rm cot}~{\theta_2\over 2}~{\rm sin}~\theta_2\nonumber\\
&& - k\alpha_1 ~{\bf A_4}~{\bf B_4}~{\rm sin}~\theta_2
- k\alpha_2 ~{\bf A_5} ~{\rm cot}~{\theta_2\over 2}~{\rm cos}~\theta_2
+ k\alpha_2 ~{\bf A_5}~{\rm cot}~{\theta_2\over 2} + {\cal O}(r_h, M^3, g_sN_f) = 0\nonumber\\
&& (vi)~~ \beta_1 ~{\partial {\bf B_2}\over \partial r}~{\rm cos}~{\theta_2}~{\rm sin}~\theta_1
- \beta_2 ~{\partial {\bf A_2}\over \partial \theta_1}~{\rm cot}~{\theta_2\over 2}~{\rm sin}~\theta_1 ~{\rm cos}~\theta_1
+ \beta_2 ~{\bf B_2}~ {\partial {\bf A_2}\over \partial \theta_1}~{\rm cot}~{\theta_1\over 2}~{\rm sin}~\theta_1 ~
{\rm cos}~\theta_2\nonumber\\
&&- \beta_2 ~{\bf A_2}~ {\partial {\bf B_2}\over \partial \theta_1}~{\rm cot}~{\theta_1\over 2}~{\rm sin}~\theta_1~{\rm cos}~\theta_2
+ \beta_3 ~{\partial {\bf A_3}\over \partial r}~{\rm cot}~{\theta_2\over 2}
- k\alpha_1 ~{\bf A_4}~{\rm sin}~\theta_1\nonumber\\
&&~~~~~~ - k\alpha_2 ~{\bf A_5} ~{\bf B_5}~{\rm cot}~{\theta_1\over 2}~{\rm cos}~\theta_1
+ k\alpha_2 ~{\bf A_5}~{\bf B_5}~{\rm cot}~{\theta_1\over 2}~{\rm cos}~\theta_2 + {\cal O}(r_h, M^3, g_sN_f)= 0
\nonumber\\
&& (vii)~~
- \beta_1 ~{\bf B_1}~ {\partial {\bf A_1}\over \partial r}~{\rm sin}~\theta_2~{\rm cos}~\theta_1
- \beta_1 ~{\bf A_1}~ {\partial {\bf B_1}\over \partial r}~{\rm sin}~\theta_2~{\rm cos}~\theta_1
- \beta_2 ~{\partial {\bf A_2}\over \partial \theta_2}~{\rm cot}~{\theta_2\over 2}~{\rm sin}~\theta_2~{\rm cos}~\theta_1
\nonumber\\
&& - \beta_2 ~{\bf B_2}~ {\partial {\bf A_2}\over \partial \theta_2}~{\rm cot}~{\theta_1\over 2}~{\rm sin}~\theta_1~{\rm cos}~\theta_2
- \beta_2 ~{\bf A_2}~ {\partial {\bf B_2}\over \partial \theta_2}~{\rm cot}~{\theta_1\over 2}~{\rm sin}~\theta_1~{\rm cos}~\theta_2
+ \beta_3 ~{\bf B_3}~ {\partial {\bf A_3}\over \partial r}~{\rm cot}~{\theta_1\over 2}\nonumber\\
&& ~~~~~~~~+ \beta_3 ~{\bf A_3}~ {\partial {\bf B_3}\over \partial r}~{\rm cot}~{\theta_1\over 2}
 + k\alpha_2 ~{\bf A_5}~{\rm cos}~\theta_1~{\rm cot}~{\theta_2\over 2}
- k\alpha_1 ~{\bf A_4} ~{\bf B_4}~{\rm sin}~\theta_2 \nonumber\\
&&~~~~~~~~~~~~~~~~~~~- k\alpha_2 ~{\bf A_5}~{\rm cot}~{\theta_2\over 2}~{\rm cos}~\theta_2
+ {\cal O}(r_h, M^3, g_sN_f) = 0 \nonumber\\
&& (viii)~~
\beta_1 ~{\partial {\bf A_1}\over \partial \theta_2}~{\rm cos}~\theta_2
+ \beta_1 ~{\bf A_1}~ {\partial {\bf B_1}\over \partial \theta_1}~{\rm sin}~\theta_2~{\rm cos}~\theta_1
- \beta_3 ~{\bf B_3}~ {\partial {\bf A_3}\over \partial \theta_1}~{\rm cot}~{\theta_1\over 2}
+\beta_3 ~{\partial {\bf A_3}\over \partial \theta_2}~{\rm cot}~{\theta_2\over 2}\nonumber\\
&&- \beta_3 ~{\bf A_3}~ {\partial {\bf B_3}\over \partial \theta_1}~{\rm cot}~{\theta_1\over 2}
- k\alpha_3 ~{\bf A_5}~{\rm cot}~{\theta_2\over 2}~{\rm sin}~\theta_1
+ k\alpha_3 ~{\bf A_5}~{\bf B_5}~{\rm cot}~{\theta_1\over 2}~{\rm sin}~\theta_2 +
{\cal O}(r_h, M^3, g_sN_f) = 0 \nonumber\\
&& (ix)~~
\beta_1 ~{\partial {\bf A_1}\over \partial \theta_2}~{\rm sin}~\theta_1
+k \alpha_3 ~{\bf A_5}~{\rm cot}~{\theta_2\over 2}~{\rm sin}~\theta_1 + {\cal O}(r_h, M^3, g_sN_f) = 0 \nonumber\\
&& (x)~~\beta_1 ~{\bf A_1}~{\partial {\bf B_1}\over \partial \theta_1}~{\rm sin}~\theta_2
+k \alpha_3 ~{\bf A_5}~{\bf B_5}~{\rm cot}~{\theta_1\over 2}~{\rm sin}~\theta_2 + {\cal O}(r_h, M^3, g_sN_f) = 0
\nonumber\\
&& (xi)~~
- \beta_2 ~{\partial {\bf A_2}\over \partial \theta_1}~{\rm cot}~{\theta_2\over 2}~{\rm sin}~\theta_2
-k \alpha_2 ~{\bf A_5}~{\bf B_5}~{\rm cot}~{\theta_1\over 2}~{\rm cos}~\theta_2
+k \alpha_2 ~{\bf A_5}~{\bf B_5}~{\rm cot}~{\theta_1\over 2}\nonumber\\
&&~~~~~~~~~~~~~~~~~~~~~~~~~~~~~~~~~~~~+ {\cal O}(r_h, M^3, g_sN_f) = 0 \nonumber\\
&& (xii)~~
- \beta_2 ~{\bf A_2}~ {\partial {\bf B_2}\over \partial \theta_2}~{\rm cot}~{\theta_1\over 2}~{\rm sin}~\theta_1
+ \beta_2 ~{\partial {\bf A_2}\over \partial \theta_2}~{\rm cot}~{\theta_2\over 2}~{\rm sin}~\theta_2
-k \alpha_2 ~{\bf A_5}~{\rm cot}~{\theta_2\over 2}\nonumber\\
&&~~~~~~~~~~~~~~~~~~
-k \alpha_2 ~{\bf A_5}~{\rm cot}~{\theta_2\over 2}~{\rm cos}~\theta_1 + {\cal O}(r_h, M^3, g_sN_f) = 0
\nd
where we have already defined $\alpha_k, {\bf A_n}$ and ${\bf B_m}$. The $\beta_i$ are now defined as:
\bg\label{adevin}
\beta_1 = M\left(1+ {3g_sN_f\over 2\pi}{\rm log}~r\right), ~~~~~~ \beta_2 = -{3g_sMN_f\over 4\pi r} = {2\beta_3\over r}
\nd
Note that in \eqref{3cond} and \eqref{condon3} we have separated the ${\cal O}(r_h, M^3, g_sN_f)$ corrections from the
resolution parameter $a^2$ in the fluxes. We may also {\it absorb} these corrections to the resolution parameter and
write the three-form fluxes completely in terms of ${\cal O}(M^3)$ corrections to the ${\cal O}(M)$ terms in the
original Ouyang solution. This may also be interpreted as though every flux components sees a {\it different}
resolution parameter $a_k^2$. Note also that we don't have an ${\cal O}(M^2)$ corrections to the Ouyang three-form
fluxes. However
from \eqref{resfac} we do expect an ${\cal O}(M)$ term for $a^2$, unless of course $b_0 = 0$ or $a_0$ is
proportional to $M$. Comparing \eqref{resfac} with \eqref{EQA}, \eqref{owf} and \eqref{respagi} may imply $b_0 = 0$.
Additionally, the scenario with $a^2_0$ being of ${\cal O}(M/N)$, is more likely, as in the absence of wrapped D5-branes the gauge theory is
conformal with the gravity dual given by $AdS_5 \times T^{1,1}$ \cite{klebwit}. Only in the presence of
wrapped D5-branes the gravity dual becomes a resolved warped-deformed conifold so the resolution parameter $a_0$
should depend on $M$. This will be consistent with \eqref{EQA} and \eqref{h1h0} as discussed earlier.
In either case, it is clear that the fluxes that we take contribute the ${\cal O}(M^3)$
terms to the original Ouyang fluxes. This helped us to get a consistent background in the presence of
fluxes and a black-hole as we saw in the previous subsection. A more elaborate study will be delegated to \cite{fangmiamik}.

Before we end this section, let us also see how the squashing factors in the three-form fluxes behave in the
light of the result \eqref{respagi}. To ${\cal O}(g_sM^2/N)$ the resolution parameter $a^2$ is only a function of the
radial coordinate $r$. This means that ${\bf A_n}, {\bf B_m}$ can be written as functions of $r$ to this order
satisfying the closure conditions \eqref{3cond} and \eqref{condon3}. For example, combining \eqref{respagi} and
\eqref{3cond}, ${\bf A_5}$ takes the following integral form:
\bg\label{intfora5}
{\bf A_5} = {1\over {\rm sin}~\theta_1~{\rm cot}~{\theta_2\over 2}}\int dr {a_5\over \alpha_3}
\nd
where $a_5$ is a function of the angular $\theta_i$ and the radial $r$ variable such that
${\partial {\bf A_5}\over \partial \theta_i} = 0$. Similarly:
\bg\label{b5what}
{\bf B_5} = {{\rm sin}~\theta_1~{\rm cot}~{\theta_2\over 2} \over {\rm sin}~\theta_2~{\rm cot}~{\theta_1\over 2}}
~{\int dr ~ c_5 \alpha_3^{-1}(r) \over \int dr' ~ a_5 \alpha_3^{-1}(r')}
\nd
where again $c_5$ is like $a_5$ discussed above. Once ($a_5, c_5$) are determined the two integral forms
\eqref{intfora5} and \eqref{b5what} not only satisfy the closure conditions \eqref{3cond} but also the
necessary EOM. These two integral forms are also consistent with the conditions ($ix$) and ($x$) of \eqref{condon3}
because $\alpha_3 \equiv -{3g_s^2MN_f\over 16\pi}$ is a constant.
Note however that, to this order, the integral form for (${\bf A_4}, {\bf B_4}$) cannot be
determined by this method,
although we will know (${\bf A_4}, {\bf B_4}$) in terms of the resolution parameter $a^2$
up to the ${\cal O}(r_h, M^3, g_sN_f)$ corrections.

On the other hand both (${\bf A_1}, {\bf B_1}$) do have an integral representation if we {\it assume} that
(${\bf A_4}, {\bf B_4}$) have some integral representation (which in turn will be determined in a different way
from the one that we have followed here). If this is the case, then:
\bg\label{a1b1}
&&{\bf A_1} = \int {dr\over \beta_1} ~\left({a_1\over {\rm sin}~\theta_1} + k\alpha_1 ~{\bf A_4}
+ k\alpha_2 ~{\bf A_5}~ {{\bf B_5}}~{\rm cot}~{\theta_1\over 2}~{\rm cot}~\theta_1
+ k\alpha_2 ~{\bf A_5}~ {{\bf B_5}}~{\rm csc}^2~{\theta_1\over 2}\right) \nonumber\\
&& {\bf B_1} = {\bf A_1} \int {dr \over \beta_1 {\bf A_1}^2}~\left({b_1 \over {\rm sin}~\theta_2}
+ k\alpha_1 ~{\bf A_4}~{\bf B_4}
+ k\alpha_2 ~{\bf A_5} ~{\rm cot}~{\theta_2\over 2}~{\rm cot}~\theta_2
- k\alpha_2 ~{\bf A_5}~{\rm csc}^2~{\theta_2\over 2}\right)\nonumber\\
\nd
where ($a_1, b_1$) are functions of $r$ and $\theta_i$ such that ${\partial{\bf A_1}\over \partial\theta_i}
= {\partial{\bf B_1}\over \partial\theta_i} = 0$ in the same sense as mentioned earlier for the other cases.
The other squashing factors in \eqref{condon3} do not however have such simpler integral forms.

Another interesting thing to note is that from condition ($xii$) of \eqref{condon3} we might get a simpler
form for ${\bf A_5}$, namely:
\bg\label{a5now}
{\bf A_5} = - {ra_7\over 2k\alpha_3 {\rm cot}~{\theta_2\over 2}(1+ {\rm cos}~\theta_1)}
\nd
This may seem to be different from \eqref{intfora5} that we derived earlier. This is however not the case because
the coefficient $a_7$ is related to $a_5$ in the following way:
\bg\label{a7a5}
a_5 = {1\over k}\left({\partial a_7\over \partial \theta_1} + r {\partial^2 a_7\over \partial \theta_1 \partial r}\right)
\nd
One may also cook up somewhat similar relation for ${b_5}$ and $b_7$ from condition ($xi$) of \eqref{condon3} as
above. The final result will again be consistent with what we got earlier, establishing the fact that the system is
well defined with the given set of boundary conditions.
Therefore these analyses complete the side of story that we expected from \cite{Mia:2009wj} and \cite{jpsi1}
in a satisfactory manner.

\subsection{The behavior of the coupling constants \label{chap2.3}}

In the above sub-section we computed the background more or less exactly up to order ${g_sM^2\over N}$. To this order
we see that the corrections to the resolution parameter $a^2$ is only functions of $r$, the radial coordinate. If we
go beyond this order, the angular dependences start showing up.

One other thing along the same line would be to study the behavior of the two gauge coupling constants of the boundary theory. For
example a crucial question would be to ask how the RG flows of the coupling constants change when the thermal effects are turned on. In the
literature there have been many confusing statements on this.
In the following we will argue that the RG flows or more appropriately the thermal beta functions should be properly interpreted and
the correct picture, in our opinion, is that the thermal beta functions do not change, but the coupling constants themselves get
renormalised. Let us elaborate the story below.

Once we know the NS $B$-field $B_2$ and the string coupling $e^{\Phi}$ then it is easy
to determine the gauge couplings at the UV of the dual gauge theory. The resulting relations are:
\begin{eqnarray}
\frac{8\pi^2}{g_1^2}&=&e^{-\Phi}\Big[\pi-\frac{1}{2}+\frac{1}{2\pi}\Big(\int_{S^2}B_2\Big)\Big]\nonumber\\
\frac{8\pi^2}{g_2^2}&=&e^{-\Phi}\Big[\pi+\frac{1}{2}-\frac{1}{2\pi}\Big(\int_{S^2}B_2\Big)\Big]
\end{eqnarray}
Now note that
when $a^2$ is a constant the string coupling and the $B_2$ field
were obtained in \cite{Mia:2009wj} as
\begin{eqnarray}\label{B2}
e^{-\Phi}&=&\frac{1}{g_s}-\frac{N_f}{8\pi}\log(r^6+9a^2r^4)-\frac{N_f}{2\pi}\log\Big(\sin\frac{\theta_1}{2}\sin\frac{\theta_2}{2}\Big) \nonumber\\
 B_2&=&\Big( b_1(r)
\cot\frac{\theta_1}{2}d\theta_1 +
b_2(r)\cot\frac{\theta_2}{2}d\theta_2\Big)\wedge e_{\psi}\nonumber\\
&+& \Big[ \frac{3g^2_s MN_f}{4\pi}(1 + \log(r^2 +
9a^2))\log\Big(\sin\frac{\theta_1}{2}\sin\frac{\theta_2}{2}\Big)+b_3(r)\Big]\sin\theta_1d\theta_1\wedge
d\phi_1\nonumber\\
&-&\Big[\frac{g_s^2MN_f}{12\pi r^2}(-36a^2+9r^2+16r^2\log
r+r^2\log(r^2+9a^2))\log\Big(\sin\frac{\theta_1}{2}\sin\frac{\theta_2}{2}\Big)+b_4(r)\Big]\nonumber\\
&&\quad\quad\quad\quad\times \sin\theta_2 d\theta_2\wedge d\phi_2
\end{eqnarray}
where we have shown the $a^2$ dependences in $B_2$. The other $a^2$ dependences come from the implicit coefficients
$b_i(r)$.
These dependences in the
first two coefficients $b_1(r), b_2(r)$ are given by\footnote{The $b_i(r)$ here are the same $b_i(r)$ that we encountered in
\eqref{H33} in the limit where the fluxes become ISD.}:
\begin{eqnarray}\label{defb}
  b_1(r) &=& \frac{g_S^2MN_f}{24\pi(r^2+6a^2)}\big(18a^2+(16r^2-72a^2)\log r+(r^2+9a^2)\log(r^2+9a^2)\big)\nonumber\\
  b_2(r) &=& -\frac{3g_s^2MN_f}{8\pi r^2}\big(r^2+9a^2\big)\log(r^2+9a^2)
\end{eqnarray}
Similarly,
other dependences on the resolution parameter
would come from the ($b_3, b_4$) coefficients. We have determined these
coefficients in terms of first-order differential equations. They are now given by:
\bg\label{gdas}
  b_3'(r) &=& \frac{3g_sMr}{r^2+9a^2} + \frac{g_s^2MN_f}{8\pi r(r^2+9a^2)}\Big[-36a^2-18a^2\log \vert a^2\vert
+34 r^2\log r\\
  && \qquad\qquad\qquad\qquad\qquad\qquad+(10 r^2+81a^2)\log(r^2+9a^2)\Big]\nonumber\\
  b_4'(r) &=& -\frac{3g_sM(r^2+6a^2)}{\kappa r^3} - \frac{g_s^2MN_f}{8\pi\kappa r^3}\Big[18a^2-18(r^2+6a^2)\log
 \vert a^2 \vert\nonumber\\
  && \qquad\qquad\qquad+(34 r^2+36a^2)\log r +(10r^2+63a^2)\log(r^2+9a^2)\Big]\nonumber
\nd
As all the above coefficients are given in terms of the resolution parameter
derived in \eqref{respagi}, they
should then be functions of $r_h$ and other radial and angular variables\footnote{Note the appearance of ${\rm log}~\vert a^2\vert$ so that
$a^2 < 0$ will not lead to any inconsistencies, as we explained earlier.}.
This means that the exact coupling should be determined in terms of the NS $B$-field $\widetilde{B}_2$ that is of the
form, up to the order that we had in \eqref{respagi}:
\bg\label{bhusla}
\widetilde{B}_2 = B_2(r, \theta_i) + (g_SM)\cdot(g_sN_f)\cdot \left({g_sM^2\over N}\right){\cal B}_2(r) +
(g_SM)\cdot (g_sN_f)^2\left({g_sM^2\over N}\right) {\cal C}_2(r, \theta_i)\nonumber\\
\nd
and not $B_2$ that we mentioned in \eqref{B2}. Also we expect
$d\widetilde{B}_2=H_3$ with $H_3$ given
as in \eqref{H33}. However, notice that the extra terms in \eqref{bhusla} are suppressed by $g_SMN_f$ over and
above the ${g_sM^2\over N}$ suppression. Therefore if we follow the limit
shown in \eqref{limits}, we can easily infer that this makes \eqref{bhusla} and the
resolution parameter $a^2$ \eqref{respagi} to have the following $\epsilon$ expansion
\bg\label{bhexp}
&&\widetilde{B}_2 = B_2(r, \theta_i) + \epsilon^{11/2}~ {\cal B}_2(r) + \epsilon^{7} ~{\cal C}_2(r, \theta_i)\nonumber\\
&&a^2 = a_0^2 + \epsilon^{9/2}~ r_h^2  + \epsilon^{9/2}~{a}_1^2(r) + \epsilon^{6}~{a}_2^2(r, \theta_i)
\nd
that basically tells us that $\widetilde{B}_2 \approx B_2$ which is consistent with our assumptions in \cite{Mia:2009wj}.
To this limit then the running of the couplings do not change, as one would have expected. This gives us:
\begin{eqnarray}\label{betaft}
\frac{\partial}{\partial\log
\Lambda}\Big[\frac{4\pi^2}{g_1^2}+\frac{4\pi^2}{g_2^2}\Big]&=&-\frac{3N_f}{4}\Big(\frac{r^2+ 6a_0^2}{r^2+9a_0^2}\Big)\nonumber\\
\frac{\partial}{\partial\log
\Lambda}\Big[\frac{4\pi^2}{g_1^2}-\frac{4\pi^2}{g_2^2}\Big]&=&3M\left[1+\frac{3g_sN_f}{4\pi}\log(r^2+9a_0^2)+...\right]
\end{eqnarray}
where $\Lambda$ is the energy scale in the gauge theory side. In fact the LHS of \eqref{betaft} is written in terms of
gauge theory variables whereas the RHS is written in terms of gravity variables.

One important question now is to ask what happens when we consider ${g_sM^2\over N}$ corrections. Clearly now we
need to consider the corrections to $a^2$ \eqref{respagi}. What does this imply for the running couplings?
Saying that the coupling constants run at a {different} rate would probably not be a meaningful
statement. The correct thing
to say at this stage would be to allow for new {\it effective} couplings $\widetilde{g}_1$ and $\widetilde{g}_2$ that
again flow at the same rate as before, i.e the effective couplings have the same beta functions as the original theory.
Alternatively this means that between the two scales, energy and temperature, we {\it fix} the energy scale and
define effective couplings for any given temperature. Once we change the energy scale, these couplings should
run exactly as before.

A way to see this would be the following illustrative example.
Imagine the complete corrections to \eqref{betaft}, from the changes in the
fluxes and the resolution parameters, may be represented as:
\begin{eqnarray}\label{betaftnow}
\frac{\partial}{\partial\log
\Lambda}\Big[\frac{4\pi^2}{g_1^2}+\frac{4\pi^2}{g_2^2}\Big]&=&-\frac{3N_f}{4}\Big(\frac{r^2+ 6a_0^2}{r^2+9a_0^2}\Big)
 + \sum_{n = 1}^\infty r_h^n {\cal F}_n(r, \theta_i)\\
\frac{\partial}{\partial\log
\Lambda}\Big[\frac{4\pi^2}{g_1^2}-\frac{4\pi^2}{g_2^2}\Big]&=&3M\left[1+\frac{3g_sN_f}{4\pi}\log(r^2+9a_0^2)+...\right]
+ \sum_{m = 1}^\infty r_h^m {\cal G}_m(r, \theta_i)\nonumber
\end{eqnarray}
where ${\cal F}_n(r, \theta_i)$ and ${\cal G}_n(r, \theta_i)$ capture all the corrections. For simplicity we have
considered the corrections to depend only on ($r, \theta_i$). Of course more generic corrections could also be
entertained here but it would only make the analysis involved without changing the underlying physics. Therefore
here we will stick to the simplest scenario.

The above corrections to our earlier set of equations can now be re-arranged to redefine two new set of
couplings $\widetilde{g}_1$ and $\widetilde{g}_2$ that are related to $g_1$ and $g_2$ in the following way:
\bg\label{newcouplings}
&&{1\over \widetilde{g}^2_1} = {1\over g_1^2} -
{1\over 32\pi^4} \int_{a_0}^\Lambda {dr\over r} \int_{0}^{2\pi}\prod_i d\theta_i\left(\sum_{n=1}^\infty r_h^n {\cal F}_n
+ \sum_{m=1}^\infty r_h^m {\cal G}_n\right)\nonumber\\
&&{1\over \widetilde{g}^2_2} = {1\over g_2^2} -
{1\over 32\pi^4} \int_{a_0}^\Lambda {dr\over r} \int_{0}^{2\pi}\prod_i d\theta_i\left(\sum_{n=1}^\infty r_h^n {\cal F}_n
- \sum_{m=1}^\infty r_h^m {\cal G}_n\right)
\nd
where we have taken an average over the angular directions so that the couplings are defined only in terms
of $r_h$ and $r$ (or temperature and the energy scale, in the language of gauge theory). With this,
one can now easily see that the two effective couplings $\widetilde{g}_1$ and $\widetilde{g}_2$ flow exactly as
\eqref{betaft} and therefore the theory has the same behavior, which in turn means the
same renormalisation group flows, in terms of these couplings. This would probably be the right way to analyse thermal
beta functions.

We can perform a few checks to justify, at least to some extent, the results got from the gravity side. Firstly, the two running equations \eqref{betaft}
are easy to justify. When the bare resolution parameter $a_0$ is small then \eqref{betaft} combine precisely to reproduce the NSVZ beta function \cite{NSVZ}.
Secondly, in the presence of a non-zero temperature $T$ in field theory, the thermal loops will renormalize the two YM couplings $g^{(i)}_{\rm YM}$ to take the
following form in the planar limit (see for example \cite{lebellac}):
\bg\label{renorcup}
{1\over {\widetilde g}^{2(i)}_{\rm YM}} ~ = ~{1\over g^{2(i)}_{\rm YM}}\left[1 + \sum_n g^{2n(1)}_{\rm YM} {\cal H}_n^{i1}(T)
+ \sum_m g^{2m(2)}_{\rm YM}{\cal K}_m^{i2}(T)\right]
\nd
where the coeffcients ${\cal H}_n^{i1}(T)$ and ${\cal K}_m^{i2}(T)$ could be determined from evaluating the thermal loops. One thing is clear: to have the same
NSVZ beta function these coefficients are related as:
\bg\label{hukla}
{\cal H}_n^{11}(T) ~ = ~ {\cal H}_n^{21}(T), ~~~~~~~~~ {\cal K}_m^{12}(T) ~ = ~ - {\cal K}_m^{22}(T)
\nd
With \eqref{hukla}, although the above YM couplings have surprising
resemblance to the analysis that we did from the gravity side in \eqref{newcouplings}, this
mapping can be made precise if we could identify the coefficients on both sides of the dictionary. This is presently work in progress and more details will be
elaborated in a forthcoming work.


\subsection{Short detour on dualities and dipole deformations \label{chap2.4}}

Our final aim of this section would be to take a short detour and study the
effect of the dipole deformations on the flavor seven-branes in the gravity picture. This dipole deformation, since
it affects the seven-branes, should also have some effect on the fundamental quarks in the gauge theory. We will
make some speculations how the dipole deformations effect the
far IR picture.

Our starting assumption would be that the solutions presented in the earlier subsections have isometries along $\phi_1, \phi_2$ and $\psi$
directions. This in particular means that the coefficients appearing in \eqref{ansatz-1A} i.e ($A_i, F_i, G_i, g_i$) are all functions of
($r, \theta_i$) only and not of ($\phi_i, \psi$). This is not a strong assumption as we saw earlier that even to ${\cal O}(g_s^2 M^2 N_f/N)$
the ($\phi_i, \psi$) dependences do not show up. It could be that the background retains its isometry along ($\phi_i, \psi$) directions to all orders
in $g_sN_f$ and $g_sM^2/N$, but we haven't shown this here.

Before moving ahead let us clarify a point here. Dipole (or non-commutative) deformations can be studied in two possible ways. In the conformal case,
one takes the D3-brane metric written in terms of its harmonic functions, and then use
$TsT$ (T-duality, followed by a shift $s$, and then another T-duality) to generate new solution. The new solution is still given in terms
of D3-branes and harmonic functions, but now there is a background $B_{\rm NS}$ field. One then takes the near horizon limit to determine the
gravity dual of this scenario. The gravity dual has no D3-branes, but both $F_5$ as well as $H_3 = dB_{\rm NS}$ fluxes are still present. The
near horizon geometry do not change the internal metric too much, and therefore analysis on both sides of the story is somewhat similar.

The above criteria changes quite a bit once we go to the non-conformal case. The gravity dual is not simply given by taking the
near-horizon limits of the D3 and the wrapped D5-branes. To avoid naked singularities of the Klebanov-Tseytlin form, one now has to
deform the internal space also. This means making a $TsT$ transformation on the brane side, one may not necessarily get the full gravity dual
picture easily. This is also clear in the geometric transition set-up, whose supergravity solution is developed in \cite{gtpapers, Chen:2010bn}.
So we could do $TsT$ transformations on two sides of the picture, leading to two possible different interpretations.

Thus, once we have solutions for both sides, namely the
gauge-theory and the gravity sides, we can use $TsT$ transformations to deform them into various
different solutions. In this paper we will not consider the dipole (or non-commutative)
deformations on the gauge-theory side of the
story\footnote{The dipole deformations on the gauge theory side, at least in the far IR and in the {\it local} case,
has been discussed earlier in \cite{gtpapers2}. The readers may refer to those papers for more details on the multiply allowed dipole
deformations.},
but concentrate only on
the gravity side.
This means,
given the background metric \eqref{bhmet1} with fluxes, five-branes and seven-branes, the $TsT$ transformed backgrounds will be related to
some interesting deformations of the four-dimensional thermal gauge theories. These deformations can be classified to fall into
four categories. They are listed
as follows\footnote{We will use ($x_0, x_1, x_2, x_3$) as a convenient reparametrization of ($t, x, y, z$) used earlier. The former will be more
convenient for the next couple of sections.}:

\vskip.1in

\noindent $\bullet$ T-dualize along one space
direction say $x_3$ then shift along another space direction say $x_2$ mixing ($x_2, x_3$)
and then T-dualize back along $x_3$ direction.

\vskip.1in

\noindent $\bullet$ T-dualize along $x_3$ and then shift\footnote{Again mixing $x_3$ with one of the internal directions.}
along one of the
internal directions that are isometries of the background, namely along $\phi_1, \phi_2$ or $\psi$
directions\footnote{For simplicity we will only consider the isometry directions.},
and then T-dualize back along $x_3$ direction.

\vskip.1in

\noindent $\bullet$ T-dualize, shift and then T-dualize along internal
directions. The shift will mix two of the internal directions in some appropriate way.

\vskip.1in


\noindent The first operation will lead to a non-commutative gauge
theory on the D7-branes with $[x_2, x_3] = i B_{23}$ as our algebra.
The second one is more interesting.
T-dualizing along $x_3$ but making a shift on the directions along which the D7-branes
are oriented i.e along $\phi_1, \theta_1$ and $\psi$
(recall that the D7-branes wrap the two-sphere parametrised by ($\theta_1, \phi_1$) and are spread along
($r, \psi$) directions) will lead again to a non-commutative gauge theory on the
D7-branes. On the other hand, if we make a shift along the orthogonal direction parametrised by
$\phi_2$, then the theory on the D7-branes will be a
dipole gauge theory.
For the last case, one is T-dualizing and shifting along the directions of the D7-branes. This will again lead to non-commutative theory on the
D7-branes. On the other hand if we shift along $\phi_2$ but T-dualise along the D7-brane directions, we will get dipole theory on the
D7-branes.

To analyse these in case-by-case basis, let
us study the first kind of deformation first. We choose the shift to
be
\begin{eqnarray}
x_2\mapsto \frac{x_2}{\cos\theta}+\sin\theta x_3,\quad x_3\mapsto
\cos\theta x_3
\end{eqnarray}
After the series of transformations discussed above, i.e $TsT$, the metric \eqref{bhmet1}, becomes\footnote{For this section we will ignore the
${\cal O}(g_sM^2/N, r_h^4/r^4)$ corrections to the internal metric \eqref{inmate}. A more precise result will not change the physics to the order that we are studying here.}:
\begin{eqnarray}
ds^2=\frac{1}{\sqrt{h}}\left[-g_1dx_0^2+dx_1^2+J(dx_2^2+dx_3^2)\right] + \sqrt{h}(g_2^{-1}dr^2+d\mathcal{M}^2_5)
\end{eqnarray}
with the Lorentz breaking deformations along ($x^2, x^3$) directions specified by $J$. There is also
a background $B_{\rm NS}$ field that accounts for the non-commutativity. Both $J$ and the $B_{\rm NS}$ field are defined as:
\bg\label{jandb}
J^{-1} ~= ~ \sin^2\theta h^{-1}+\cos^2\theta, ~~~~~~~ B_{23}~ = ~ \tan\theta h^{-1}J
\nd
The metric has the same form as in
\cite{Maldacena:1999mh} and the gauge theory on the D7-branes
become non-commutative in the $x^2$ and $x^3$ directions.

For the second kind of deformation we follow similar procedure as above except that now we shift along $\psi$ direction and T-dualise along
$x^3$ direction.
The resulting metric
is
\begin{eqnarray}
ds^2&=&\frac{1}{\sqrt{h}}\left(-g_1dx_0^2+dx_1^2+dx_2^2+\frac{9}{9\cos^2\theta+r^2\sin^2\theta}dx_3^2\right)\\
&&+\sqrt{h}\left[\frac{r^2 (d\psi+\cos\theta_1d\phi_1+\cos\theta_2d\phi_2)^2}{9\cos^2\theta+r^2\sin^2\theta}+...\right]\nonumber
\end{eqnarray}
where note that the $x^3$ and the $\psi$ circle is non-trivially warped. The dotted terms are unchanged from the original metric \eqref{bhmet1}.
However the $B_{\rm NS}$ field now is non-trivial because of the $\psi$ fibration structure:
\bg\label{bfib}
B=\frac{r^2\tan\theta}{9\cos^2\theta ~+ ~ r^2\sin^2\theta} ~dx_3\wedge
(d\psi+\cos\theta_1d\phi_1+\cos\theta_2d\phi_2)
\nd
The scenario now is interesting because we have three components of the $B_{\rm NS}$ field, with two of the components
$B_{3\psi}$ and $B_{3\phi_1}$ parallel to the D7-branes and one component $B_{3\phi_2}$ having one leg orthogonal to the D7-branes.
Existence of these three components would lead to a complicated theory on the D7-branes that in some limit may be
considered as a combination of both dipole and non-commutative deformations of the world-volume theory on the D7-branes.

As an example for the third kind of
deformation\footnote{Note that we cannot construct another theory by shifting along $\theta_2$ direction and then T-dualing along
$\phi_2$ direction because $\theta_2$ is
not an isometric direction. A shift along the non-isometry directions, like the $\theta_2$ direction, will destroy the existing isometry directions
making the T-duality operations highly non-trivial.}
 we first T-dualize
along $\psi$ direction and then shift along $\phi_2$ direction. The resulting
metric, keeping both the squashing factors ($F, {\cal G}$) in \eqref{bhmet1}, will look like
\begin{eqnarray}
ds^2&=&\frac{1}{\sqrt{h}}(-g_1dt^2+dx^2+dy^2+dz^2)+\sqrt{h}\Big[r^2J(d\psi+\cos\theta_1\cos\theta
d\phi_1+\cos\theta_2 d\phi_2)^2\nonumber\\
&&+\frac{3}{2}r^2J(1+F)(1+ {\cal G})\sin\theta_2^2d\phi_2^2+...\Big]
\end{eqnarray}
with the dotted terms being the terms unchanged from the original metric \eqref{bhmet1}. As expected, note that both the $\psi$ fibration structure
as well as the $\phi_2$ directions get warped by $J$. As before there is also a $B_{\rm NS}$ field. Both $J$ and $B_{\rm NS}$ are given by:
\bg\label{jbnow}
&& B_{\psi\phi_2}~ = ~ \frac{1}{6}\tan\theta~\sin^2\theta_2~(1+F)(1+{\cal G})J\nonumber\\
&& J^{-1}~ = ~ 9\cos^2\theta~+~ \frac{1}{6}hr^2(1+F)(1+ {\cal G})\sin^2\theta_2\sin^2\theta
\nd
This solution shows that the gauge theory on the D7-branes has become a
non-local dipole theory\footnote{On the other hand, the shifts and the duality directions that we choose are not the most generic ones. We can make
numerous other shifts. One simple example could be as follows: we T-dualize along
space direction $x_3$, then shift as $z\mapsto
z+\lambda\theta_2^2/2$ and finally T-dualize back to generate a non-trivial background with the metric
$$ds^2~ = ~ \frac{1}{\sqrt{h}}(-g_1dt^2+dx^2+dy^2+dz^2)+\sqrt{h}(g_2^{-1}dr^2+d\mathcal{M}_5^2+\lambda^2~
\theta_2^2~d\theta_2^2)$$
and a $B_{\rm NS}$ field, $B_{3\theta_2}~=~\lambda \theta_2$. This would generate dipole deformation on the D7-branes.}.

All these solutions generated from our $TsT$ duality operations lead to new
gauge theories on the D7-branes. As mentioned earlier, it is not clear to us whether these deformations are the
corresponding gravity duals of the respective deformations on the gauge theory side of the picture. One thing however is clear: due to the
dipole deformations on the D7-branes, the KK
masses of the fluctuations are different from the original theory. In fact the dipole deformations (along appropriate directions) tend to make the
KK states {\it heavier} \cite{gtpapers2}. Therefore we would expect
operator dimensions on the field theory side to also change accordingly.

The $TsT$ duality operation doesn't change the warp factor nor the BH factor. Naively
applying the criteria from \cite{jpsi1} one would think this
doesn't change the thermal behavior of the theory. However, notice
that now there is an extra non-constant $B_{\rm NS}$ field that cannot be
gauged away. This means when we write down the Nambo-Goto action for the strings,
the effect of the $B_{\rm NS}$ field can no longer be ignored and it'll definitely change the
criteria of the confinement/deconfinement transition studied in \cite{jpsi1, Mia:2010zu}.
This shouldn't be surprising because one would expect the thermal behavior of the dipole deformed quarks to
be different from the un-deformed ones.

\section{Background chemical potential and backreactions \label{chap3}}

After carefully working out all the details about the background, it is now time to study various applications.
In earlier papers \cite{Mia:2009wj, jpsi1, Mia:2010zu, Mia:2011iv} we managed to study aspects of phase transitions, quarkonium meltings,
viscosity to entropy ratio and many other things despite lacking a precise knowledge
of the background. Of course our ignorance about the
precise details of the background, first proposed in \cite{Mia:2009wj}, wasn't quite a handicap as the knowledge of the
{\it existence} of an UV completion
was enough to get results
for many of the above mentioned computations. For those computations that needed specific details, for example
the question of how much we deviate
from the well-known viscosity to entropy bound, were left in their functional forms. Once precise background informations become available, these
functional forms may be replaced by their exact values.

In the following sections we will however {\it not} re-address those questions here. Instead we will ask what happens if we switch on a
chemical potential in our theory. In fact, even before we study the consequences of switching on a chemical potential, we should ask {\it how}
to generate a chemical potential in our set-up. This is unfortunately not so simple as the AdS case. Part of the reason is that,
a-priori there are two known ways by which we could generate chemical potential: one via duality chasing, and the other via the D7-brane gauge fields.
Both give different result, and we will argue below that the former, i.e the duality chasing method, is inherently flawed due to certain subtleties
associated with the curvature of the background. The latter will turn out to be more useful.

The second reason that makes the story a little more involved has to do with the underlying RG flows in the dual gauge theory.
Due to this, we need to understand the effect of the chemical potential
at a {\it given} renormalization scale. In the gravity side, this means that we have to study the background at a chosen radial coordinate.  We will
discuss this below, and also point out the subtleties with the duality chasing idea that make it difficult to use this as a viable method
of generating chemical potential in our theory. But first let us give a brief discussion of the concept of chemical potential in thermal
field theories.

\subsection{Chemical potential in thermal field theories \label{chap3.1}}

Chemical potential is associated with conserved current. Once there
is a conserved charge $Q=\int  j^0 d^3x$, there is a chemical
potential $\mu$ conjugate to $Q$, and the grand partition function
for the thermal field becomes
\begin{eqnarray}
Z&=&\textrm{Tr} e^{-\beta(H-\mu Q)}\nonumber\\
&=&\int D\psi \exp\Big[-\int_0^{\beta} d^4x (\mathcal{L}_E-\mu
j^0)\Big]
\end{eqnarray}
where $H$ is the Hamilton and $\mathcal{L}_E$ is the Lagrangian
density after Wick rotation. We have omitted other gauge group
indices for simplification.

In the field theory the only field naturally couples to $j^0$ is
$A_0$ so the chemical potential in the thermal field theory is
defined as the time component of the gauge field $A_0$. Loosely
speaking, according to the AdS/CFT correspondence this is the value
of the bulk gauge field $A_t(r)$ at boundary, i.e. radial infinity.
If we have a gauge field $A_t$ in the bulk its value at
infinity will introduce a chemical potential associated with some
conserved charge for the gauge theory living on the boundary. In the following section we
will first try to generate such a bulk field by duality chasing.

A more interesting thing to do is to study the chemical potential
associated with the quark number. In our case we have $M$
coincident fractional D3 branes (in the far IR)
and $N_f$ coincident D7 branes so the $SU(M)$
gauge theory has a global $U(N_f)\simeq SU(N_f)\times U(1)$
symmetry. The fundamental matter $\psi$ and $\bar{\psi}$, which are
quarks, transform under the global $U(1)$ symmetry with charges $1$
and $-1$ respectively. Thus the $U(1)$ charge counts the number of
quark numbers $n_q$. In terms of partition function it can be
written as
\bg \label{part_func}
Z=\exp(-\beta W )=\exp\left[-\beta (H-\mu n_q)\right]
\nd
where $W$ is the Gibbs free energy which satisfies
\begin{eqnarray}\label{W2}
\frac{\delta W}{\delta \mu}=-n_q
\end{eqnarray}
From the string point of view, the Gibbs free energy is proportional
to the on shell D7 brane action,
\begin{eqnarray}\label{W3}
\delta W=\int_{r_0}^{\infty} d^4x ~dr ~d\Omega_3
\delta\mathcal{L}_E=\int_{r_0}^{\infty} d^4x ~dr~
d\Omega_3\frac{\delta\mathcal{L}_E}{\delta\partial_r
A_0(r)}\partial_r\delta A_0(r)
\end{eqnarray}
where $r_0$ is some scale that will be made precise later and $\Omega_3$ is the internal three-cycle along which D7 branes
are extended. On the other hand
$$\int d\Omega_3 \frac{\delta\mathcal{L}_E}{\delta\partial_r
A_0(r)}=-n_q$$ The reason is the following. The $U(1)$ field on the
D7-brane worldvolume can be thought of as arising from fundamental
strings dissolved into the D7-branes. Thus the density of the
quarks is precisely the density of the strings. Since the
fundamental strings source the NS two-form $B_{\rm NS} \equiv B_{0r}$, so its density can be
determined from the local charge density of the $B_{\rm NS}$-field.
Furthermore, the gauge invariance requires that the D7-brane action
only involves the combination $B_{0r}+2\pi l_s^2 F_{0r}$ so that
\begin{eqnarray}\label{cd4}
\int d\Omega_3 \frac{\delta\mathcal{L}_E}{\delta\partial_r A_0(r)}=-2\pi
l_s^2\int d\Omega_3 \frac{\delta\mathcal{L}_E}{\delta B_{0r}}=-n_q
\end{eqnarray}
Plugging \eqref{cd4} into \eqref{W3} we find
\begin{eqnarray}\label{W5}
\delta W=-n_q \left[\delta A_0(\infty)-\delta A_0(r_0)\right]
\end{eqnarray}
since $A_0(r_0)$ always vanishes, comparing \eqref{W5} and
\eqref{W2} we get $\mu=A_0(\infty)$. We
will calculate this chemical potential in the next subsection after we clarify the subtleties associated with duality chasing method to generate
chemical potential in our model.

\subsection{Chemical potential using duality chasing \label{chap3.2}}

The duality chasing idea is somewhat similar to the technique that we used to generate non-commutative and dipole theory on the D7-branes.
Instead of using shift, we will apply boost and then T-dualize. We call this technique as $TbT$ where $b$ stands for boost. The
boosting mixes $x_0$ with the internal coordinate, say $y^m$, to generate a $B_{\rm NS}$ field of the form $B_{0m}$. This means from the
five-dimensional point-of-view this would indeed be a gauge field $A_0 \equiv B_{0m}$. Since the duality chasing preserves\footnote{Both sides are
weakly coupled, and so no non-perturbative effects could enter here.}
EOMs, our method should give us a background with a vector potential $A_0$. Unfortunately, as we show below, this fails in a rather subtle way.

Our starting point would be to consider the metric and the fluxes presented in section 2 but ignoring the ${\cal O}(g_sM^2/N, r_h^4/r^4)$ corrections (this means we are
in the range \eqref{rrange}).
Since our background is divided into three Regions, namely
1,2 and 3, the easiest way would be to keep all the warp factors undetermined so that
after $TbT$ we can get the results for the three Regions simultaneously. This is easier said than done because the background also has $B_{\rm NS}$
fields that start decaying faster as we approach Region 3 \cite{jpsi1}.

To avoid these subtleties, let us study the effect of the $TbT$ operations only in Region 1. The $B_{\rm NS}$ field in Region 1 will be of the
form \eqref{bhusla}. However due to the scaling arguments given in \eqref{bhexp}, the $B_{\rm NS}$ field components take the form \eqref{B2}.
We can simplify this a little further by making the components independent of $\phi_2$ direction with a gauge transformation:
\begin{eqnarray}
{B}_{\rm now} ~= ~ B_2~ + ~ dC_1, \quad \textrm{with} \quad C_i=\int B_{\phi_2 i}~ d\phi_2
\end{eqnarray}
where $B_2$ is given in \eqref{B2} and ${B}_{\rm now}$ is the $B_{\rm NS}$ field that will enter the duality relations. Taking this into account,
after $TbT$ the metric will be:
\begin{eqnarray}\label{tbtmeto}
ds^2&=&\frac{1}{\sqrt{h}}\left(-\frac{\mathcal{A}g_1}{\mathcal{A}\cosh^2\beta-\frac{g_1}{\sqrt{h}}\sinh^2\beta}dt^2+dx_{123}^2\right)+\frac{\sqrt{h}}{g_1}dr^2\nonumber\\
&&+\sqrt{h}\,r^2\left[\frac{1}{9}-\frac{\sqrt{h}\,r^2}{81}\cos^2\theta_2\mathcal{A}\left(1-\frac{\mathcal{A}\cosh^2\beta}{\mathcal{A}\cosh^2\beta-\frac{g_1}{\sqrt{h}}\sinh^2\beta}\right)\right]
(d\psi+\cos\theta_1d\phi_1+\cos\theta_2d\phi_2)^2\nonumber\\
&&+\frac{\sqrt{h}\,r^2}{6}(d\theta_1^2+\sin^2\theta_1d\phi_1^2)
+\frac{\sqrt{h}\,r^2\,(1+F)(1+ {\cal G})}{6}\left(\frac{d\theta_2^2}{1+ {\cal G}}+\sin^2\theta_2 d\phi_2^2\right)\nonumber\\
&&+\frac{h\,r^4}{9}\mathcal{A}\left(1-\frac{\mathcal{A}\cosh^2\beta}{\mathcal{A}\cosh^2\beta-\frac{g_1}{\sqrt{h}}\sinh^2\beta}\right) \times \nonumber\\
&&\left[\frac{1}{9}\cos\theta_1\cos^3\theta_2
d\phi_1d\phi_2 -\left(\frac{1}{4}\sin^4\theta_2+\frac{1}{3}\sin^2\theta_2\cos^2\theta_2\right)d\phi_2^2\right]\nonumber\\
&&-\frac{\sqrt{h}\,r^2}{9}\left(1-\frac{\mathcal{A}\cosh\beta}{\mathcal{A}\cosh^2\beta-\frac{g_1}{\sqrt{h}}\sinh^2\beta}\right)\times\nonumber\\
&&\left[\frac{\sqrt{h}\,r^2\,(1+F)(1+{\cal G})}{6}\cos\theta_2\sin^2\theta_2\mathcal{A}d\psi d\phi_2
+\cos\theta_1\cos\theta_2d\phi_1\phi_2\right]\nonumber\\
&& +\frac{h\,r^4}{81}\frac{\mathcal{A}\cosh\beta
\,\cos^3\theta_2(1-\cosh\beta)}{\mathcal{A}\cosh^2\beta-\frac{g_1}{\sqrt{h}}\sinh^2\beta}\,d\psi
d\phi_2
\end{eqnarray}
where more details are given in {\bf Appendix \ref{TBT}}. The background now indeed has a gauge flux $A_0$ coming from \eqref{b2com} along with a
dilaton $\widetilde\phi$. The various components are given by:
\bg\label{varcoshay}
&&\mathcal{A} ~ = ~ \frac{18}{\sqrt{h}\,r^2\,\left[2\cos^2\theta_2 +3(1+F)(1+ {\cal G})\sin^2\theta_2\right]}\nonumber\\
&& \widetilde{\phi}~ = ~ \phi-\frac{1}{2}\ln
\left(\cosh^2\beta-\frac{g_1\sinh^2\beta}{\mathcal{A}\sqrt{h}}\right) \nonumber\\
&& A_0 ~ = ~ B_{0\phi_1} d\phi_1 ~ + ~ B_{0\phi_2} d\phi_2 ~ + ~ B_{0\psi} d\psi ~ + ~ B_{0\theta_1} d\theta_1
\nd
with $B_{0\phi_1}, B_{0\phi_2}, B_{0\psi}$ and $B_{0\theta_1}$ are the four-components given in \eqref{b2com}. There are also the RR fields that we
haven't discussed here. They do not mix with the metric and the $B_{\rm NS}$ fields and therefore do not change the chemical potential.
This is because the gauge field in \eqref{varcoshay} come from the cross-terms in the metric generated by the boost operation and not from the fluxes.

Unfortunately the solution presented in \eqref{varcoshay} and \eqref{tbtmeto} has numerous issues.
We notice that the dilaton becomes imaginary and $g_{tt}$ becomes
infinity and changes sign at $r=r_1$, where
\begin{eqnarray}\label{cdt}
r_1  ~= ~ \frac{1}{\sqrt{g_1}}\cdot \frac{3\sqrt{2} ~\coth~\beta}{\sqrt{2 \cos^2 \theta_2 + 3 (1+F)(1 + {\cal G})\sin^2\theta_2}}
\end{eqnarray}
and becomes negative when $r>r_1$. We will asume that $r_1 >> (N/M)^{1/4} r_h$. When
dilaton becomes imaginary that means the theory is not well defined
in this region and signals a serious problem for the $TbT$
done for this metric. Actually after the boost in type IIA the
metric already have $g_{tt}$ and $g_{\phi_2\phi_2}$ changing sign at
$r=r_1$ which means at that point $t$ becomes spacelike while
$\phi_2$ becomes timelike. In \cite{Hull:1998br} the authors
discussed the dimensional reduction on a timelike direction, but for our case its not clear
whether the dimensional reduction or T-dualities along
the half spacelike half timelike direction makes sense. Of course one might try to do the $TbT$ operation
along flat space-time directions but this will {\it not} generate any new
solutions. In the literature $TbT$ was generally done on asymptotically flat directions,
see for example \cite{Peet:2000hn} and references therein, however, when one
tries to do the $TbT$ operation along very curved directions as in our case, one will
generally get into the same problem as we do.

The above issues tell us that the duality chasing idea may be a futile endeavor to look for chemical potential from the gravity side. There exists a
much better way to determine the source of chemical potential involving considerable less effort. We turn to this in the following section.

\subsection{Chemical potential from D7 brane world-volume gauge theory \label{chap3.3}}

Another way to introduce chemical potential is to study the $U(1)$ gauge field on the D7 brane as in \cite{Mateos:2007vc} (see also \cite{cotbeta}).
To study this case thoroughly one has to take into account not only the backreaction of the D7 brane but also the backreaction
of the gauge field on the brane.

We first solve for the gauge field. To do this, we consider the solution from
\cite{Mia:2009wj} where we already took into account the backreactions
from the seven-brane as well as from
the fluxes and black hole. This way the metric ansatze will be more generic than what we took here. Of course in \cite{Mia:2009wj}, the
warp factor and the $h_i$ coeffcients of the internal metric were left undetermined, so the chemical potential that we will get using this background will be
left in terms of these coefficients. Our aim then would be to plug in the values of the coefficients that we determined here by considering the backreaction from
the black hole and compute a representative chemical potential.

Our starting point then is the metric\footnote{Note that in this section we are taking a metric ansatze more general than the one considered in \eqref{inmate}. Therefore
we will assume that the radial range is $r \ge r_h$ and not \eqref{rrange}. Of course the metric of the internal space could still change to ${\cal O}(r_h^4/r^4)$ but we
will not address the issue here. A more elaborate analysis will be presented elsewhere.}:
\begin{eqnarray}
ds_{10}^2&=&\frac{1}{\sqrt{h}}(-g_{tt}dt^2+dx_{123}^2)+\sqrt{h}(g_{rr}dr^2+r^2dM_5^2)
\end{eqnarray}
where $dM_5^2$ is the same as $ds^2_6$ in \eqref{tbtmetric}. We take the
simplest D7 brane embedding profile which is $\phi_2=0$,
$\theta_2=\theta_2(r)$, so the induced metric for D7 brane is
\begin{eqnarray}
ds_{8}^2&=&\frac{1}{\sqrt{h}}(-g_{tt}dt^2+dx_{123}^2)+\sqrt{h}\Big[(g_{rr}+r^2h_3h_4(\partial_r \theta_2)^2)dr^2+r^2h_1(d\psi+\cos\theta_1d\phi_1)^2\nonumber\\
&&+r^2h_2(d\theta_1^2+\sin^2\theta_1d\phi_1^2)+h_5\cos\psi(\partial_r \theta_2)drd\theta_1+h_5\sin\psi\sin\theta_1(\partial_r\theta_2)drd\phi_1\Big]\nonumber\\
\end{eqnarray}
In the UV region the NS and RR two-forms,
$B_2$ and $C_2$ respectively, can be neglected. If we allow the seven-brane embedding to be wholly in Region 3 of \cite{jpsi1},
then the scenario will be no
different from the ones in \cite{Mateos:2007vc}. Therefore by the same
argument in \cite{Mateos:2007vc} we can only turn on the $A_t$
component on the brane which will be a function of the radial coordinate. The DBI action for
the D7 brane is,
\begin{eqnarray}
\mathcal{I}&=&T_{D7}\int d^8x\, e^{-\phi}r^3\sqrt{h_1h_2}\sin\theta_1\Big[g_{tt}g_{rr}h_2+g_{tt}h_2\partial_r\phi^i\partial_r\phi^i\sqrt{h}+h_2F_{tr}^2\nonumber\\
&&+g_{tt}r^2(h_2h_3h_4-h_5^2/4)(\partial_r\theta_2)^2\Big]^{1/2}\label{DBI}
\end{eqnarray}
Neglecting the fluctuations in $\phi^i$ i.e keeping
$\partial_r\phi^i\partial_r\phi^i=0$, the equation of motion for
$A_t$ is
\begin{eqnarray}
\partial_r \,\left\{\frac{e^{-\phi}r^3\sqrt{h_1h_2}\,h_2\,\partial_r A_t}{ \sqrt{g_{tt}g_{rr}h_2+h_2F_{tr}^2+
g_{tt}r^2(h_2h_3h_4-h_5^2/4)(\partial_r\theta_2)^2}}\right\}=\partial_r
D=0
\end{eqnarray}
At $r\rightarrow \infty$, one can find $A_t=\mu+\frac{a}{r^2}$ where
$\mu$ is the chemical potential on the field theory side. And $\mu$
can be described as
\begin{eqnarray}\label{cpot}
\mu=\int_{r_0}^{\infty} dr
D\sqrt{\frac{g_{tt}g_{rr}h_2+g_{tt}r^2(h_2h_3h_4-h_5^2/4)(\partial_r\theta_2)^2}{e^{-2\phi}r^6h_1h_2^3-h_2D^2}}
\end{eqnarray}
where $r_0 >> r_h$ with $r_0$ to be a point in Region 3 determined by the minimum {\it dip} of the seven-brane in Region 3.
For simplicity, we have also assumed the chemical potential for $A_t(r_0)=0$.

Note that the chemical potential $\mu$ in \eqref{cpot} depends on the precise embedding $\theta_2(r)$. To eliminate $\theta_2(r)$ in \eqref{cpot} we need to
determine the equation of motion for this. For this we
first construct the Legendre transform of eq. \eqref{DBI} with respect
to $D$ to eliminate $A_t$ in the action. The new action is,
\begin{eqnarray}
\mathcal{I}=T_{D7}\int d^8x e^{-\phi}r^3\sin\theta_1\sqrt{\frac{h_1h_2\left(g_{tt}g_{rr}h_2+g_{tt}r^2(h_2h_3h_4-h_5^2/4)(\partial_r\theta_2)^2\right)}
{1-\frac{D^2}{e^{-2\phi}r^6h_1h_2^2}}}\nonumber\\
\end{eqnarray}
With this the equation of motion for $\theta_2(r)$ is,
\begin{eqnarray}
\partial_r\left(e^{-\phi}r^5\sqrt{\frac{h_1h_2}{1-\frac{D^2}{e^{-2\phi}r^6h_1h_2^2}}}\frac{g_{tt}(h_2h_3h_4-h_5^2/4)\partial_r\theta_2}{\sqrt{g_{tt}g_{rr}h_2+g_{tt}r^2(h_2h_3h_4-h_5^2/4)(\partial_r\theta_2)^2}}\right)\equiv\partial_r K=0\nonumber\\
\end{eqnarray}
After simplification we find that the chemical potential can be
expressed in terms of two constants $D$ and $K$ in the following way:
\begin{eqnarray}\label{chempot}
\mu=\int_{r_0}^{\infty} dr ~ {\cal F}_\mu\left(r; \{h_i\}\right)
\end{eqnarray}
where we have defined
\bg\label{frdefined}
{\cal F}_\mu\left(r; \{h_i\}\right) = \sqrt{\frac{g_{rr}g_{tt}^2(h_2h_3h_4-h^2_5/4)r^{14}h_1^2h_2^3D^2}{(r^6h_1h_2^2-D^2e^{2\phi})\Big(e^{-2\phi}g_{tt}(h_2h_3h_4-h^2_5/4)r^{14}h_1^2h_2^3-K^2(r^6h_1h_2^2-D^2e^{2\phi})\Big)}}\nonumber\\
\nd
The above formula for the chemical potential although elegant, is unfortunately not very illuminating for us because the warp factors $h_i$ are now all
in Region 3 whereas in this paper we have concentrated ourselves only in Region 1. It could be that the $h_i$ continue to remain the same from IR to UV, but
we don't have any {\it a-priori} information on this. In addition to that we expect the seven-brane to go all the way to $r = r_h$ in Region 1. Therefore a
more generic formula for the chemical potential, that takes into account various regimes of validity of our construction, can be expressed succinctly
as\footnote{In deriving this formula we have ignored the pull-backs from the NS and RR three-form fluxes from Regions 1 and 2. It will be interesting to see
how the chemical potential depends on these informations. Here, however, we will suffice ourselves with approximate estimates, and a more detailed exposition will
be addressed elsewhere.}:
\bg\label{cpnow}
\mu &= & \int_{r_h}^{\infty} dr ~ {\cal F}_\mu\left(r; \{h_i\}\right)\\
&= & \int_{r_h}^{r_{\rm min}} dr ~ {\cal F}_{\mu, 1}\left(r; \{h^{(1)}_i\}\right)
+ \int_{r_{\rm min}}^{r_0} dr ~ {\cal F}_{\mu, 2}\left(r; \{h^{(2)}_i\}\right)
+ \int_{r_0}^{\infty} dr ~ {\cal F}_{\mu, 3}\left(r; \{h^{(3)}_i\}\right)\nonumber
\nd
where $h_i^{(n)}$ are the warp factors of the internal space in Region $n$ with the corresponding function ${\cal F}_{\mu, n}\left(r; \{h^{(n)}_i\}\right)$.
One interesting thing about \eqref{cpnow}, and also of \eqref{chempot},
is that the warp factor $h$ doesn't appear in the final formula (at least for the embedding that we have chosen). We also note that, in the limit:
\bg\label{intlim}
K = 0, ~~~~~~~ \phi ~ \ll ~ {\rm log}\left(\frac{r^3h^{(n)}_2\sqrt{h^{(n)}_1}}{D}\right)
\nd
which means that the value of the dilaton at any point in the radial direction is bounded above by the log function,
the formula for the chemical potential drastically simplifies to:
\bg\label{chemchem}
\mu = D\left( \int_{r_h}^{r_{\rm min}} dr ~\frac{e^\phi}{r^3 h^{(1)}_2\sqrt{h^{(1)}_1}} + \int_{r_{\rm min}}^{r_0} dr ~\frac{e^\phi}{r^3 h^{(2)}_2\sqrt{h^{(2)}_1}} +
\int_{r_0}^{\infty} dr ~ \frac{e^\phi}{r^3 h^{(3)}_2\sqrt{h^{(3)}_1}}\right)
\nd
The actual value of the chemical potential in this limit now depends on the value of the dilaton in the three regions as well on the internal warp factors
$h^{(n)}_i$. To estimate the value of the chemical potential let us now consider a toy example where the dilaton is approximately constant
over the three regions and the internal warp factors remain
{\it same} in Regions 1, 2 and 3, i.e:
\begin{eqnarray}\label{simchoice}
&& e^\phi ~ \approx ~ g_s, ~~~~~~~~ D ~ \equiv ~ c_0 r_h^3\\
&&h^{(n)}_1=\frac{1}{9},\quad h^{(n)}_2=\frac{1}{6}, \quad h^{(n)}_3=\frac{1}{1+{\cal G}},
\quad h^{(n)}_4=\frac{(1+F)(1+ {\cal G})}{6}, \quad h^{(n)}_5=0\nonumber
\end{eqnarray}
where $c_0$ is a constant\footnote{$c_0$ can be thought of as a free parameter of our theory, not determined by the equation of motion. In
fact using (\ref{part_func}), one readily gets that $c_0\sim r_h^{-1}(W-H)/n_q$. Thus $c_0$ is related to quark density.} and the scaling of $D$ is motivated from \cite{Mateos:2007vc}. Plugging \eqref{simchoice} in \eqref{chemchem} we get our final
estimate for the chemical potential as:
\bg\label{finest}
\mu ~ = ~ -9c_0 r_h g_s
\nd
where the sign of $\mu$ will be determined by the sign of $c_o$. Note that our estimate says that the chemical potential is very small (provided
$\vert c_o\vert \ll 1$) and is proportional to $r_h$, the horizon radius. In the dual field theory, the temperature is related to the horizon
in the following way
\bg
T=\frac{g'(r_h)}{4\pi \sqrt{h(r_h)}}
\nd
This means that in terms of temperature, the chemical potential is given by
\bg \label{mu}
\mu= c_0 T r_h^2\sqrt{h(r_h)}
\nd
where we have taken the appoximation that $g(r)=1-r_h^4/r^4$. As seen from our numerical analysis, $g_0\ll 1$ and we can effectively ignore
the modification of $g(r)$ due to the presence of fluxes, while $A_0$ near the horizon is not negligible. Thus we have to consider
$h(r_h)=h^0(r_h)+h^1(r_h)$ in the above expression. From (\ref{mu}), we see that chemical potential is approximately linear in terms of
temperature, however as
\bg\label{linerh}
r_h^2 \sqrt{h(r_h)}~\sim ~ L^2 \sqrt{1+ \frac{3g_sM^2}{2\pi N} {\rm log}\left(\frac{r_h}{r_*}\right)+A_0(r_h)}
\nd
with $r_*$ being another relevant scale,
(\ref{mu}) also involes
terms $\sim {\rm log}(T)$. But this is precisely consistent with the logarithmic running of the gauge coupling with temperature and hence our
analysis reveals that chemical potential is sensitive to the underlying  structure of the gauge theory, which by construction is non-conformal.

\section{Discussions and Conclusions \label{cono}}

In this paper we have addressed one of the crucial issue left undetermined in our earlier papers
\cite{Mia:2009wj, jpsi1, Mia:2010zu, Mia:2011iv}, namely the backreactions from black hole, branes and fluxes on the background geometry and on the
various UV completions. We found that in certain limit \eqref{limits} the background EOMs allow for a perturbative suppressions \eqref{suppression}
under which the backreactions from branes and fluxes can be consistently ignored. This means we only had to study the backreaction from black hole on the
geometry, a task, which in the same limit, simplifies immensely to a set of three
second-order partial differential equations \eqref{BHfactorA}, \eqref{TrRmnA} and \eqref{EQA} provided we are in the radial range \eqref{rrange}.
Solutions to these equations are provided in {\bf Appendix}
\ref{bgcomp}. Of course the analysis that we present here is only meant for the IR regime of the theory, but it would probably be possible
to extend this
to Region 3 where UV caps were introduced in \cite{Mia:2009wj, jpsi1}. Two challenges still remain: one, to study the equations at $r = r_h$ in Region 1 and
two, to study them in the intermediate {\it buffering} region i.e Region 2. In both cases the analysis may get very involved because for the first case one
would now have to solve all the twenty internal Einstein's equations; and for the second case
the ($p, q$) five-brane sources and fluxes will further complicate the scenario. These details are left for future works.

We took few other directions in this paper too, many of which are {\it not} restricted by the constraint \eqref{rrange}.
We gave a detailed analysis of the backreactions on the fluxes from the black hole and the flavor seven-branes.
These backreactions tend to make the three-form fluxes non-ISD. Interesingly the effect of the horizon radius $r_h$ on the fluxes is, to the order that we
present here, implicit: the appearance is via the resolution parameter $F$ (or $a^2$). The fact that the resolution parameter would have dependence on the
black hole radius was anticipated in \cite{Mia:2009wj, jpsi1}. Here we confirm that prediction.

Another direction that we took here is related to the study of the running couplings. Our analysis predicts that the two couplings change under various
backreactions from the dual gravity side, but their RG runnings remain the {\it same} as for the non-thermal case. We believe this should be the correct way
to interpret thermal beta functions. An alternative interpretation where the thermal beta function is assumed to be {\it different} from the non-thermal case
is probably not a very meaningful conclusion.

Our final analysis, ignoring the detour that we took to study dipole deformed quarks, is the study of chemical potential. We showed how the chemical potential
should be studied in a theory with an underlying RG flow. Our analysis yields, by ignoring certain subtleties, a result of the form
\eqref{cpnow}. Under simplifying assumtions this gives us \eqref{finest} which turns out to be a reasonably good estimate. A full analysis however will
require us to compute, among other things, the internal warp factors $h_i^{(n)}$. Furthermore, going beyond $g_sN_f$ and $g_sM^2/N$ orders, and estimating the
background would yield interesting results for the IR regime of large $N$ thermal QCD. All of these are of course very challenging questions but, as we saw
in this paper, there may exist corners of the solution spaces where one may find unexpected simplifications. It is worth at least, if not for anything else, to
look for these regimes of simplicity in a subject that is notorious for its unyielding complexity.


\vskip.2in

\centerline{\bf Acknowledgement}

\noindent We would like to thank Miklos Gyulassy and
Guy David Moore for helpful discussions. The work of M. M is supported in part by the Office of Nuclear Science of the US
Department
of Energy under grant No. DE-FG02-93ER40764,
the work of K. D is supported in part by NSERC grant, the work of F.C and P. F is
supported in parts by Schulich scholarships and FQRNT grants. S. V would like to thank the High
Energy Group at McGill University for financial support.


\appendix

\section{An exercise in duality chasing \label{TBT}}

In this section our aim would be to understand how to switch on a gauge field, $A_0$, along a five-dimensional space in type IIB string theory.
The five-dimensional space will be the dual
gravitational description of a gauge theory with running couplings. This would mean that all the type IIB fields, namely the three and the five-forms
including the axion-dilaton are all switched on. We will generate this solution using {\it two} stages of duality transformations. The first stage of duality
transformations would convert a simple torsional background to another background that has both the three-forms including five-forms and axion-dilaton switched on.
The second stage of duality transformations would further convert this background to the one that we want.

Our starting point would then be to take the following torsional background:
\bg\label{tbtmetric}
ds_{10}^2=h^{\frac{1}{2}}e^{\phi}h_0ds_{4}^2+h^{-\frac{1}{2}}e^{\phi}ds_6^2
\nd
where $\phi$ is the dilaton and we have defined $ds^2_6$ and $ds^2_4$ in the following way:
\bg\label{ds6ds4}
&&ds_4^2 = -g_{tt} dt^2 + dx^2_{123}\nonumber\\
&&ds_6^2=h_1dr^2+h_2(d\psi+\cos\theta_1d\phi_1+\cos\theta_2d\phi_2)^2+h_{3}(d\theta_1^2+\sin^2\theta_1d\phi_1^2)
+h_{4}(h_6\,d\theta_2^2+\sin^2\theta_2d\phi_2^2)\,\nonumber\\
&&\quad\quad\quad+h_5\cos\psi\,(d\theta_1d\theta_2-\sin\theta_1\sin\theta_2d\phi_1\phi_2)+h_5\sin\psi\,(\sin\theta_1d\theta_2d\phi_1-\sin\theta_2d\theta_1d\phi_2)\quad\quad\quad\quad
\end{eqnarray}
with ($h, h_i$) being the required warp factors, and the torsion $H_3$ is given by the usual relation:
\begin{eqnarray}\label{torsion}
H_3=e^{2\phi}\star d(e^{-2\phi}J)
\end{eqnarray}
$J$ is the fundamental form that we will compute separately in {\bf Appendix} \ref{funJ}. Now using our first stage of duality transformations,
as described in details in \cite{Chen:2010bn}, we can easily get the following background from \eqref{tbtmetric}, \eqref{torsion} and the dilaton $\phi$:
\begin{eqnarray}\label{mmbackground}
&&F_3=h\,h_0^2\cosh\beta e^{2\phi}\star d(e^{-2\phi} J),\nonumber\\
&&H_3=-h\,h_0^2\sinh\beta e^{2\phi}d(e^{-2\phi}J), \nonumber\\
&&F_5=-\frac{1}{4}(1+\star)dA_0\wedge dx^0\wedge dx^1 \wedge dx^2
\wedge dx^3, \nonumber\\
&&ds^2=h_0ds_{4}^2+h_1dr^2+h_2(d\psi+\cos\theta_1d\phi_1+\cos\theta_2d\phi_2)^2+h_{3}(d\theta_1^2+\sin^2\theta_1d\phi_1^2)
+h_{4}(h_6\,d\theta_2^2\,\nonumber\\
&&\quad\quad\quad+\sin^2\theta_2d\phi_2^2)+h_5\cos\psi\,(d\theta_1d\theta_2-\sin\theta_1\sin\theta_2d\phi_1\phi_2)+h_5\sin\psi\,(\sin\theta_1d\theta_2d\phi_1-\sin\theta_2d\theta_1d\phi_2)\quad\nonumber\\
\end{eqnarray}
For this solution to be supersymmetric, the minimum requirement, in the {\it absence} of a black hole, is $h\,h_0^2e^{2\phi}=1$. Using this we can compute the
NS $B$-field from the three-form $H_3 = dB_2$ as:
\begin{eqnarray}
B_2&=&-e^{-2\phi}\sinh\beta J\nonumber\\
&=&-e^{-2\phi}\sinh\beta (\sqrt{F_1F_2}dr\wedge (d\psi+\cos\theta_1d\phi_1+\cos\theta_2d\phi_2)\nonumber\\
&&\,+AB\sin\psi\sin\theta_1\sin\theta_2(w+za)d\phi_1\wedge d\phi_2-ABz\sin\theta_1d\phi_1\wedge d\theta_1\nonumber\\
&&\,+AB\cos\psi\sin\theta_2(w+za)d\theta_1\wedge d\phi_2\nonumber\\
&&\, -[ABa(w+za)-C\sqrt{F_4-a^2A^2}(z-wa)]\sin\theta_2d\theta_2\wedge d\phi_2\nonumber\\
&&\,-[ABaz-C\sqrt{F_4-a^2A^2}w]\cos\psi\sin\theta_1d\theta_2\wedge d\phi_1\nonumber\\
&&\,+[ABaz-C\sqrt{F_4-a^2A^2}w]\sin\psi d\theta_2\wedge d\theta_1])
\end{eqnarray}
where $A, B, C, a, w, z, J$ are worked out in {\bf Appendix} \ref{funJ}.

In our second stage of duality on \eqref{mmbackground},
we will do a T-duality followed by a Boost and then another T-duality ($TbT$) along $\phi_2$ direction. To make the $TbT$ operation simpler,
we can do a gauge transformation to eliminate the $B_{\phi_1 i}$ components in the following way:
\begin{eqnarray}
\widetilde{B}_2=B_2+dC_1,\quad \textrm{with} \quad C_i=\int B_{\phi_2 i}\,d\phi_2
\end{eqnarray}
Now that the $B_2$ field has no $B_{\phi_2 i}$ components, $TbT$ is straightforward. The resulting metric is:
\begin{eqnarray}\label{mettbt}
ds^2&=&h_0\left(-\frac{\mathcal{A}g_{tt}}{\mathcal{A}\cosh^2\beta-h_0g_{tt}\sinh^2\beta}dt^2+dx_{123}^2\right)+h_1dr^2\nonumber\\
&&+\left(h_2-h_2^2\cos^2\theta_2\mathcal{A}(1-\frac{\mathcal{A}\cosh^2\beta}{\mathcal{A}\cosh^2\beta-h_0g_{tt}\sinh^2\beta})\right)
(d\psi+\cos\theta_1d\phi_1+\cos\theta_2d\phi_2)^2\nonumber\\
&&+h_{3}(d\theta_1^2+\sin^2\theta_1d\phi_1^2)
+h_{4}(h_6\,d\theta_2^2+\sin^2\theta_2 d\phi_2^2)\nonumber\\
&&+\mathcal{A}\left(1-\frac{\mathcal{A}\cosh^2\beta}{\mathcal{A}\cosh^2\beta-h_0g_{tt}\sinh^2\beta}\right)\left(h_2h_5\sin\psi\cos\theta_2\sin\theta_2d\psi d\theta_1
\right.\nonumber\\
&&-\frac{h_5^2\sin^2\psi\sin^2\theta_2}{4}d\theta_1^2-\left.\frac{h_5^2\cos^2\psi\sin^2\theta_1\sin^2\theta_2-h_2h_5\cos\psi\sin\theta_1\cos\theta_1\sin\theta_2\cos\theta_2}{4}d\phi_1^2\right.\nonumber\\
&&-(h_4^2\sin^4\theta_2+2h_2h_4\sin^2\theta_2\cos^2\theta_2)d\phi_2^2-(2h_2\cos\theta_1\cos\theta_2-h_5\cos\psi\sin\theta_1\sin\theta_2)\nonumber\\
&&\left.\times \frac{h_5\sin\psi\sin\theta_2}{2}d\phi_1d\theta_1+h_2h_5\cos\psi\sin\theta_1\sin\theta_2\cos\theta_2d\psi d\phi_1+h_2^2\cos\theta_1\cos^3\theta_2 d\phi_1d\phi_2\right)\nonumber\\
&&-\left(1-\frac{\mathcal{A}\cosh\beta}{\mathcal{A}\cosh^2\beta-h_0g_{tt}\sinh^2\beta}\right)(h_2\cos\theta_1\cos\theta_2d\phi_1\phi_2+h_2h_4\cos\theta_2\sin^2\theta_2\mathcal{A}d\psi d\phi_2)\nonumber\\
&&-\frac{\mathcal{A}\cosh\beta}{\mathcal{A}\cosh^2\beta-h_0g_{tt}\sinh^2\beta}\Big(h_5\cos\psi\sin\theta_1\sin\theta_2d\phi_2d\phi_1+h_5\sin\psi\sin\theta_2d\phi_2d\theta_1 \Big)\nonumber\\
&&+\frac{\mathcal{A}\cosh\beta}{\mathcal{A}\cosh^2\beta-h_0g_{tt}\sinh^2\beta}h_2^2\cos^3\theta_2(1-\cosh\beta)d\psi d\phi_2+h_5(\cos\psi d\theta_1d\theta_2+\sin\psi\sin\theta_1 d\theta_2 d\phi_1)\nonumber\\
\end{eqnarray}
where $\beta$ is the boost parameter, and we have defined $\mathcal{A}$ as:
\bg\label{lvarga}
\mathcal{A} ~= ~ \frac{1}{h_2\cos^2\theta_2+h_4\sin^2\theta_2}
\nd
The metric \eqref{mettbt} is accompanied by all the type IIB RR and NS form fields. These can be easily worked out. Here we will just quote the $B_2$ field and the
dilaton $\widetilde\phi$:
\begin{eqnarray}\label{b2com}
&&\widetilde{\phi}=\phi-\frac{1}{2} {\rm log}(\cosh^2\beta-h_0g_{tt}\sinh^2\beta) + \frac{1}{2}{\rm log}~\mathcal{A}\nonumber\\
&&B_{0\phi_2}=\frac{(\mathcal{A}-h_0g_{tt})\sinh\beta\cosh\beta}{\mathcal{A}\cosh^2\beta-h_0g_{tt}\sinh^2\beta},\quad\quad\quad B_{0\psi}=\mathcal{A}h_2\cos\theta_2\sinh\beta\nonumber\\
&&B_{0\phi_1}=\frac{\mathcal{A}}{2}(2h_2\cos\theta_1\cos\theta_2-h_5\cos\psi\sin\theta_1\sin\theta_2)\sinh\beta,\, B_{0\theta_1}=\frac{\mathcal{A}}{2}h_5\sin\psi\sin\theta_2\sinh\beta  \nonumber\\
\end{eqnarray}
It is now easy to see what the five-dimensional gauge-field would look like. Its simply the following wedge product:
\bg\label{a0wedge}
A_0~ \equiv ~ B_{0\phi_2} d\phi_2 +  B_{0\psi} d\psi + B_{0\phi_1} d\phi_1 + B_{0\theta_1} d\theta_1
\nd
Notice that for the metric and dilaton to be regular we require the metric component $g_{tt}$ and the warp factors to satisfy:
\begin{eqnarray}
h_0g_{tt} (h_2\cos^2\theta_2+h_4\sin^2\theta_2)\leqslant \coth^2\beta
\end{eqnarray}
If we demand this requirement to hold for all the values of $\theta_2$ and $\beta$, then we need the following two scenarios:
\begin{eqnarray}
&& \textrm{If}\quad h_4>h_2,\quad \textrm{then}\quad h_0h_4g_{tt}\leqslant 1 \nonumber\\
&& \textrm{If}\quad h_2>h_4,\quad \textrm{then}\quad h_0h_2g_{tt}\leqslant 1
\end{eqnarray}

\newpage

\section{The fundamental form $J$ \label{funJ}}
We first define the vielbeins as follows,
\begin{eqnarray}
&&e^1=\sqrt{h_1}dr,\quad\quad e^5=\sqrt{h_4h_6-a^2A^2}d\theta_2\nonumber\\
&&e^2=\sqrt{h_2}(d\psi+\cos\theta_1d\phi_1+\cos\theta_2d\phi_2)\nonumber\\
&&e^3=A(\sin\psi\sin\theta_1d\phi_1+\cos\psi d\theta_1-a d\theta_2)\nonumber\\
&&e^4=B\Big(w\sin\theta_2d\phi_2+z(\cos\psi\sin\theta_1 d\phi_1-\sin\psi d\theta_1+a \sin\theta_2 d\phi_2)\Big)\nonumber\\
&&e^6=C\Big(z\sin\theta_2 d\phi_2-w(\cos\psi\sin\theta_1 d\phi_1-\sin\psi d\theta_1+a\sin\theta_2 d\phi_2)\Big)
\end{eqnarray}
where $A$, $B$, $C$, $a$, $w$, $z$ must satisfy,
\begin{eqnarray}
&&A^2=h_3, \quad a=-\frac{h_5}{2h_3},\quad B^2zw+C^2zw=\frac{3}{2}h_5 \nonumber\\
&&B^2z^2+C^2w^2=h_3, \quad B^2z-C^2w=h_3, \quad B^2w^2+C^2z^2=h_4-\frac{h_5^2}{4F_3}
\end{eqnarray}
The fundamental form $J$ is,
\begin{eqnarray}
J&=&e^1\wedge e^2+e^3\wedge e^4+e^5\wedge e^6\nonumber\\
&=&\sqrt{h_1h_2}dr\wedge (d\psi+\cos\theta_1d\phi_1+\cos\theta_2d\phi_2)\nonumber\\
&&\,+AB\sin\psi\sin\theta_1\sin\theta_2(w+za)d\phi_1\wedge d\phi_2-ABz\sin\theta_1d\phi_1\wedge d\theta_1\nonumber\\
&&\,+AB\cos\psi\sin\theta_2(w+za)d\theta_1\wedge d\phi_2\nonumber\\
&&\, -[ABa(w+za)-C\sqrt{h_4h_6-a^2A^2}(z-wa)]\sin\theta_2d\theta_2\wedge d\phi_2\nonumber\\
&&\,-[ABaz-C\sqrt{h_4h_6-a^2A^2}w]\cos\psi\sin\theta_1d\theta_2\wedge d\phi_1\nonumber\\
&&\,+[ABaz-C\sqrt{h_4h_6-a^2A^2}w]\sin\psi d\theta_2\wedge d\theta_1
\end{eqnarray}

\newpage

\section{More details on the squashing and the warp factor computations \label{bgcomp}}
 In the absence of blackhole that is $e^{2B}=1$, the warp factor
 $\alpha=e^{4A}=1/h$ only depends on $r$,
 $\theta_1$ and $\theta_2$ even when D7 back reaction is taken into account by considering the running axion-dilaton field \cite{Ouyang:2003df}. In this extremal limit $r_h=0$, we have ISD
 three-form fluxes $G_3$ and the internal metric $\widetilde{g}_{mn}$ describes a Ricci flat deformed cone.
 For the non-extremal case, we will demand similar behavior for the warp factor $h$  and will find that such solutions do exist.
Using $h\equiv h(r,\theta_1,\theta_2)$ only, in the non-extremal case we get,

\bg \label{F_5e1}
d\widetilde{F}_5&=&d\left(\left[\partial_rh \;\zeta +e^{-2B} \left(\bar{g}^{\theta_1\theta_1}\partial_{\theta_1}
h\;\eta_1+\bar{g}^{\theta_2\theta_2}\partial_{\theta_2} h\;\eta_2\right) \right]\frac{r^5(1+F+ {\cal G}/2)}{108} {\rm sin}\theta_1 {\rm sin}\theta_2\right)\nonumber\\
&\equiv& d {\cal D}
\nd
where we have used our metric ansatz (\ref{bhmet1}, \ref{inmate}) and
definition of the five-form flux (\ref{F5}). We have only kept
linear
terms in $F, {\cal G}$ and this is justified as we look for solutions $F,
{\cal G}\ll 1$ and ignore terms higher order term. In the above we
have also defined
\bg\label{bargpq}
\bar{g}_{pq}=e^{-2B}\widetilde{g}_{pq}
\nd
where $p,q$ run over the compact directions and thus (\ref{bargpq}) implies $\bar{g}_{pq}$ is indepndent of $B$. Here
$\zeta,\eta_i$ are five-forms given by
\bg
\zeta&=& d\psi\wedge d\phi_1\wedge d\phi_2\wedge d\theta_1\wedge d\theta_2\nonumber\\
\eta_1&=&d\psi\wedge dr\wedge d\phi_1\wedge d\phi_2\wedge d\theta_2\nonumber\\
\eta_2&=&d\psi\wedge dr\wedge d\phi_1\wedge d\phi_2\wedge d\theta_1
\nd

 Just like the  extremal case, we will assume that
$\partial_{\theta_i} h\sim {\cal O}(g_s^2N_fM^2/N)$ and we will find that this choice is consistent with all the Einstein equations and equations
for the fluxes  . With this assumption, we readily get
up to  ${\cal O}(g_s M^2/N)$, and ignoring ${\cal O}(F,G,{\cal G} ){\cal O}(g_s
M^2/N) $ (since $F,G, {\cal G}\ll 1 $)
\bg
\bar{g}^{\theta_i\theta_i}\partial_{\theta_i} h= \bar{g}_0^{\theta_i\theta_i}\partial_{\theta_i} h
\nd
where $i=1,2$ and $\bar{g}_0^{pq}$ is zeroth order in $M,N_f$. But at zeroth order in $M,N_f$, the compact five dimensional internal space
${\cal M}_5$ is exactly the deformed cone and thus $\bar{g}_0^{pq}$ is precisely the metric of deformed $T^{1,1}$.
Our ansatz for the black hole factor $e^{2B}$ is given in (\ref{BHfactora})
where $G$ is at least ${\cal O}(M/N,g_sM^2/N,g_sN_f)$. This is a sufficient condition as in the absence of five-branes and seven branes,
we have  $AdS\times T^{1,1}$ with black hole  where  $e^{2B}=1-\frac{\bar{r}_h^4}{r^4}$ and $\bar{r}_h\gg b$. This is because for large $r$, the deformed
cone becomes the regular cone and considering $\bar{r}_h\gg a$, we are effectively  putting a black hole in a regular cone. In other words, the
non-extremal limit of the geometry only `sees' the regular cone and deformation of the cone is hidden behind  the black hole horizon. This
also means our non-extremal solution is valid only for large horizon,  that is the non-extremal solution only captures the large temperature
deconfined chirally symmetric phase of the gauge theory. The extremal solution without any black hole   is dual to the confined
phase.

Now using (\ref{BHfactora}) in  ${\cal D}$ reads
\bg \label{D}
{\cal D}&=&\left[\partial_r h\; \zeta +\frac{1}{1-\frac{r_h^4}{r^4}} \left(\bar{g}^{\theta_i\theta_i}_0\partial_{\theta_i}h \;\eta_i\right)\right]
r^5 \frac{{\rm sin}\theta_1 {\rm sin}\theta_2}{108} \left(1+F+G/2\right)\nonumber\\
&=& \left[\partial_r h^0 \;\zeta+\bar{g}^{\theta_i\theta_i}_0\partial_{\theta_i}h^0 \;\eta_i\right]r^5 \frac{{\rm sin}\theta_1 {\rm
sin}\theta_2}{108}
+\Bigg[\partial_r h^1\;\zeta+ \frac{1}{1-\frac{r_h^4}{r^4}}\bar{g}_0^{\theta_i\theta_i}\partial_{\theta_i} h^1 \;\eta_i\nonumber\\
&+& \frac{r_h^4/r^4}{1-\frac{r_h^4}{r^4}}\bar{g}_0^{\theta_i\theta_i}\partial_{\theta_i} h^0 \;\eta_i\Bigg]r^5 \frac{{\rm sin}\theta_1 {\rm
sin}\theta_2}{108} +r^5 \frac{\left(F+G/2\right) {\rm sin}\theta_1 {\rm sin}\theta_2}{108}\partial_r h^0 \;\zeta
\nd
where we have only considered up to ${\cal O}(g_sM^2/N)$ terms. Here $h^0$ is the Ouyang solution and $h^1$ is the correction due to the black hole
which alters the internal compact space ${\cal M}_5$.
But the Ouyang solution satisfies Bianchi identity exactly as:
\bg
d\left[\left(\partial_r h^0 \zeta+\bar{g}^{\theta_i\theta_i}_0\partial_{\theta_i}h^0\eta_i\right)r^5 \frac{{\rm sin}\theta_1 {\rm
sin}\theta_2}{108}\right]= H_3^{(0)}\wedge F_3^{(0)}
\nd
Using this in (\ref{bianchi}) we get
\bg
d\left[\Bigg(\partial_r h^1\zeta+ \frac{1}{g}\bar{g}_0^{\theta_i\theta_i}\partial_{\theta_i} h^1 \eta_i
+ \frac{r_h^4/r^4}{g}\bar{g}_0^{\theta_i\theta_i}\partial_{\theta_i} h^0\eta_i\Bigg)r^5 \frac{{\rm sin}\theta_1 {\rm
sin}\theta_2}{108} +r^5 \frac{\left(F+G/2\right) {\rm sin}\theta_1 {\rm sin}\theta_2}{108}\partial_r h^0\zeta\right]=0\nonumber
\nd
which gives us (\ref{EQA}). The derivations of (\ref{BHfactorA}), (\ref{TrRmnA}) and (\ref{GKP_BH_M^2A}) have already been discussed in section
2.2.

We will now solve the four equations  (\ref{EQA}), (\ref{BHfactorA}), (\ref{TrRmnA}) and (\ref{GKP_BH_M^2A}) by ignoring all terms of ${\cal
O}(g_s N_f)$. In this limit, all angular dependences vanish and all the functions $A,F$ and $G$ are only functions of the radial coordinate $r$.
This also means we are ignoring the back reaction of the seven branes and our solution should be considered as the non-extremal generalization
of Klebanov-Strassler theory with modified UV behavior. For $N_f=0$, with $e^{-4A}=h=h^0+h^1$, we take the following ansatz
\bg \label{ansatz-1}
h^1&=& \frac{L^4}{r^4}\left(A_0(r)+A_1(r) {\rm log}~r+A_2(r) {\rm log}^2r\right)\nonumber\\
e^{2B}&\equiv&g=1-\frac{r_h^4}{r^4} +g_0(r)+ g_1(r) {\rm log}~r+g_2(r) {\rm log}^2r\nonumber\\
F&=& F_0(r)+F_1(r) {\rm log}~r+F_2(r){\rm log}^2r\nonumber\\
\nd
With our ansatz, only taking up to linear order terms in $A_i,F_i $ and $g_i$
one obtains that the equation derived from (\ref{GKP_BH_M^2A}) is
trivial. Also up to linear order, $A_1=A_2=F_1=F_2=g_1=g_2=0$ is a solution
with $A_0, F_0, g_0$ being the only non-trivial functions.  The equations resulting from
(\ref{EQA}),(\ref{BHfactorA}) and (\ref{TrRmnA}) are as follows
\bg\label{Bianchi_an}
&&(i) ~ r A_0'' - 3  A_0' -4 F_0'=0\nonumber\\
&&(ii) ~ 5 r^4 g_0'+ 4 \bar{r}_h^4F_0'
+ r^5 g_0''=0\nonumber\\
&&(iii)~  \frac{6 g_sM^2}{N\pi} \bar{r}_h^4+56 r^4g_0+16r^4 F_0+4r \bar{r}_h^4A_0'+49 r^5 g_0'+24 r^5 F_0'+12 r r_h^4 F_0' \nonumber\\
&&~~~~ +7r^6 g_0''+4r^6F_0''-4r^2r_h^4 F_0''=0
\nd

To solve these second order differential equations, all we need to do now is specify the boundary conditions. As we have second order
differential equations, we can choose two boundary conditions. A priori we do not know where the horizon is, that is we do not $r_h$ such that
$e^{2B(r_h)=0}$, so we cannot specify the boundary condition at the horizon. Additionally we cannot take
$r$ to be smaller than the range \eqref{rrange}. However, since we are looking for solution such that asymptotically we
recover the extremal geometry, we can impose the following boundary conditions:
\bg \label{bcond1}
&&{\rm lim}_{r\rightarrow \infty}\; A_0(r)=0\nonumber\\
&&{\rm lim}_{r\rightarrow \infty}\; g_0(r)=0\nonumber\\
&&{\rm lim}_{r\rightarrow \infty}\; F_0(r)=0
\nd
From the form of the equations in (\ref{Bianchi_an}), we see that  inverse power series in $r$ is a
possible candidate for the solutions that obey the boundary conditions (\ref{bcond1}). On the other hand, as already discussed in section 2.1,
we expect $A_i, g_i$ and $F_i$ to be proportional to the horizon $r_h$. Thus our anstaz is
\bg \label{Ansatz}
&&A_0(r)=\bar{a}_k^0\left(\frac{r_h}{r}\right)^k, ~~~~~F_0(r)=\bar{f}^0_k\left(\frac{r_h}{r}\right)^k\nonumber\\
&&g_0(r)=\bar{\zeta}^0_k\left(\frac{r_h}{r}\right)^k,
\nd
where $\bar{a}_k^0,\bar{f}^0_k$ and $\bar{\zeta}^0_k$ are  atleast ${\cal O}( M/N, g_s M^2/N)$, and the radial coordinate $r$ is
assumed in the range \eqref{rrange}. The boundary condition (\ref{bcond1}) implies
\bg
\bar{a}_0^0=\bar{f}_0^0=\bar{\zeta}_0^0=0
\nd
We can further choose three other boundary conditions. Again since (a) we do not know where the horizon is and (b) the radial coordinate is constrained by \eqref{rrange},
we will choose the following boundary conditions:
at $r=\infty$ and choose
\bg \label{bcond2}
&&{\rm lim}_{r\rightarrow \infty}\; A_0'(r)=0\nonumber\\
&&{\rm lim}_{r\rightarrow \infty}\; g_0'(r)=0\nonumber\\
&&{\rm lim}_{r\rightarrow \infty}\; F_0'(r)=0
\nd
which is automatically solved by our ansatz (\ref{Ansatz}).
With the set of boundary conditions (\ref{bcond1}) and (\ref{bcond2}), we solve (\ref{Bianchi_an})
numerically. The exact solution (whose validity should be considered for $r > (N/M)^{1/4} r_h$)
is plotted in {\bf Figures 1, 2, 3}. Observe that the numerical solutions are consistent with the analytic behavior in
(\ref{Ansatz}). When $M=0$, equations  (\ref{Bianchi_an}) imply that we have the trivial solution, i.e. $A_0=g_0=F_0=0$.
But since $M\neq 0$, we must have non-trivial solutions to satisfy  (\ref{Bianchi_an}).

\end{document}